\documentclass[preprints, article, accept, moreauthors, pdftex]{Definitions/mdpi} 

\usepackage{aas_macros}
\usepackage{subcaption}
\usepackage{longtable}
\usepackage{lscape}
\usepackage{array}
\usepackage{booktabs}

\usepackage{makecell}

\graphicspath{{./}{figs/}}

\newcommand{\nuc}[2]{$^{#1}$#2}

\newcommand{\ttherm}[1]{$T_{\rm therm}^{#1}$}

\history{}

\pdfoutput=1

\Title{Sensitivity of Neutron-Rich Nuclear Isomer Behavior to Uncertainties in Direct Transitions\footnote{Intended for unlimited release under LA-UR-21-21559.}}

\newcommand{\orcidGWM}{0000-0002-0637-0753} % Misch
\newcommand{\orcidTMS}{0000-0002-4375-4369} % Sprouse
\newcommand{\orcidMRM}{0000-0002-9950-9688} % Mumpower
\newcommand{\orcidAC}{0000-0002-0861-3616}  % Couture
\newcommand{\orcidCLF}{0000-0002-7827-2247} % Fryer
\newcommand{\orcidBSM}{0000-0001-6307-9818} % Meyer
\newcommand{\orcidYS}{0000-0002-1411-4135}  % Sun

\Author{
G.~Wendell~Misch $^{1,2}$\orcidGWM{}, 
Trevor~M.~Sprouse $^{1}$\orcidTMS{}, 
Matthew~R.~Mumpower $^{1,2}$\orcidMRM{}, 
Aaron~Couture$^{3}$\orcidAC, 
Chris~L.~Fryer$^{2,4}$\orcidCLF{}, 
Bradley~S.~Meyer$^{5}$\orcidBSM{}, and 
Yang~Sun$^{6}$\orcidYS{}
}

\AuthorNames{G.~Wendell~Misch, Trevor~M.~Sprouse, Matthew~R.~Mumpower, Aaron~Couture, Chris~L.~Fryer, Bradley~S.~Meyer, and Yang~Sun}

\address{%
$^{1}$ \quad Theoretical Division, Los Alamos National Laboratory, Los Alamos, NM, 87545, USA \\
$^{2}$ \quad Center for Theoretical Astrophysics, Los Alamos National Laboratory, Los Alamos, NM, 87545, USA \\
$^{3}$ \quad Physics Division, Los Alamos National  Laboratory, Los Alamos, NM, 87545, USA \\
$^{4}$ \quad Computational Division, Los Alamos National Laboratory, Los Alamos, NM, 87545, USA \\
$^{5}$ \quad Department of Physics and Astronomy, Clemson University, Clemson, SC 29634, USA \\
$^{6}$ \quad School of Physics and Astronomy, Shanghai Jiao Tong University, Shanghai 200240, China
}

\corres{Correspondence: wendell@lanl.gov}

\abstract{
Nuclear isomers are populated in the rapid neutron capture process ($r$ process) of nucleosynthesis. 
The $r$ process may cover a wide range of temperatures, potentially starting from several tens of GK (several MeV) and then cooling as material is ejected from the event. 
As the $r$-process environment cools, isomers can freeze out of thermal equilibrium or be directly populated as astrophysical isomers (astromers).
Two key behaviors of astromers---ground state $\leftrightarrow$ isomer transition rates and thermalization temperatures---are determined by direct transition rates between pairs of nuclear states.
We perform a sensitivity study to constrain the effects of unknown transitions on astromer behavior.
We also introduce a categorization of astromers that describes their potential effects in hot environments. 
We provide a table of neutron-rich isomers that includes the astromer type, thermalization temperature, and key unmeasured transition rates.
}


\begin{document}
%%%%%%%%%%%%%%%%%%%%%%%%%%%%%%%%%%%%%%%%%%

%%%%%%%%%%%%%%%%%%%%%%%%%%%%%%%%%%%%%%%%%%
\section{Introduction}

Nuclear isomers are excited states of atomic nuclei with half-lives longer than the typical half-lives of picoseconds or femtoseconds \citep{Soddy1917}.  These metastable states exhibit inhibited transitions to lower-lying levels due to structural dissimilarities between them; large differences in nuclear deformation, spin, and the projection of spin along the symmetry axis can each be responsible for nuclear isomerism \cite{walker1999energy, aprahamian2005long, Dracoulis2016}.

\citet{Hahn1921} was the first to experimentally verify isomers in a 1921 study of uranium.  Since then, hundreds of isomeric states have been identified in nuclei across the chart of nuclides from mass number $A=12$ (the 2.251 MeV state in \nuc{12}{Be}) up to mass number $A=277$ (the uncertain state of \nuc{277}{Hs}) as of the time of this writing\footnote{From ENSDF database as of June 29th, 2020.  Version available at \url{http://www.nndc.bnl.gov/ensarchivals/}} \cite{jain2015atlas}.  Measurements of isomers and their properties continue to be a point of experimental interest for a variety of reasons including not only as astrophysical model inputs, but also for applications in industry, medicine, and tests of fundamental nuclear physics \citep{Langanke2000, Brown2014, masuda2019x, zhang2019isomer, liu2020isomeric, nesterenko2020isomer, orford2020isomer, sikorsky2020measurement, walker2020properties, izzo2021mass, gombas2021beta}.  For a recent review, see Ref.~\cite{Walker2020}. 

Although many nuclei possess isomeric states, not all isomers are of interest in astrophysical environments.  They may readily reach thermal equilibrium, or they may decay at a rate similar to the ground state (GS).  Those isomers that lead to other-than-thermal behavior in an astrophysical environment are known as ``astromers''~\citep{Misch2020a}.

Astrophysical nucleosynthesis simulations usually take one of two approaches to the distribution of nuclear states: either they assume a thermal equilibrium distribution, or they use only the ground state properties.  The presence of an isomer complicates calculations because it can render invalid both of these assumptions.  Inhibited communication between the GS and isomer can hinder thermal equilibration as well as trap the nucleus in an excited state that might behave radically different from the GS; this is exemplified by \nuc{26}{Al}.  The \nuc{26}{Al} isomer decays faster than it can be thermally repopulated from the GS when the ambient temperature is below about 35 keV; the implication is that the isomer becomes depopulated relative to a thermal distribution.  Furthermore, at those low temperatures, \nuc{26}{Al} that is produced in the isomeric state will $\beta$ decay before it can be thermally driven to the GS.  Therefore, neither the GS-only assumption nor the thermal-equilibrium assumption holds \citep{Coc1999, Gupta2001, Runkle2001, iliadis2011effects, Banerjee2018}.  Consequently, the study of element formation via nucleosynthesis must treat certain nuclei with isomers explicitly as two distinct species: a ground-state species, and a separate astromer species such as in Ref.~\cite{Reifarth2018}. 

The 228 keV state in \nuc{26}{Al} is the best-known astromer.  The ground state of \nuc{26}{Al} has a $\beta$-decay half-life of $\sim$700 kyr, while its isomer decays with a half-life of $\sim$6 s.  This nuclide---the first radioisotope to be observed in the heavens---is an important tracer of star formation \citep{Mahoney1982, Diehl1995, lugaro200826al}.  Other well-known nuclei with astromers include \nuc{34}{Cl} (possibly visible in nova bursts, isomer at 146 keV \cite{Coc1999, Parikh2009}) and \nuc{85}{Kr} (in the slow neutron capture ($s$)~process a branch point, in the $r$~process a $\beta$-decay accelerant and possible electromagnetic source; isomer at 305 keV \cite{Abia2001, Misch2020b}).  Isomers, including the 130 keV isomer in \nuc{38}{K}, may play a role in the rapid proton capture process ($rp$ process) \cite{novikov2001isomeric, grineviciute2014role, chipps2018k, pain2020exp}.  \nuc{176}{Lu} can be an $s$-process thermometer \cite{doll1999lifetime}, and geochemists use it as a chronometer \cite{soderlund2004176lu, albarede2006gamma, shafer2010trace, bloch2015176}, both of which are influenced by its isomer at 123 keV.

Despite being known in the contexts discussed above, isomers have only recently been included in larger networks such as those that describe the $r$ process that is believed to occur in explosive environments \citep{thielemann2011astrophysical}.  \citet{Fujimoto2020} included the direct population of nuclear isomers in the $r$ process by replacing GS properties with isomer properties.  In their study, several hand-picked isomers in the second $r$-process peak were shown to impact the radioactive heating of a kilonova. \citet{Misch2020b} recently studied the dynamic population, de-population, and decay of nuclear isomers in the $r$ process by including in a radioactive-decay network all isomers in the ENSDF database with a half-life greater than 100 $\mu$s.  As the temperature of an $r$-process event drops below each isomer's thermalization temperature (temperature below which the nuclide cannot reach thermal equilibrium), the isomer will freeze out as an astromer and affect the subsequent heating and evolution of isotopic abundances.

In this paper, we treat nuclear isomers as in Refs.~\cite{sprouse2021jade, Misch2020b} and extend the results of those works.  In particular, we examine the impact of the unknown properties of the intermediate states that facilitate GS $\leftrightarrow$ isomer transitions in neutron-rich $r$-process nuclei.  The GS $\leftrightarrow$ isomer transition rates set the thermalization temperature \ttherm{}, which in turn governs astromer freeze-out.  We show that the missing data have a large effect on our computed rates and thermalization temperatures.  We use the pathfinding technique of \citet{Misch2020a} to identify key nuclear states and transitions for experimental campaigns to target.

%%%%%%%%%%%%%%%%%%%%%%%%%%%%%%%%%%%%%%%%%%
\section{Methods}

% Definitions
In cold environments (e.g. terrestrial), nuclear isomers---by definition---do not readily transition to lower-lying states.  However, in a hot environment (e.g. stellar interiors, explosive astrophysical events, etc.), the ground state and isomer can more readily communicate by transitions through other intermediate excited nuclear states.  We use the terms ``direct transition'' or ``state-to-state transition'' (interchangeably) to refer to transitions directly from one nuclear state to another.  ``Effective transitions'' between long-lived states (GS and isomers) include direct GS $\leftrightarrow$ isomer transitions as well as chains of thermally mediated direct transitions through intermediate states.

We employ the formalism of \citet{Misch2020a} to compute the effective transition rates between long-lived nuclear states via intermediate states.  This method takes as inputs temperature and spontaneous nuclear transition rates, uses them to compute thermally enhanced direct transition rates (both exothermic ``down'' transitions and endothermic ``up'' transitions), and uses the results to derive effective transition rates.
We restrict our rate enhancements to transitions driven by a thermal photon bath and do not include e.g. electron collisions.  The thermal direct transition rates between a higher-energy state $h$ and a lower-energy state $l$ are then given by

\begin{align}
    \lambda_{hl} &= \lambda_{hl}^s (1+u) \label{eq:lambda_hl} \ , \\
    \lambda_{lh} &= \frac{2J_h+1}{2J_l+1}\lambda_{hl}^s u \ , \label{eq:lambda_lh} \\
    u &= \frac{1}{e^{(E_h-E_l)/T}-1} \ . \label{eq:u}
\end{align}
In these equations, $E$ and $J$ are the energy and spin of the indicated nuclear level, $T$ is the temperature, and $\lambda_{hl}^{s}$ is the spontaneous transition rate. 

With the direct rates in hand, the next step is to compute the probability $b_{st}$ that nuclear state $s$ goes to state $t$ when it transitions.  This is the fraction of the total transition rate out of $s$ that is to $t$ and is subject to the constraint $\sum\limits_t b_{st}=1$. 

\begin{align}
    b_{st} = \frac{\lambda_{st}}{\sum\limits_f \lambda_{sf}}
    \label{eq:b_definition}
\end{align}

We now use the $b$'s to calculate the probability $P_{iB}$ that a nucleus in an intermediate state $i$ follows a chain of transitions that takes it to long-lived state $B$ without passing through long-lived state $A$.  This quantity can be computed from a recursive relationship that forms a system of coupled linear equations.

\begin{align}
    P_{iB} = b_{iB} + \sum\limits_j b_{ij}P_{jB}
    \label{eq:P_recursion}
\end{align}

Finally, we have all of the ingredients to compute the effective transition rate $\Lambda_{AB}$ from long-lived state $A$ to long-lived state $B$:

\begin{align}
    \Lambda_{AB} &= \lambda_{AB} + \sum\limits_i \lambda_{Ai}P_{iB}.
    \label{eq:Lam_eff}
\end{align}
This expression for $\Lambda_{AB}$ includes explicitly the direct transition rate $\lambda_{AB}$ and implicitly the rates to follow all possible chains of transitions through intermediate states (``paths'').

Naturally, this method is powered by nuclear data, which we take from ENSDF \cite{ENSDF}; we use the evaluated nuclear level energies, spins, parities, half-lives, and $\gamma$ intensities\footnote{ENSDF database as of June 29th, 2020.  Version available at \url{http://www.nndc.bnl.gov/ensarchivals/}}.  We convert half-lives and intensities into state-to-state transition rates (``measured'' rates).  We estimate unmeasured rates with the Weisskopf approximation \cite{Weisskopf1930}.  In turn, the Weisskopf approximation requires the level spins and parities as inputs; where the level spins and parities are uncertain, we average together the Weisskopf rates for all possible combinations of initial- and final-state spins and parities within the uncertainties.  States with completely unknown half-lives, spins, and parities are not included.

Apart from the excluded information-deficient states, the uncertainties in our calculations of effective GS $\leftrightarrow$ isomer transition rates lie principally in the unmeasured Weisskopf rates: most measured rates have a relative uncertainty of less than 50\%, while the Weisskopf approximation often disagrees with experiment by one or two orders of magnitude.  Therefore, in this sensitivity study, we fixed the measured rates and varied the Weisskopf rates up and down by factors of 10 and 100.  In these variations, we shifted \emph{all} of the Weisskopf rates together rather than independently.  This approach constrains the likely bounds of effective GS $\leftrightarrow$ isomer transition rates.

The constraints on the GS $\leftrightarrow$ isomer transition rates also bound the thermalization temperatures for astromers.  Astromers thermalize and behave like non-isomeric states when the transition rates dominate other reaction rates, that is, when their communication is sufficiently unhindered such that they can reach a thermal-equilibrium population.  We define the thermalization temperature \ttherm{} as the lowest temperature such that the transition rate out of each nuclear state is equal to or greater than its destruction rate.

The thermalization temperature is a key component to understanding the behavior and influence of astromers: isomers freeze out of thermal equilibrium as astromers at their thermalization temperatures.  We computed the range of \ttherm{} under the influence of our Weisskopf variations for each isotope in our study.  The thermalization temperature depends on which destruction channels are in play in a given environment, and our intent is to highlight the effects of nuclear uncertainties without the distractions of astrophysical uncertainties.  Therefore, we considered only $\beta$ decay because it is utterly essential astrophysically, straightforward to compute, and relatively insensitive to anything but temperature and electron density.  We did not include any reaction rates, and all subsequent instances of \ttherm{} should be considered to be a thermalization temperature with respect to $\beta$ decay only (\ttherm{\beta}).

Because we varied all Weisskopf rates together, our variations alone do not reveal which individual unmeasured transitions are most influential.  We addressed this by using the pathfinding method of \citet{Misch2020a} to isolate which individual state-to-state transitions contribute most to the total effective GS $\leftrightarrow$ isomer transition rate.  For unstable nuclei, we identified all unmeasured (Weisskopf) transitions which lie along paths that contribute at least 1\% of the effective transition rate at temperatures below \ttherm{}.  For stable nuclei, we performed the same analysis with a fixed cutoff temperature of 30 keV.

%%%%%%%%%%%%%%%%%%%%%%%%%%%%%%%%%%%%%%%%%%
\section{Results}
\label{sec:results}

We included in our study those nuclei with isomers in the neutron-rich $r$-process region between mass numbers $A = 69$ and $A = 209$ with half-lives $T_{1/2} > 100$ $\mu$s.  We highlight the specific astromers of likely import in the $r$ process listed in Table I of \citet{Misch2020b}.  In identifying those astromers, that work held the Weisskopf rates fixed and used a single temperature-density trajectory \citep{lippuner2015heating, zhu2018cf} in the Jade network of \citet{sprouse2021jade}.  The fixed Weisskopf rates may inadvertently over- or under-emphasize some astromers, but we nevertheless have an adequate starting set to examine.  The $r$ process is subject to astrophysical variations in environmental conditions \citep{Horowitz2019, Cote2021} and we plan to address this point in follow up work.  Here we focus our attention on the nuclear uncertainties that arise from unknown transitions.  This approach allows us not only to zoom in on potential key astromers, but also to disentangle the nuclear physics uncertainty effects on their behavior from the astrophysics.  An influential astromer with large uncertainties is then a priority for experimental and/or deeper theoretical inquiry.

We use a system of types to categorize each astromer according to the role it could play in the $r$-process decay back to stability.

Type A (``accelerant'') astromers have a $\beta$-decay rate greater than the ground state and can accelerate abundance evolution and energy release.  Even if the isotope is thermalized, the greater decay rate of the thermally populated excited state will accelerate the overall $\beta$-decay rate.

Type B (``battery'') astromers decay (via all channels, including de-excitation) slower than the ground state, storing energy and releasing it later.  Type B astromers have an associated temperature above which they are not batteries; it is the temperature above which the total destruction rate of the isomer via all channels is greater than or equal to the GS $\beta$-decay rate.  Above this temperature, the isomer is clearly not a battery, because it is not in fact storing energy for longer than the GS.

Type N (``neutral'') astromers don't fall into either of these categories and will not have a large direct impact on decay or heating under the conditions of this study.  However, type N astromers may decay to feed another more interesting astromer, and particularly long-lived type N astromers may produce an electromagnetic signal; some type N astromers in stable isotopes can play the latter role.

We assign the type of the isomer using the following procedure.  If the isotope is stable, the isomer is type N.  For unstable isotopes, we compare the total decay rates of the GS and isomer at low temperature; if the rates are not different by a threshold factor (we used a factor of 2), the isomer is type N.  If the isomer's $\beta$-decay rate is greater than the ground state rate by the threshold factor, it is a type A astromer.  If the total decay rate of the isomer is slower than the $\beta$-decay rate of the GS by the threshold factor, it is a type B astromer.  For type B astromers, we also identify the temperature above which they are not batteries, that is, the temperature above which the decay rate is no longer dissimilar from the GS rate by the threshold factor.

In what follows, we present detailed results for the nuclei appearing in Table I of \citet{Misch2020b}.  For each detailed isotope, we show a figure indicating the uncertainty bands for the effective transition rates in each of these nuclei.  The figures also show the $\beta$-decay rates and a vertical line indicating the approximate thermalization temperature; if no \ttherm{} line is shown, there is insufficient data to calculate a thermalization temperature.  In the discussion accompanying each figure, we indicate the number of measured levels appearing in ENSDF and the number of those used in our calculations; for convenience, we also include all of the information from Table \ref{tab:all}.  We follow this information with some brief comments about each isotope.  Type B astromers have the associated temperature listed along with the type.

\textbf{Important note}: Because we do not speculate about any uncertainties other than the Weisskopf approximation, the bands here should be considered \emph{lower bounds} on the effective transition rate uncertainties.  Other as yet unmeasured nuclear properties could have a dramatic effect on the rates.  There may be substantial uncertainty in a critical experimental rate, or key intermediate states may be missing or lack adequate information for a Weisskopf calculation.  A striking example of the latter situation is \nuc{126}{Sb}, shown below in figure \ref{fig:it_Sb}.  The dominant pathways consist almost entirely of measured transitions, so the bands are extremely narrow.  However, this isotope only has six measured levels, the highest of which is at 127.9 keV.  More levels would open more paths, and the effective transition rates would certainly change.

Our full sensitivity study results are provided in Table \ref{tab:all} of the appendix.  The table includes the thermalization temperature range, type, and key unmeasured transitions for each potential astromer.  

\subsection{Zn ($Z=30$) Isotopes}

\begin{figure}[H]
\centering
\begin{subfigure}{.5\textwidth}
  \includegraphics[width=\linewidth]{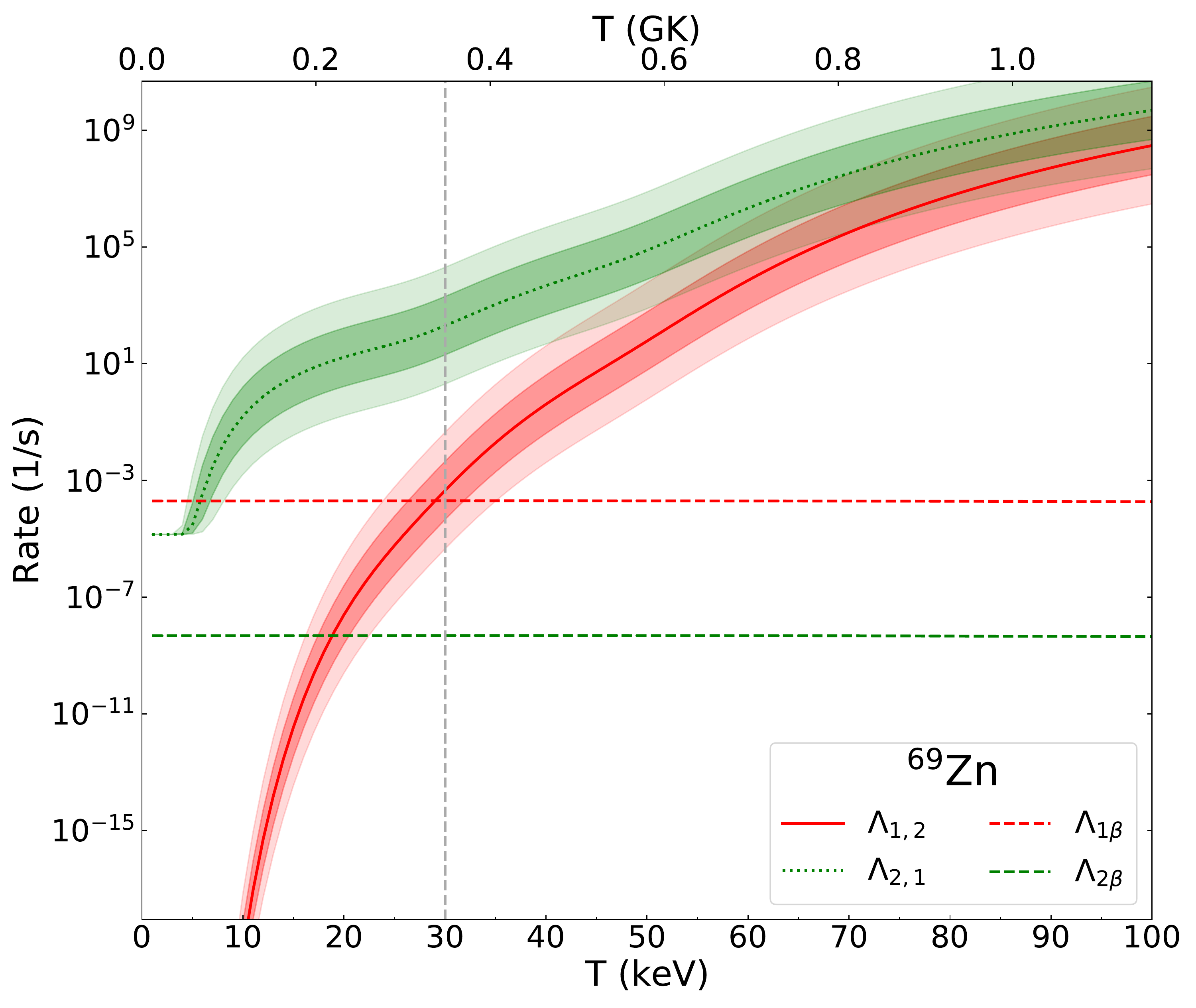}
\end{subfigure}%
\begin{subfigure}{.5\textwidth}
  \includegraphics[width=\linewidth]{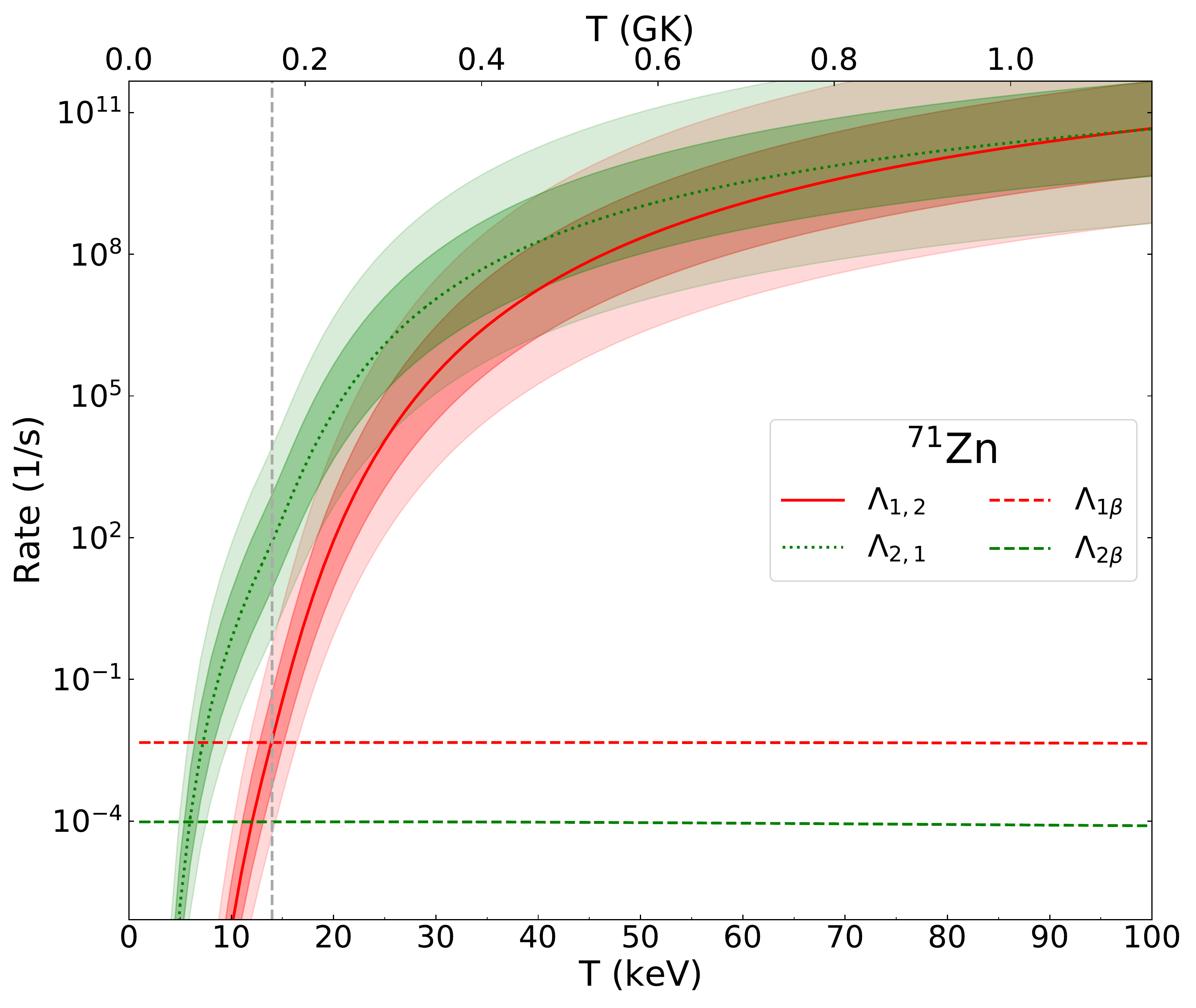}
\end{subfigure}%
\caption{Effective transition rates for Zn ($Z=30$) isotopes. Darkest shaded band shows unmeasured rates increased/decreased by one order of magnitude; light by two orders of magnitude. Thermalization temperature, \ttherm{}, estimated by dashed vertical grey line. }
\label{fig:it_Zn}
\end{figure}

\nuc{69}{Zn}: First $r$-process peak ($A\sim80$) nuclide. 73 measured levels, 30 in this calculation.  Isomer at 438.636 keV (type B, 5 keV).  Known uncertainties dominated by unmeasured ($531.3 \rightarrow 0.0$), ($531.3 \rightarrow 438.636$), ($872.0 \rightarrow 438.636$), and ($872.0 \rightarrow 531.3$) transition rates.  The isomer greatly delays $\beta$ decay, which may influence late time nucleosynthesis and/or result in a $\gamma$-ray signal shortly after an $r$-process event.

\nuc{71}{Zn}: First $r$-process peak ($A\sim80$) nuclide. 158 measured levels, 30 in this calculation.  Isomer at 157.7 keV (type B, 6 keV).  Known uncertainties dominated by unmeasured ($157.7 \rightarrow 0.0$), ($286.3 \rightarrow 0.0$), ($286.3 \rightarrow 157.7$), ($465.0 \rightarrow 157.7$), ($465.0 \rightarrow 286.3$), ($489.8 \rightarrow 0.0$), ($489.8 \rightarrow 286.3$) transition rates.  The isomer somewhat delays $\beta$ decay, which may influence late time nucleosynthesis and/or result in a $\gamma$-ray signal shortly after an $r$-process event.

\subsection{Se ($Z=34$) Isotopes}

\begin{figure}[H]
\centering
\begin{subfigure}{.5\textwidth}
  \includegraphics[width=\linewidth]{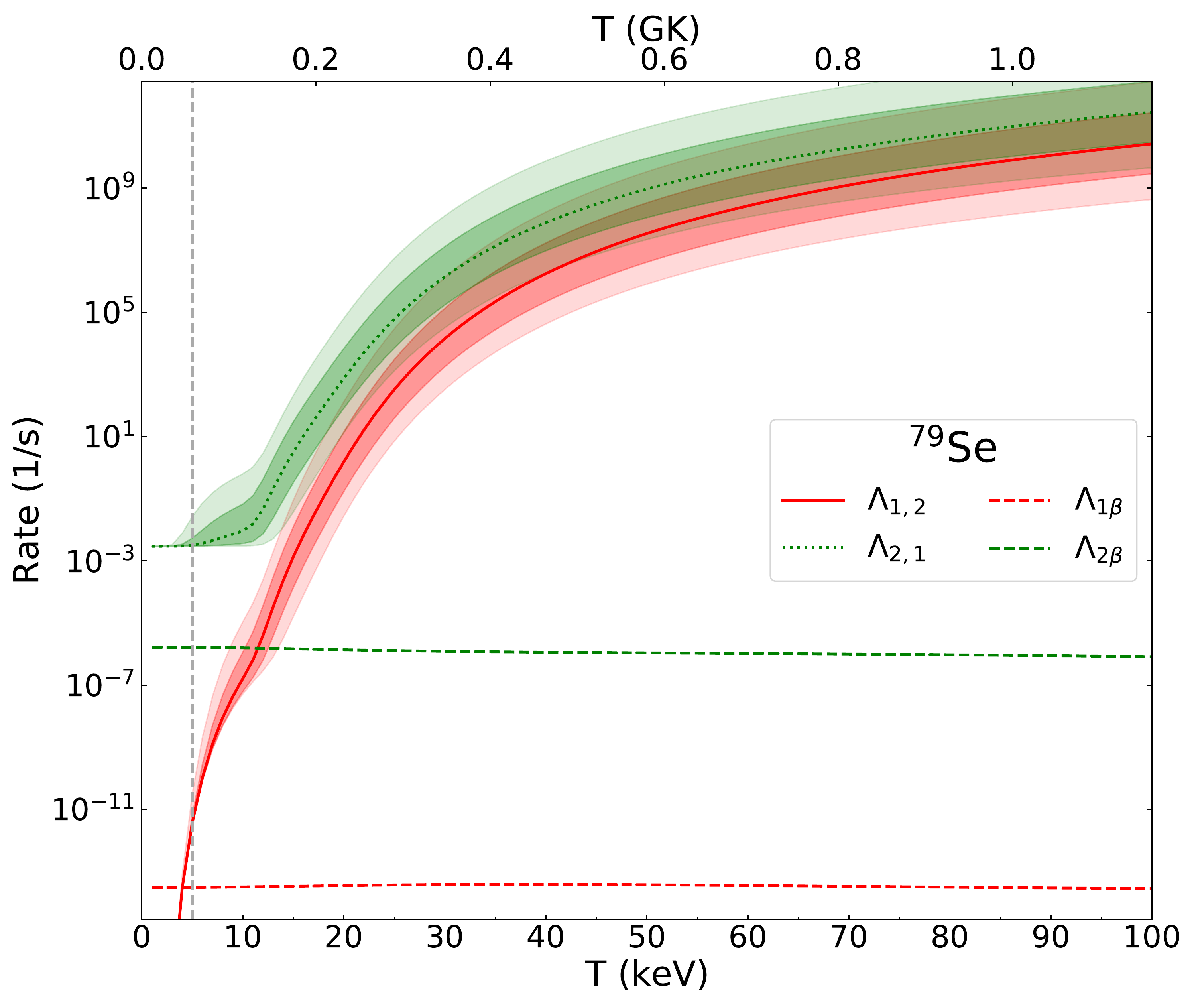}
\end{subfigure}%
\begin{subfigure}{.5\textwidth}
  \includegraphics[width=\linewidth]{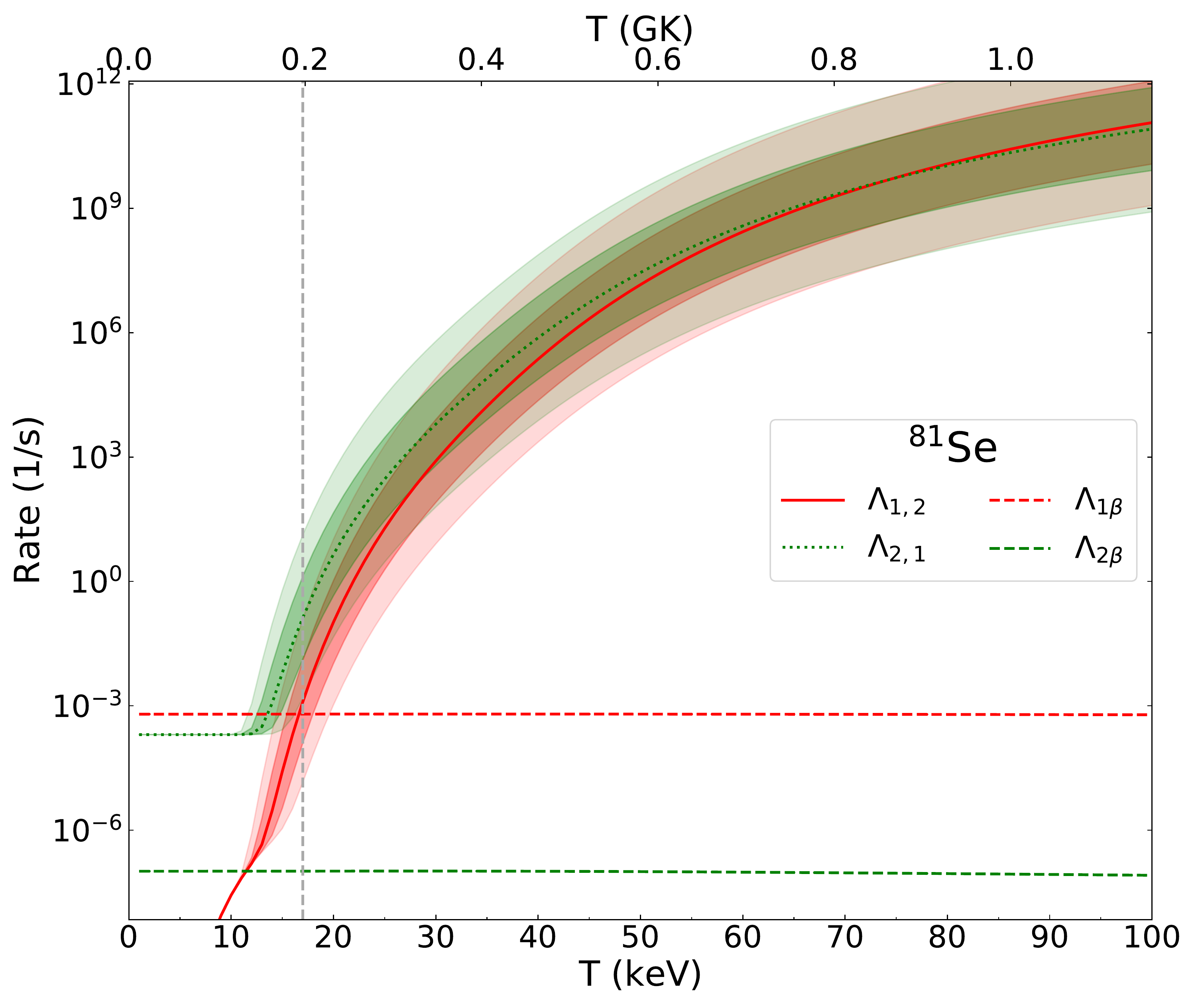}
\end{subfigure}%
\caption{Effective transition rates for Se ($Z=34$) isotopes. Darkest shaded band shows unmeasured rates increased/decreased by one order of magnitude; light by two orders of magnitude. Thermalization temperature, \ttherm{}, estimated by dashed vertical grey line. }
\label{fig:it_Se}
\end{figure}

\nuc{79}{Se}: First $r$-process peak ($A\sim80$) nuclide. 116 measured levels, 30 in this calculation.  Isomer at 95.77 keV (type A).  Known uncertainties dominated by unmeasured ($128.0 \rightarrow 0.0$), ($128.0 \rightarrow 95.77$) transition rates.  No expected new effect on the $r$ process due to its half-life being much shorter than that of its $\beta$-decay parent.

\nuc{81}{Se}: First $r$-process peak ($A\sim80$) nuclide. 77 measured levels, 30 in this calculation.  Isomer at 103.0 keV (type B, 13 keV).  Known uncertainties dominated by unmeasured ($467.74 \rightarrow 0.0$), ($467.74 \rightarrow 103.0$), ($491.06 \rightarrow 0.0$), ($491.06 \rightarrow 103.0$), ($491.06 \rightarrow 467.74$), ($624.11 \rightarrow 103.0$), ($624.11 \rightarrow 467.74$) transition rates.  The isomer significantly delays $\beta$ decay, which may influence late-time $r$-process nucleosynthesis and/or heating.

\subsection{Kr ($Z=36$) Isotopes}

\begin{figure}[H]
\centering
\begin{subfigure}{.5\textwidth}
  \includegraphics[width=\linewidth]{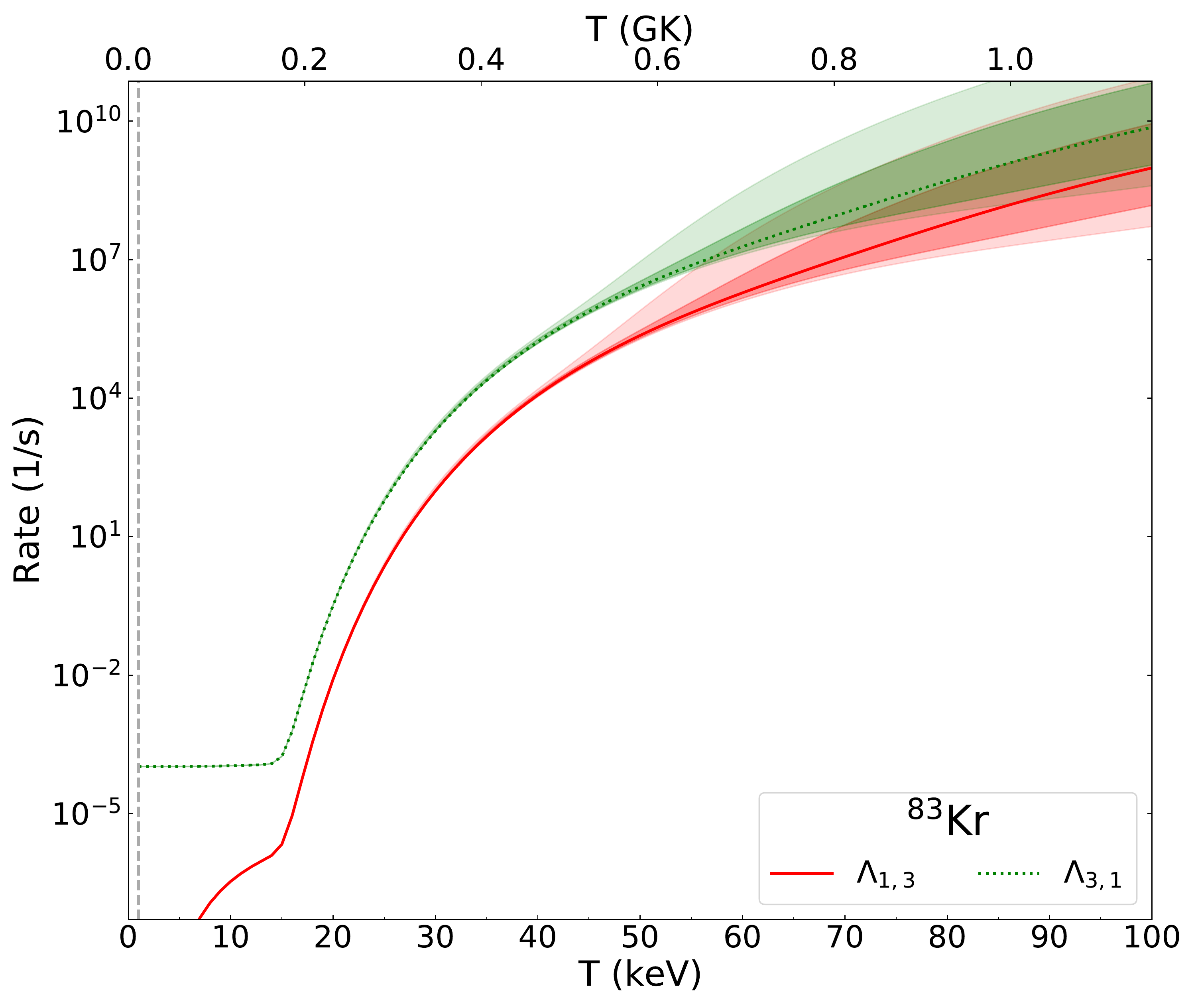}
\end{subfigure}%
\begin{subfigure}{.5\textwidth}
  \includegraphics[width=\linewidth]{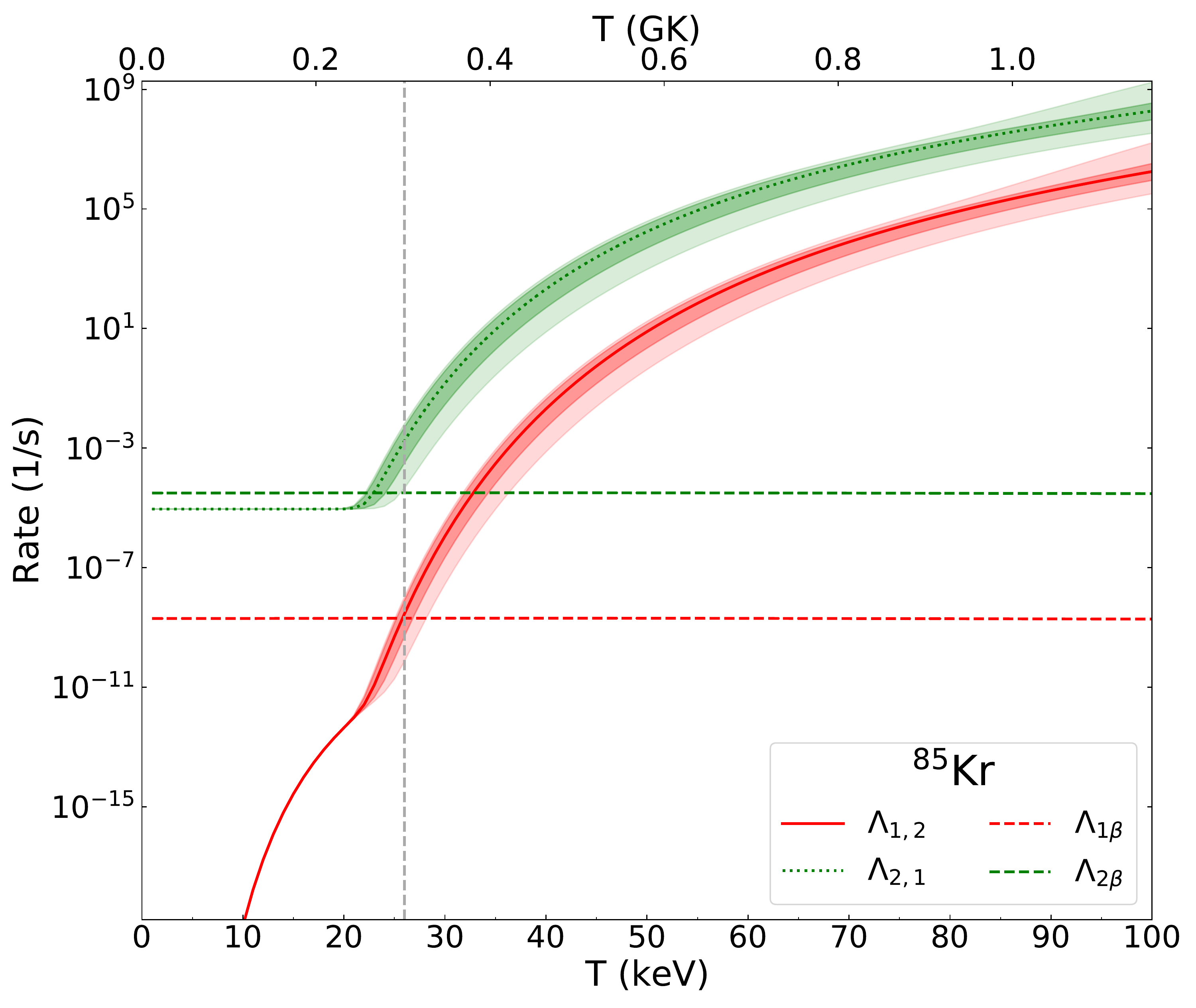}
\end{subfigure}%
\caption{Effective transition rates for Kr ($Z=36$) isotopes. Darkest shaded band shows unmeasured rates increased/decreased by one order of magnitude; light by two orders of magnitude. Thermalization temperature, \ttherm{}, estimated by dashed vertical grey line. }
\label{fig:it_Kr}
\end{figure}

\nuc{83}{Kr}: First $r$-process peak ($A\sim80$) nuclide. 74 measured levels, 30 in this calculation.  Isomer at 41.5575 keV (type N).  Known uncertainties dominated by unmeasured ($571.1538 \rightarrow 561.9585$), ($571.1538 \rightarrow 9.4057$), ($799.48 \rightarrow 561.9585$), ($799.48 \rightarrow 571.1538$) transition rates.  No expected new effect on the $r$ process due to its short half-life and lack of $\beta$ decay.

\nuc{85}{Kr}: First $r$-process peak ($A\sim80$) nuclide. 222 measured levels, 30 in this calculation.  Isomer at 304.871 keV (type A).  Known uncertainties dominated by unmeasured ($1107.32 \rightarrow 304.871$), ($1140.73 \rightarrow 1107.32$), ($1166.69 \rightarrow 1140.73$), ($1166.69 \rightarrow 304.871$), ($1223.98 \rightarrow 1140.73$), ($1223.98 \rightarrow 1166.69$), ($1416.57 \rightarrow 1107.32$) transition rates.  The isomer greatly accelerates $\beta$ decay, which may result in a $\gamma$-ray signal shortly after an $r$-process event.

\subsection{Nb ($Z=41$) and Tc ($Z=43$) Isotopes}

\begin{figure}[H]
\centering
\begin{subfigure}{.5\textwidth}
  \includegraphics[width=\linewidth]{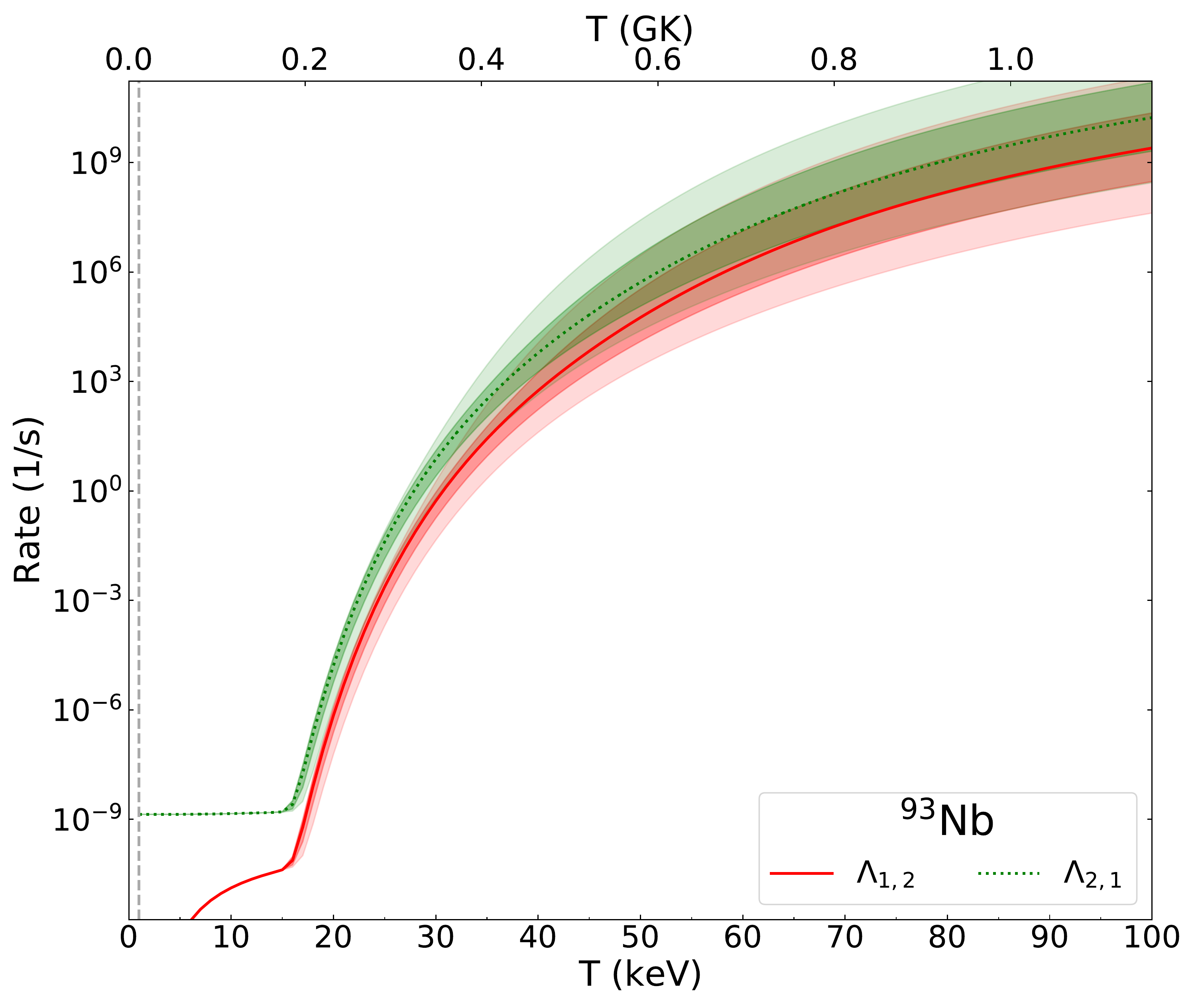}
  \includegraphics[width=\linewidth]{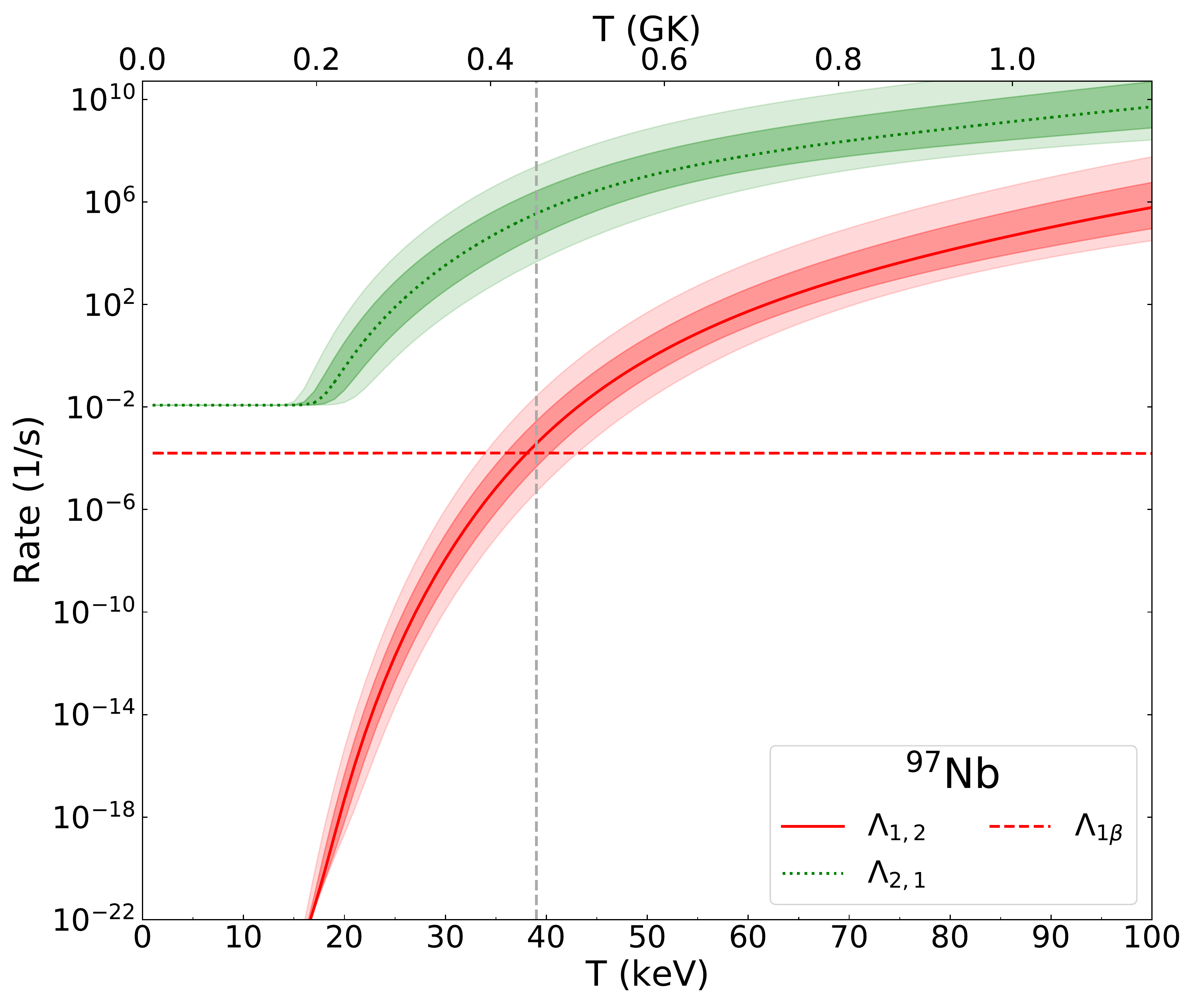}
\end{subfigure}%
\begin{subfigure}{.5\textwidth}
  \includegraphics[width=\linewidth]{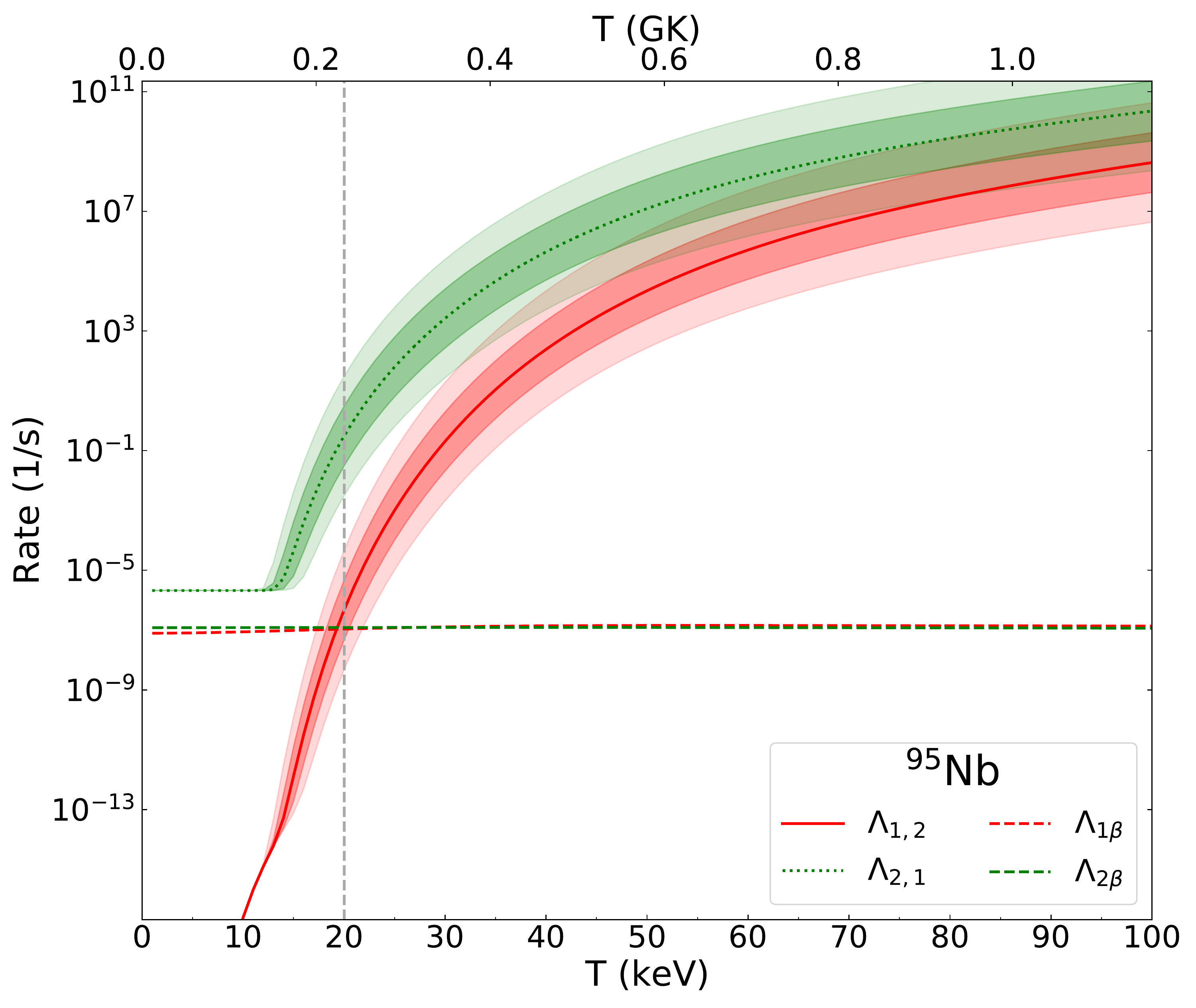}
  \includegraphics[width=\linewidth]{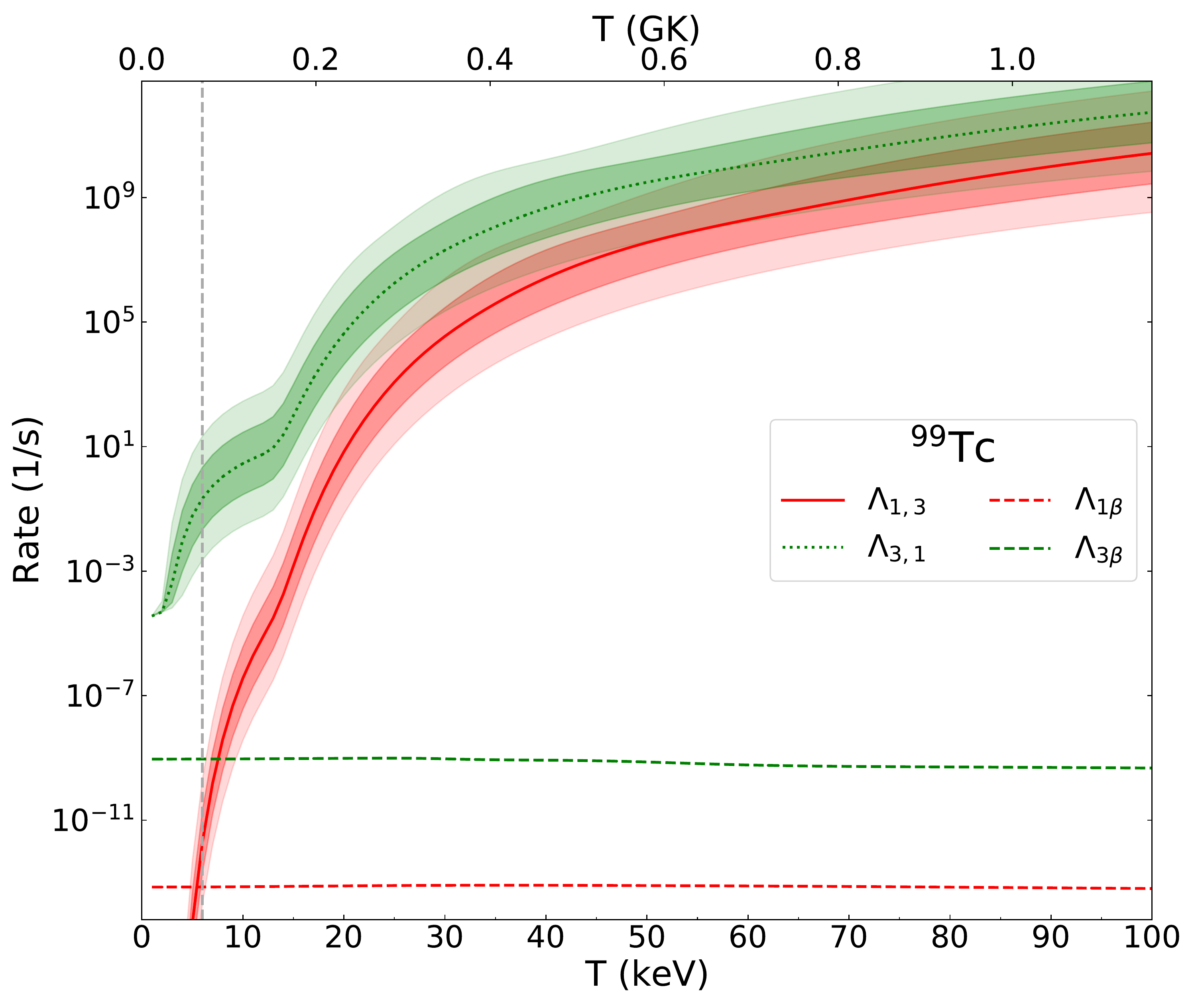}
\end{subfigure}%
\caption{Effective transition rates for Nb ($Z=41$) and Tc ($Z=43$) isotopes. Darkest shaded band shows unmeasured rates increased/decreased by one order of magnitude; light by two orders of magnitude. Thermalization temperature, \ttherm{}, estimated by dashed vertical grey line. }
\label{fig:it_Nb}
\end{figure}

\nuc{93}{Nb}: Transition region nuclide. 128 measured levels, 30 in this calculation.  Isomer at 30.77 keV (type N).  Known uncertainties dominated by unmeasured ($1127.09 \rightarrow 0.0$), ($1127.09 \rightarrow 30.77$), ($808.82 \rightarrow 686.79$), ($810.32 \rightarrow 743.95$) transition rates.  No expected new effect on the $r$ process due to its half-life being much shorter than that of its $\beta$-decay parent.

\nuc{95}{Nb}: Transition region nuclide. 119 measured levels, 30 in this calculation.  Isomer at 235.69 keV (type N).  Known uncertainties dominated by unmeasured ($730.0 \rightarrow 0.0$), ($756.728 \rightarrow 730.0$), ($799.0 \rightarrow 235.69$), ($799.0 \rightarrow 730.0$), ($805.0 \rightarrow 0.0$), ($805.0 \rightarrow 730.0$), ($877.0 \rightarrow 0.0$), ($877.0 \rightarrow 799.0$) transition rates.  No expected new effect on the $r$ process due to its half-life being much shorter than that of its $\beta$-decay parent.

\nuc{97}{Nb}: Transition region nuclide. 62 measured levels, 30 in this calculation.  Isomer at 743.35 keV (type N).  Known uncertainties dominated by unmeasured ($1160.0 \rightarrow 0.0$), ($1160.0 \rightarrow 1147.96$), ($1251.01 \rightarrow 0.0$), ($1251.01 \rightarrow 743.35$), ($1276.09 \rightarrow 0.0$), ($1276.09 \rightarrow 1147.96$), ($1276.09 \rightarrow 1251.01$), ($1433.92 \rightarrow 1147.96$), ($1433.92 \rightarrow 1251.01$), ($1433.92 \rightarrow 1276.09$), ($1548.36 \rightarrow 0.0$), ($1548.36 \rightarrow 1147.96$), ($1548.36 \rightarrow 1276.09$), ($1548.36 \rightarrow 743.35$), ($1750.43 \rightarrow 1251.01$), ($2357.0 \rightarrow 0.0$), ($2357.0 \rightarrow 743.35$) transition rates.  No expected new effect on the $r$ process due to its half-life being much shorter than that of its $\beta$-decay parent.

\nuc{99}{Tc}: Transition region nuclide. 144 measured levels, 30 in this calculation.  Isomer at 142.6836 keV (type A).  Known uncertainties dominated by unmeasured ($181.09423 \rightarrow 142.6836$) transition rate.  No expected new effect on the $r$ process due to its half-life being much shorter than that of its $\beta$-decay parent.

\subsection{Cd ($Z=48$) Isotopes}

\begin{figure}[H]
\centering
\begin{subfigure}{.5\textwidth}
  \includegraphics[width=\linewidth]{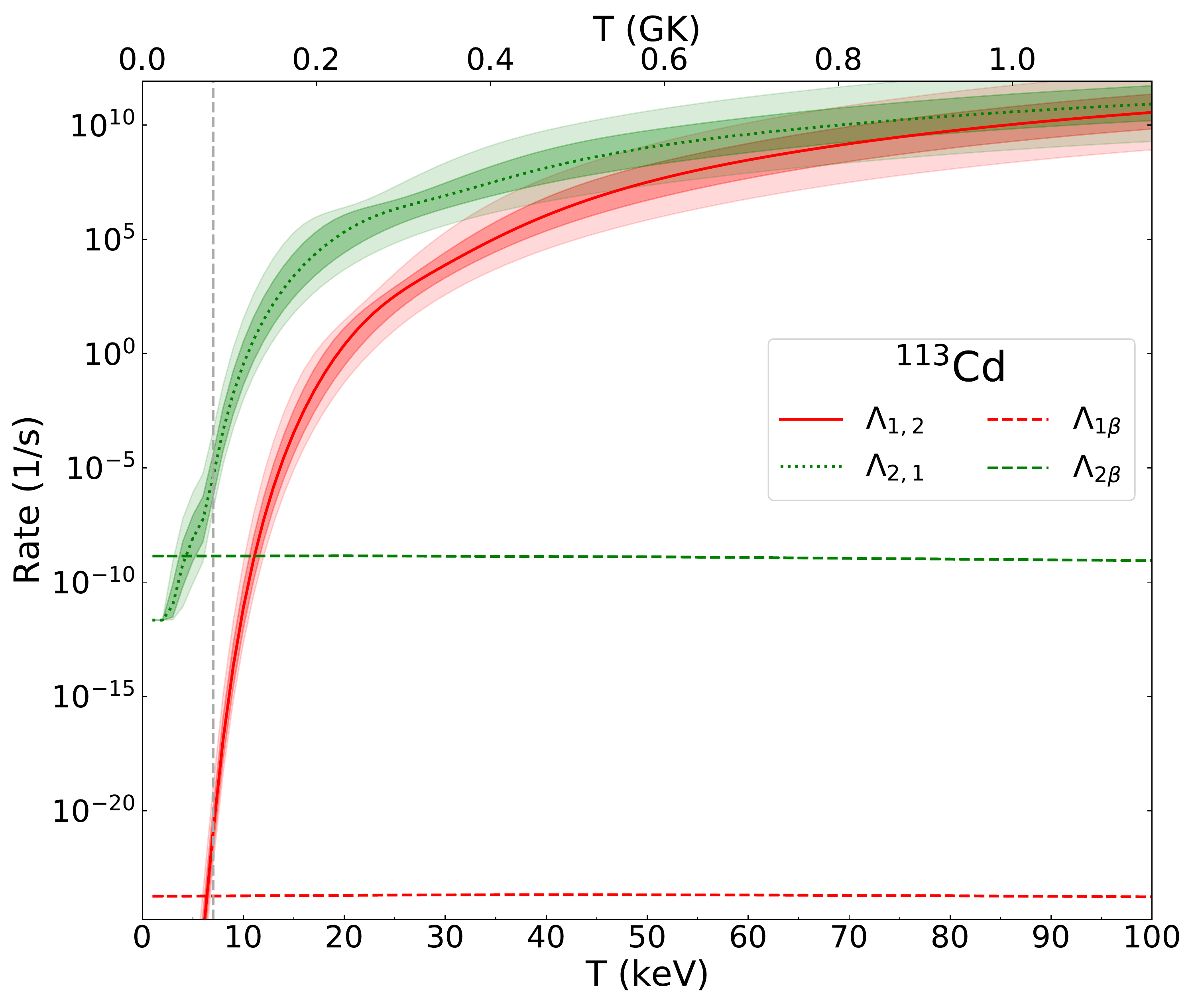}
\end{subfigure}%
\begin{subfigure}{.5\textwidth}
  \includegraphics[width=\linewidth]{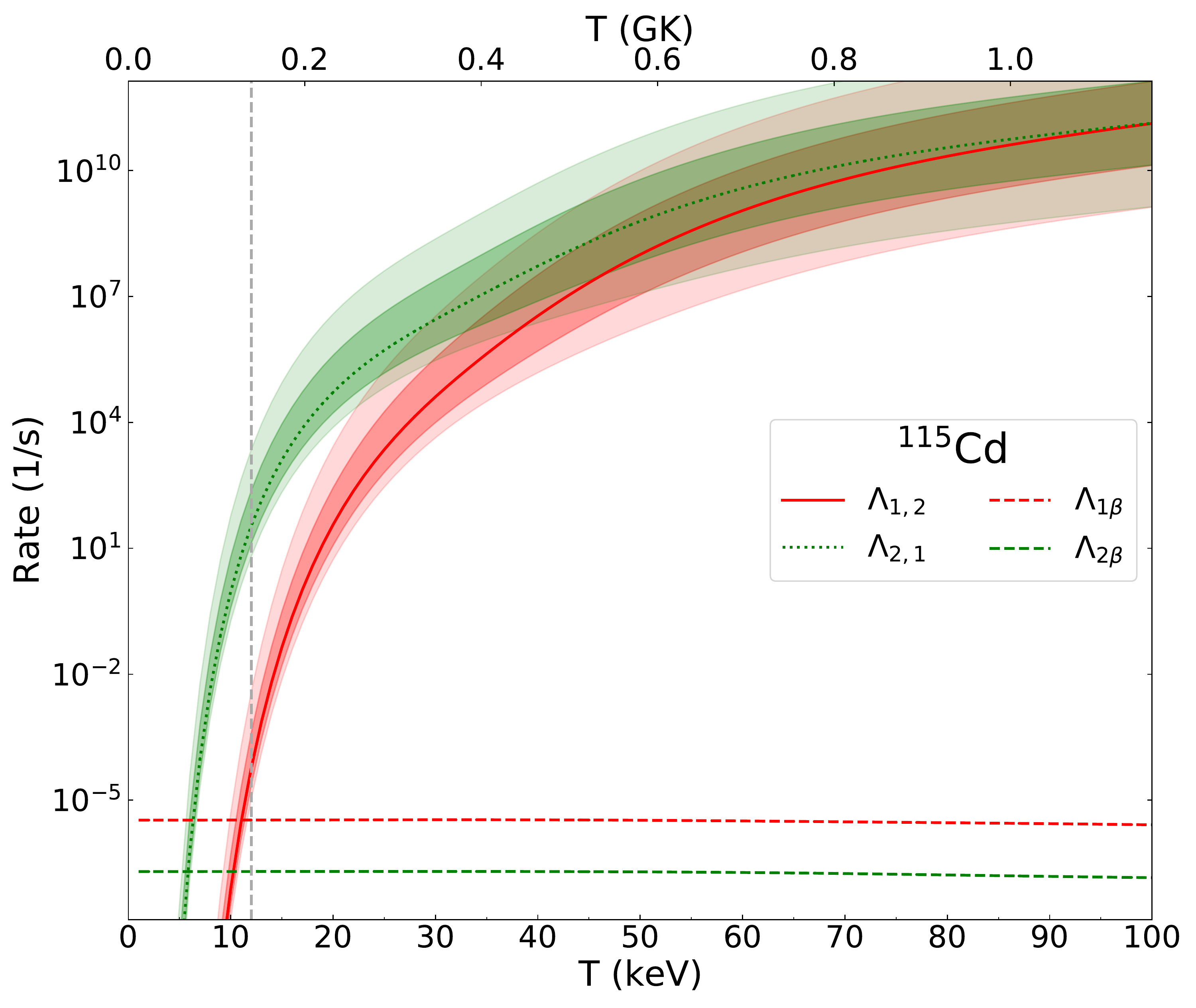}
\end{subfigure}%
\caption{Effective transition rates for Cd ($Z=48$) isotopes. Darkest shaded band shows unmeasured rates increased/decreased by one order of magnitude; light by two orders of magnitude. Thermalization temperature, \ttherm{}, estimated by dashed vertical grey line. }
\label{fig:it_Cd}
\end{figure}

\nuc{113}{Cd}: Transition region nuclide. 215 measured levels, 30 in this calculation.  Isomer at 263.54 keV (type A).  Known uncertainties dominated by unmeasured ($316.206 \rightarrow 263.54$), ($458.633 \rightarrow 263.54$), ($458.633 \rightarrow 316.206$), ($522.259 \rightarrow 458.633$), ($530.0 \rightarrow 263.54$), ($530.0 \rightarrow 316.206$), ($530.0 \rightarrow 458.633$) transition rates.  Isomer greatly accelerates $\beta$ decay, but likely unobservable because the half-life is still quite long (14 y) and the isomer population is relatively low.

\nuc{115}{Cd}: Transition region nuclide. 70 measured levels, 30 in this calculation.  Isomer at 181.0 keV (type B, 6 keV).  Known uncertainties dominated by unmeasured ($181.0 \rightarrow 0.0$), ($229.1 \rightarrow 0.0$), ($360.5 \rightarrow 229.1$), ($389.0 \rightarrow 181.0$), ($389.0 \rightarrow 360.5$), ($393.9 \rightarrow 360.5$), ($393.9 \rightarrow 389.0$), ($417.2 \rightarrow 181.0$), ($417.2 \rightarrow 393.9$) transition rates.  Isomer significantly slows $\beta$ decay with possible consequences for $r$-process heating curves.

\subsection{In ($Z=49$) Isotopes}

\begin{figure}[H]
\centering
\begin{subfigure}{.5\textwidth}
  \includegraphics[width=\linewidth]{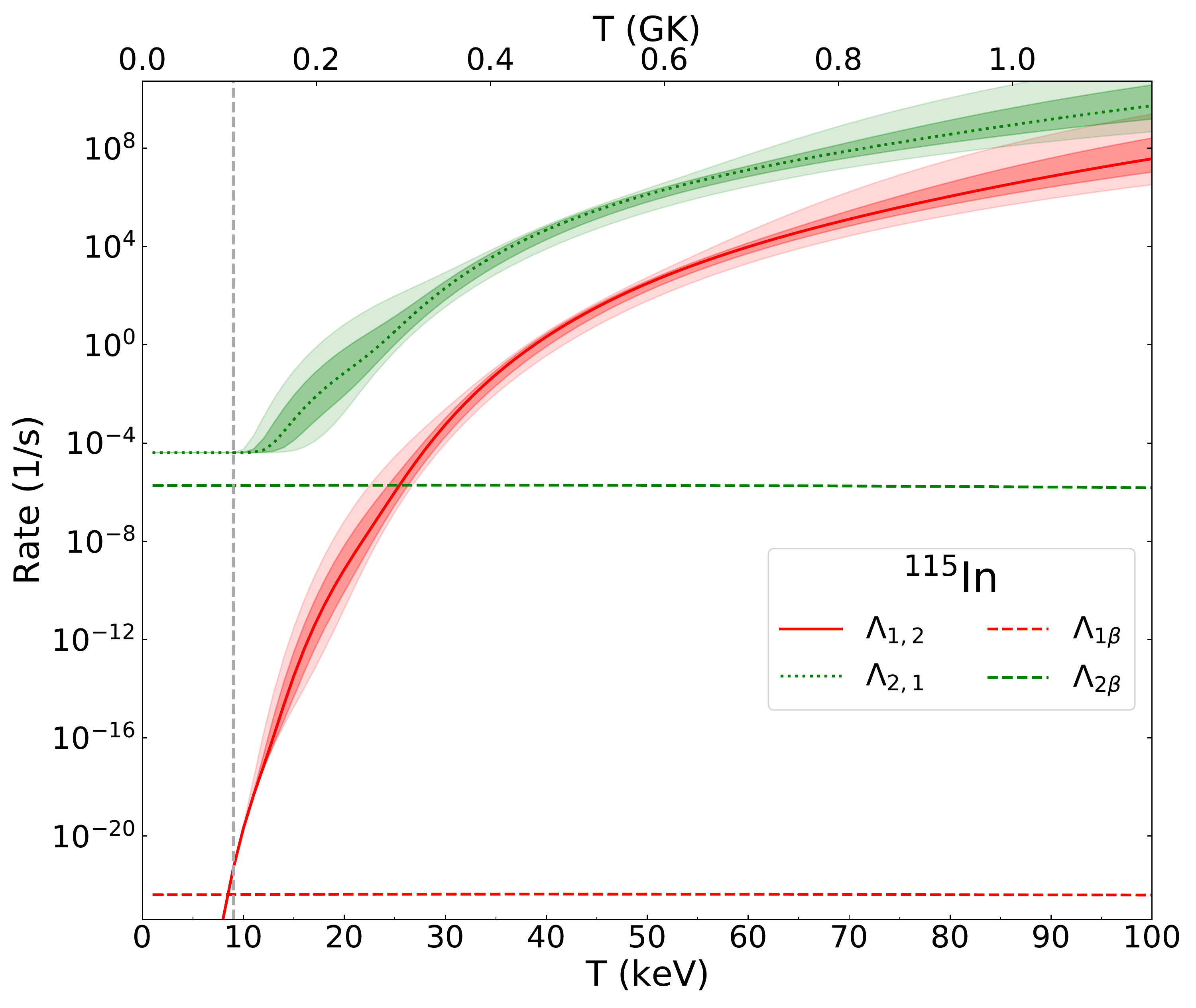}
  \includegraphics[width=\linewidth]{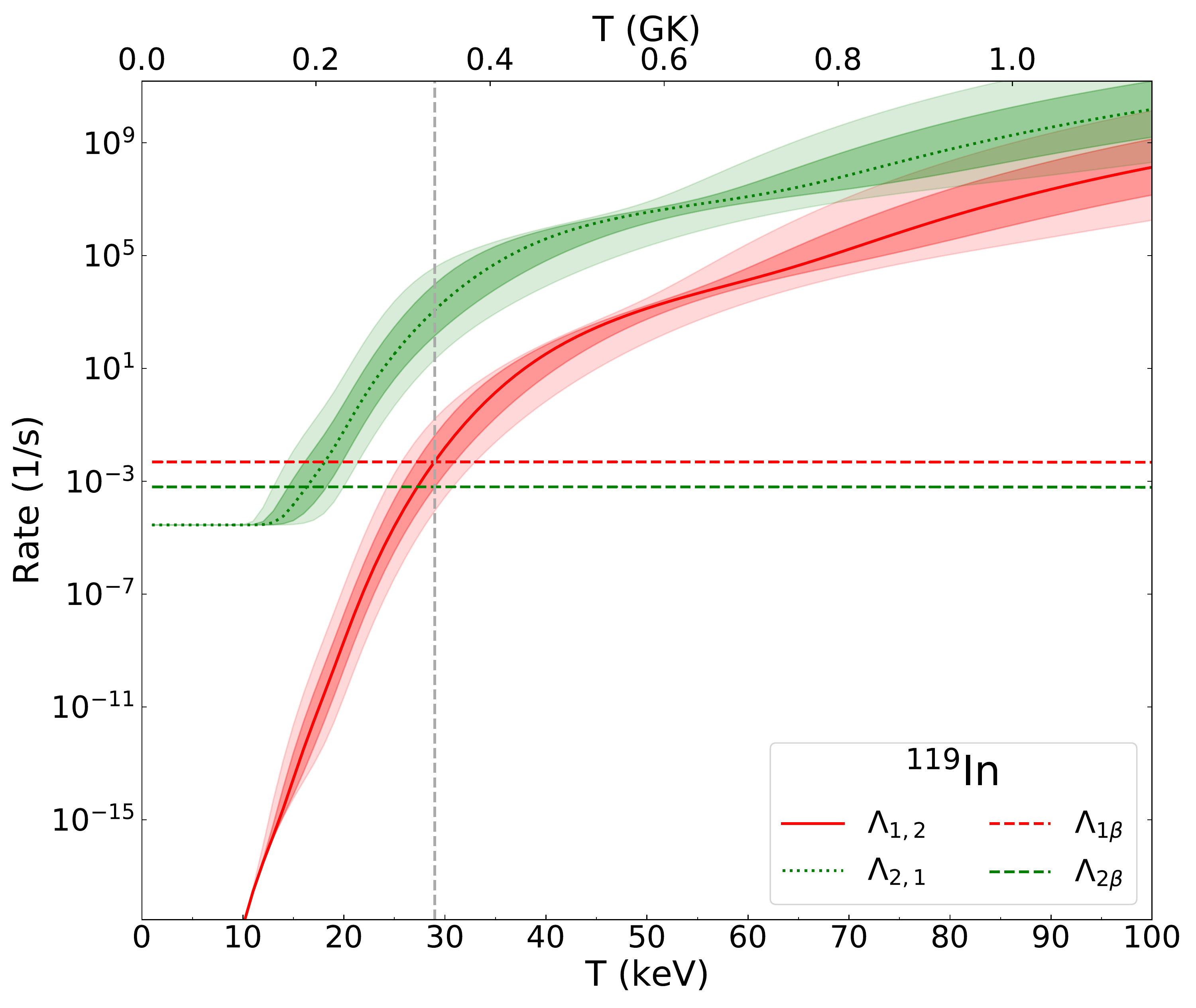}
\end{subfigure}%
\begin{subfigure}{.5\textwidth}
  \includegraphics[width=\linewidth]{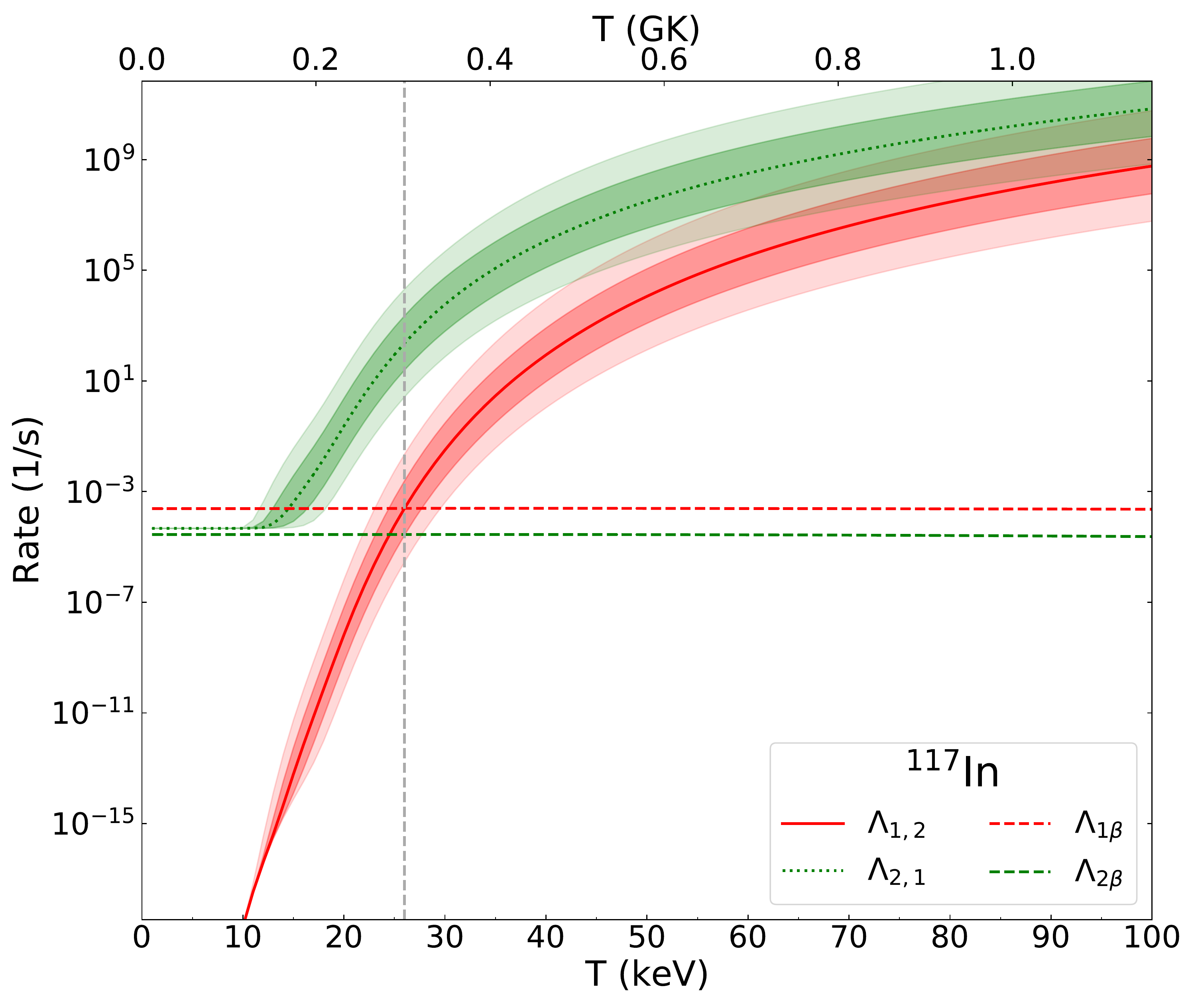}
  \includegraphics[width=\linewidth]{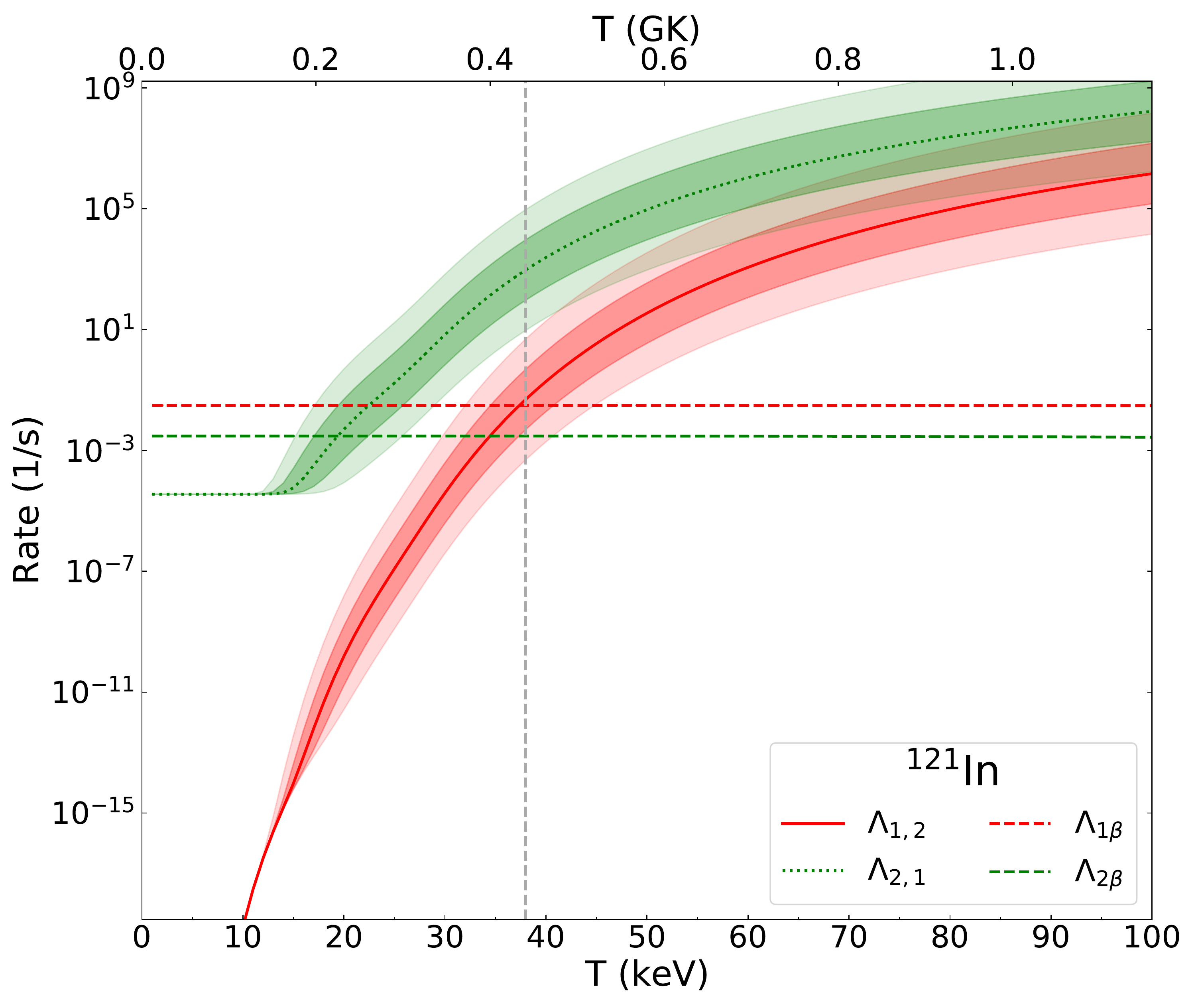}
\end{subfigure}
\caption{Effective transition rates for In ($Z=49$) isotopes. Darkest shaded band shows unmeasured rates increased/decreased by one order of magnitude; light by two orders of magnitude. Thermalization temperature, \ttherm{}, estimated by dashed vertical grey line.}
\label{fig:it_In}
\end{figure}

\nuc{115}{In}: Transition region nuclide. 92 measured levels, 30 in this calculation.  Isomer at 336.244 keV (type A).  Known uncertainties dominated by unmeasured ($597.144 \rightarrow 0.0$) transition rate.  Isomer could boost \nuc{115}{Sn} production and may generate a $\gamma$-ray signal shortly after an $r$-process event.

\nuc{117}{In}: Transition region nuclide. 79 measured levels, 30 in this calculation.  Isomer at 315.303 keV (type B, 13 keV).  Known uncertainties dominated by unmeasured ($1028.04 \rightarrow 659.765$), ($1028.04 \rightarrow 748.07$), ($1028.04 \rightarrow 880.72$), ($588.653 \rightarrow 0.0$), ($748.07 \rightarrow 0.0$), ($880.72 \rightarrow 0.0$), ($880.72 \rightarrow 588.653$), ($880.72 \rightarrow 659.765$), ($880.72 \rightarrow 748.07$) transition rates.  No expected new effect on the $r$ process due to its half-life being much shorter than that of its $\beta$-decay parent.

\nuc{119}{In}: Second $r$-process peak ($A\sim130$) nuclide. 74 measured levels, 30 in this calculation.  Isomer at 311.37 keV (type B, 17 keV).  Known uncertainties dominated by unmeasured ($1044.44 \rightarrow 654.27$), ($1044.44 \rightarrow 720.6$), ($1044.44 \rightarrow 788.26$), ($604.18 \rightarrow 0.0$), ($720.6 \rightarrow 0.0$), ($720.6 \rightarrow 654.27$), ($941.43 \rightarrow 0.0$), ($941.43 \rightarrow 604.18$), ($941.43 \rightarrow 654.27$), ($941.43 \rightarrow 720.6$) transition rates.  The isomer significantly slows $\beta$ decay with possible consequences for $r$-process heating curves.

\nuc{121}{In}: Second $r$-process peak ($A\sim130$) nuclide. 64 measured levels, 30 in this calculation.  Isomer at 313.68 keV (type B, 21 keV).  Known uncertainties dominated by unmeasured ($1040.33 \rightarrow 0.0$), ($1040.33 \rightarrow 637.9$), ($637.9 \rightarrow 0.0$), ($637.9 \rightarrow 313.68$) transition rates.  The isomer somewhat slows $\beta$ decay, but it is primarily interesting because it feeds the \nuc{121}{Sn} isomer.

\subsection{Sn ($Z=50$) Isotopes}

\begin{figure}[H]
\centering
\begin{subfigure}{.5\textwidth}
  \includegraphics[width=\linewidth]{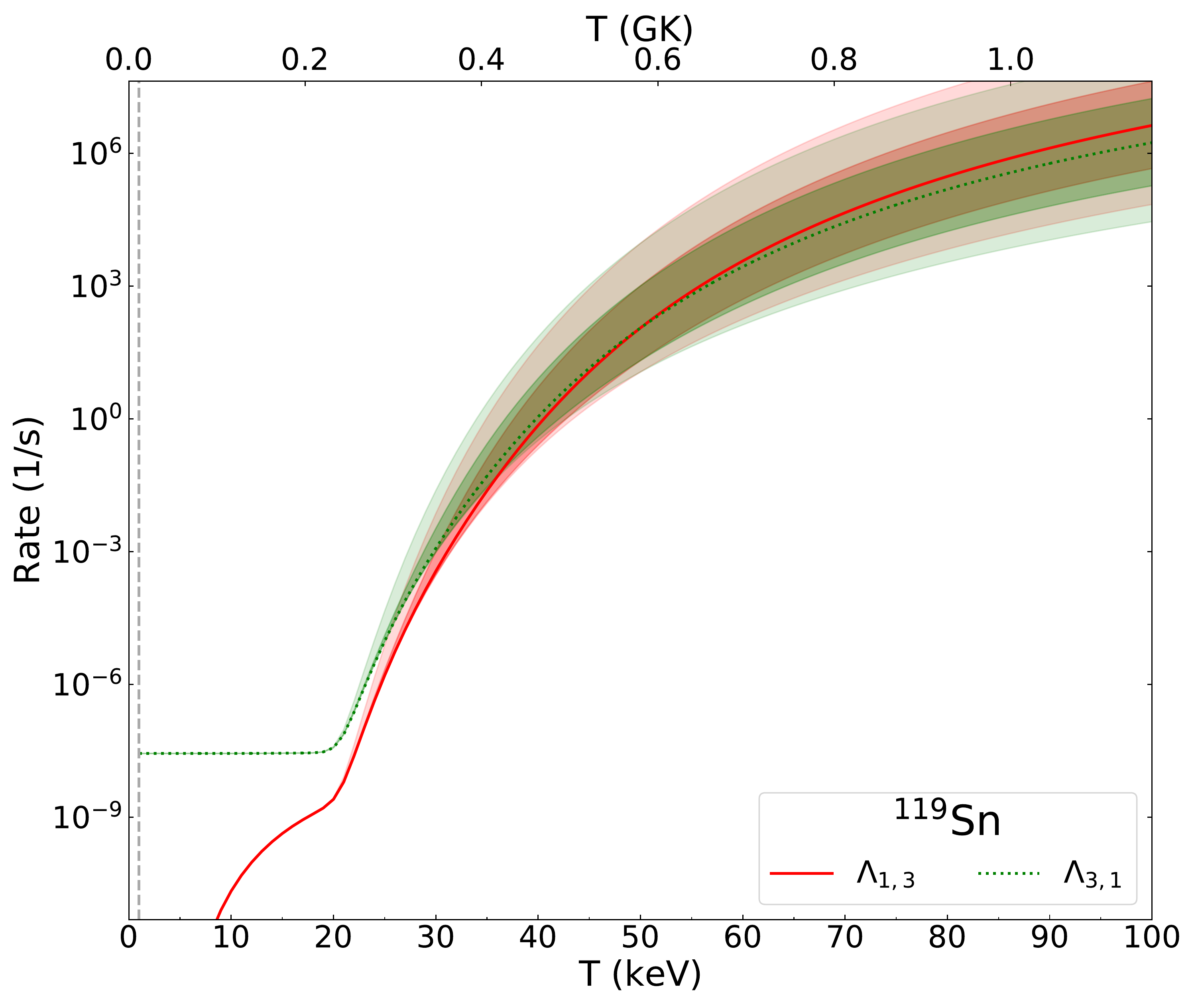}
  \includegraphics[width=\linewidth]{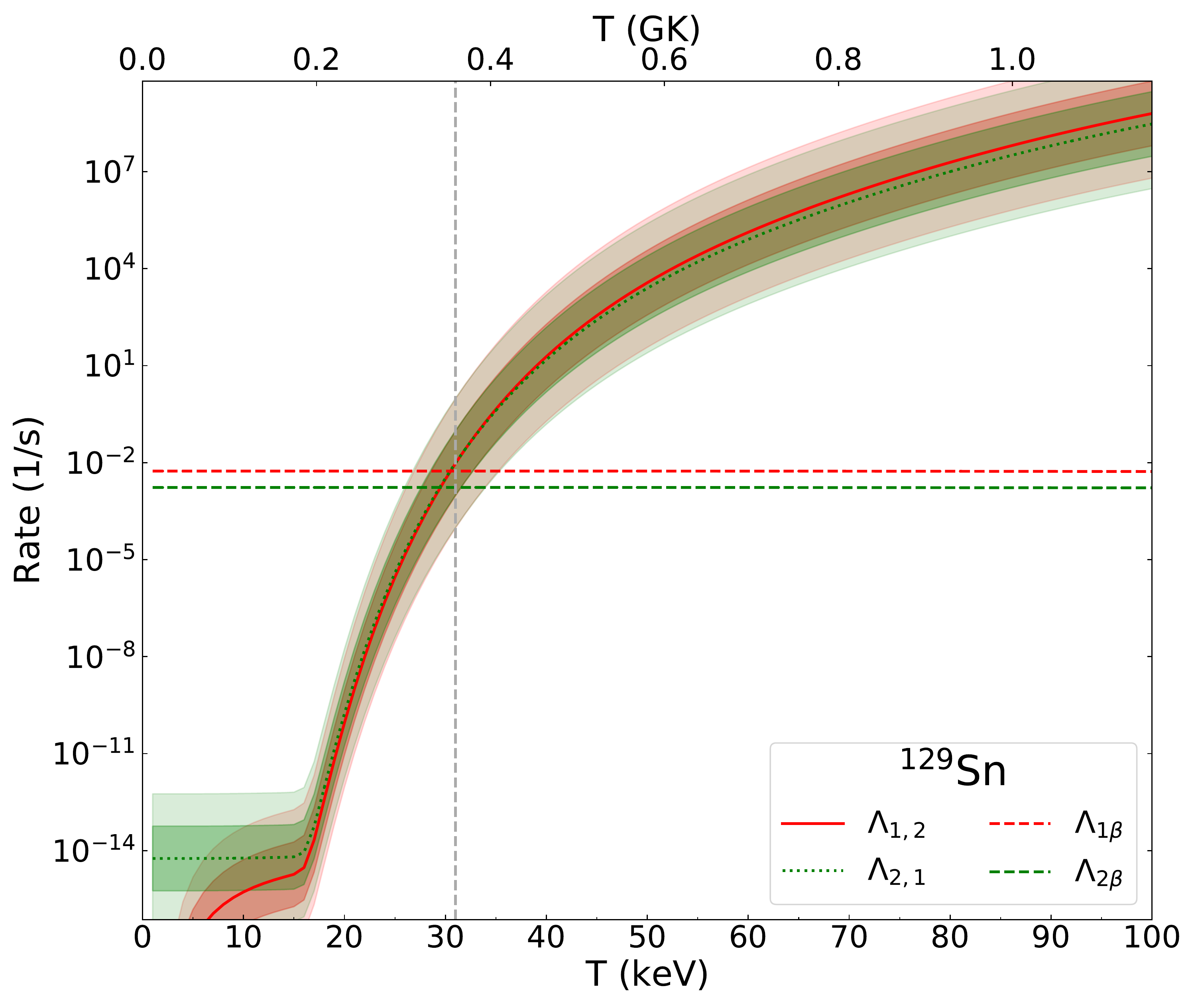}
\end{subfigure}%
\begin{subfigure}{.5\textwidth}
  \includegraphics[width=\linewidth]{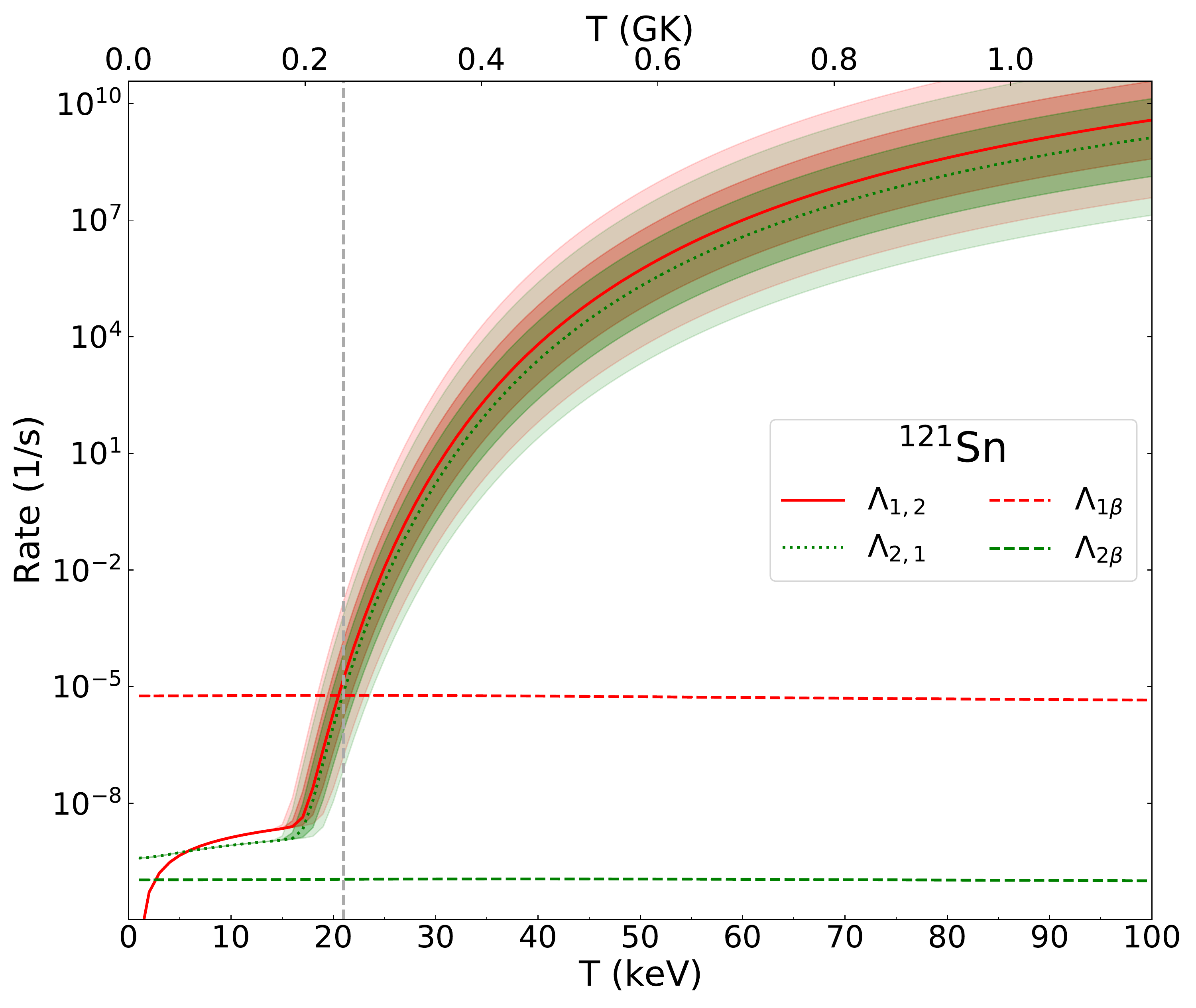}
\end{subfigure}
\caption{Effective transition rates for Sn ($Z=50$) isotopes. Darkest shaded band shows unmeasured rates increased/decreased by one order of magnitude; light by two orders of magnitude. Thermalization temperature, \ttherm{}, estimated by dashed vertical grey line.}
\label{fig:it_Sn}
\end{figure}

\nuc{119}{Sn}: Second $r$-process peak ($A\sim130$) nuclide. 157 measured levels, 30 in this calculation.  Isomer at 89.531 keV (type N).  Known uncertainties dominated by unmeasured ($1062.4 \rightarrow 787.01$), ($1062.4 \rightarrow 89.531$), ($1062.4 \rightarrow 921.39$), ($921.39 \rightarrow 787.01$) transition rates.  The relatively long half-life of $\sim$ 300 days may imply an x- or $\gamma$-ray signal from an $r$-process event.

\nuc{121}{Sn}: Second $r$-process peak ($A\sim130$) nuclide. 143 measured levels, 30 in this calculation.  Isomer at 6.31 keV (type B, 20 keV).  Known uncertainties dominated by unmeasured ($663.63 \rightarrow 0.0$), ($663.63 \rightarrow 6.31$), ($869.25 \rightarrow 0.0$), ($869.25 \rightarrow 663.63$), ($908.8 \rightarrow 0.0$), ($908.8 \rightarrow 663.63$) transition rates.  The isomer dramatically slows $\beta$ decay, which contributes to heating and a possible electromagnetic signal on the timescale of years after an $r$-process event.

\nuc{129}{Sn}: Second $r$-process peak ($A\sim130$) nuclide. 34 measured levels, 30 in this calculation.  Isomer at 35.15 keV (type B, 29 keV).  Known uncertainties dominated by unmeasured ($1043.66 \rightarrow 35.15$), ($1043.66 \rightarrow 763.7$), ($1043.66 \rightarrow 769.07$), ($1047.35 \rightarrow 0.0$), ($1047.35 \rightarrow 763.7$), ($1047.35 \rightarrow 769.07$), ($1054.21 \rightarrow 763.7$), ($1054.21 \rightarrow 769.07$), ($35.15 \rightarrow 0.0$), ($763.7 \rightarrow 0.0$), ($763.7 \rightarrow 35.15$), ($769.07 \rightarrow 0.0$), ($769.07 \rightarrow 35.15$) transition rates.  The isomer somewhat slows $\beta$ decay early in the $r$ process decay back to stability, possibly affecting the heating curve.

\subsection{Sb ($Z=51$) Isotopes}

\begin{figure}[H]
\centering
\begin{subfigure}{.5\textwidth}
  \includegraphics[width=\linewidth]{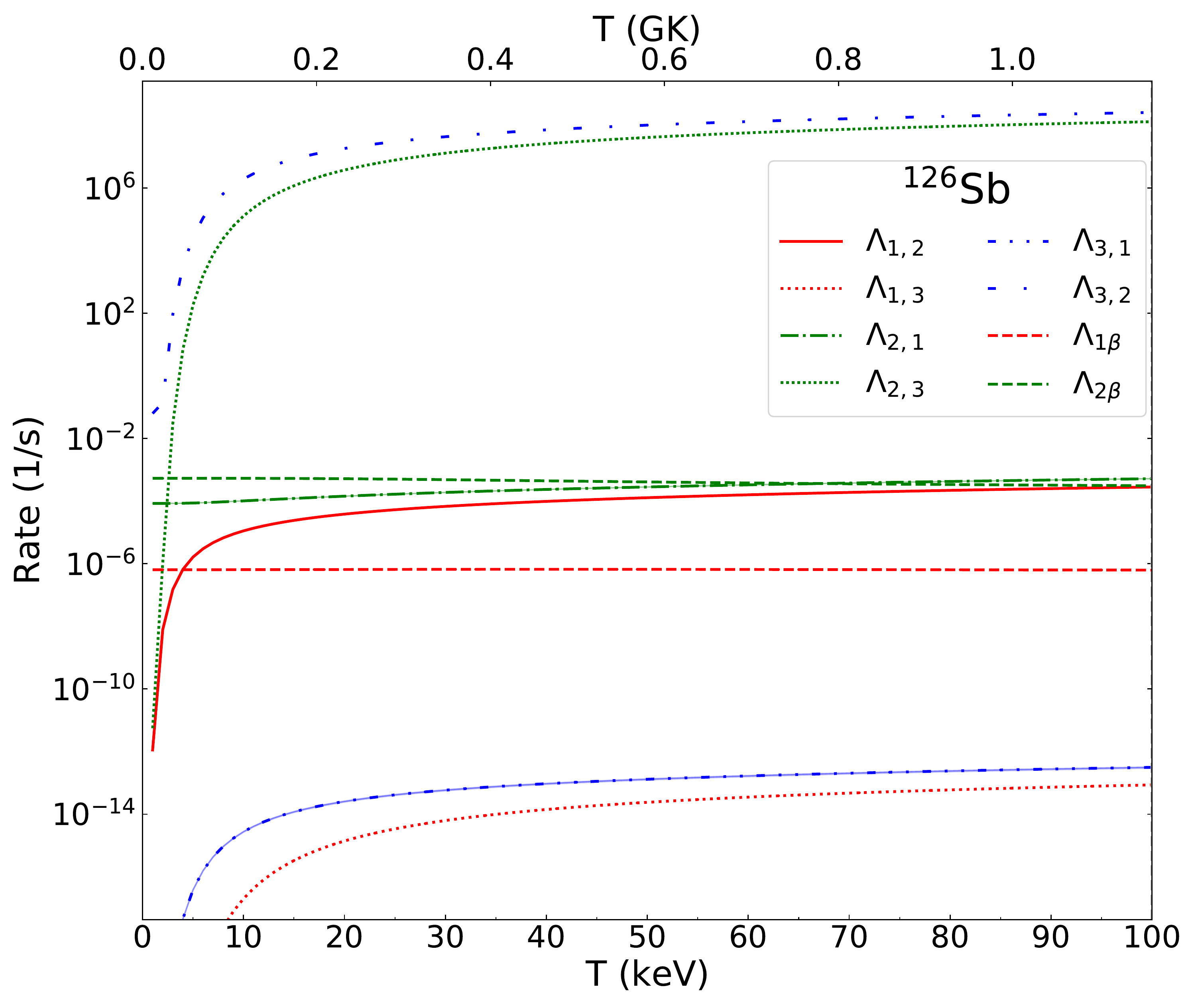}
  \includegraphics[width=\linewidth]{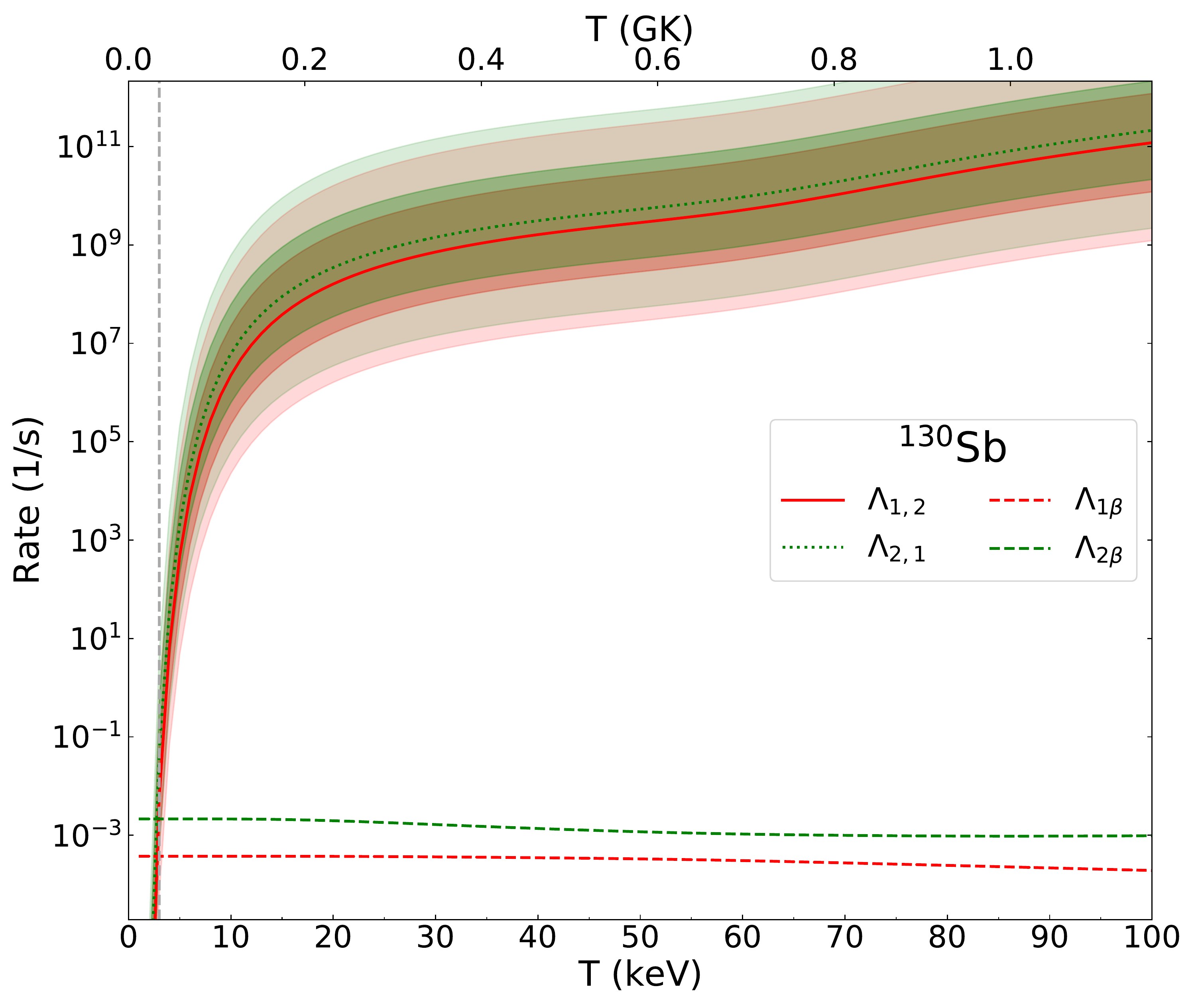}
\end{subfigure}%
\begin{subfigure}{.5\textwidth}
  \includegraphics[width=\linewidth]{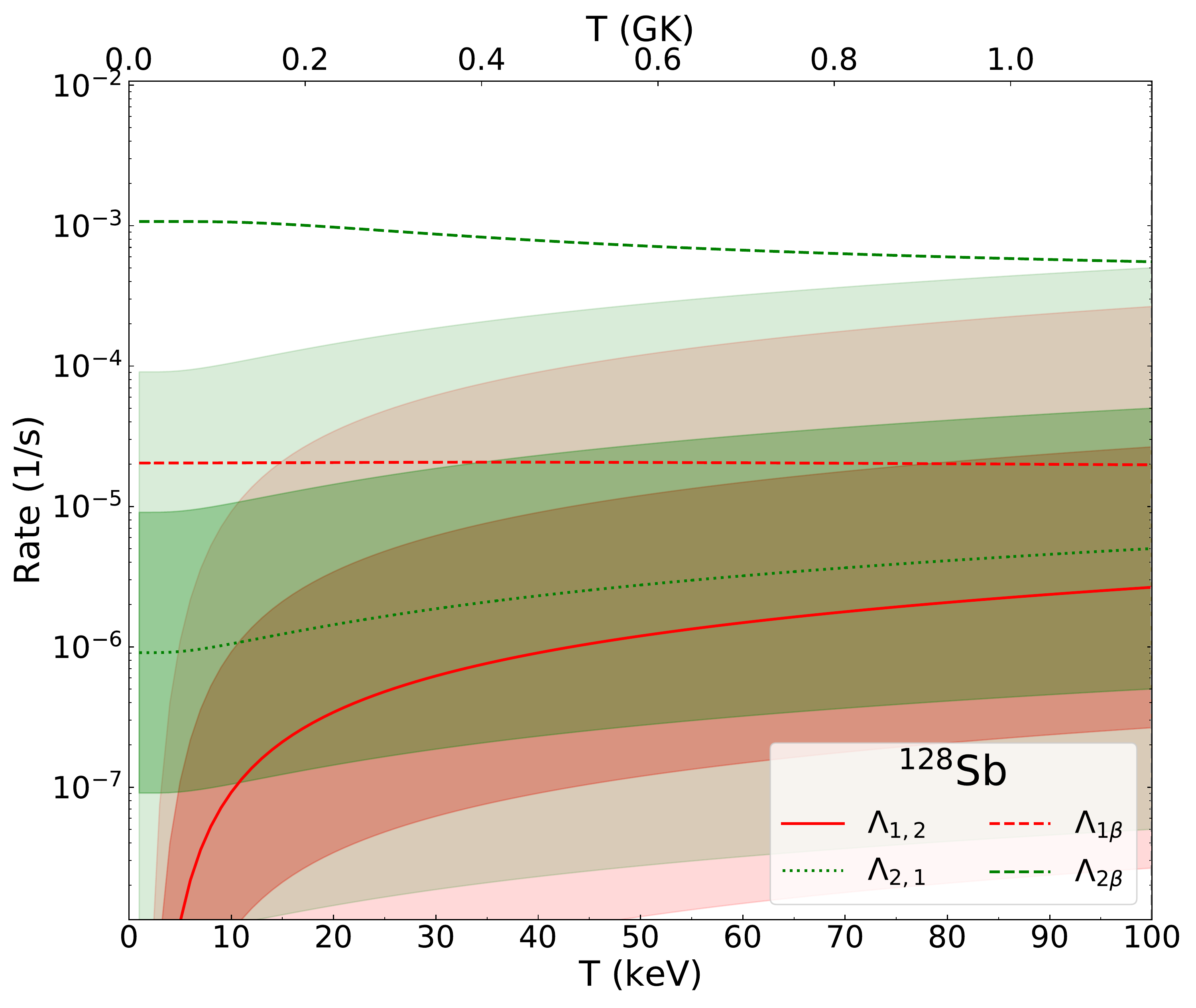}
\end{subfigure}
\caption{Effective transition rates for Sb ($Z=51$) isotopes. Darkest shaded band shows unmeasured rates increased/decreased by one order of magnitude; light by two orders of magnitude. Thermalization temperature, \ttherm{}, estimated by dashed vertical grey line. Note $^{128}$Sb never thermalizes due to lack of experimental data. }
\label{fig:it_Sb}
\end{figure}

\nuc{126}{Sb}: Second $r$-process peak ($A\sim130$) nuclide. 6 measured levels, 6 in this calculation.  Isomers at 17.7 keV (type A) and 40.4 keV (type N).  The effective transition rates are dominated at all temperatures by experimental state-to-state transition rates.  However, the  small number of measured states suggests that there are almost certainly many missing intermediate states that could dramatically change the effective rates.  Nevertheless, we do not expect new effects on the $r$ process due to the half-lives being much shorter than that of the $\beta$-decay parent.

\nuc{128}{Sb}: Second $r$-process peak ($A\sim130$) nuclide. 9 measured levels, 9 in this calculation.  Isomer at unknown energy, likely $<20$ keV (type A).  There are no measured transitions to ground and no measured intermediate state half-lives.  The isomer is potentially highly influential in the decay back to stability in the $r$ process as it seems to be heavily populated by the $\beta$-decay parent and it has a far shorter half-life than the ground state.

\nuc{130}{Sb}: Second $r$-process peak ($A\sim130$) nuclide. 61 measured levels, 30 in this calculation.  Isomer at 4.8 keV (type A).  Known uncertainties dominated by unmeasured ($4.8 \rightarrow 0.0$), ($84.67 \rightarrow 0.0$), ($84.67 \rightarrow 4.8$) transition rates.  The isomer significantly accelerates $\beta$ decay early in the $r$ process decay back to stability, possibly affecting the heating curve.

\subsection{Te ($Z=52$) Isotopes}

\begin{figure}[H]
\centering
\begin{subfigure}{.5\textwidth}
  \includegraphics[width=\linewidth]{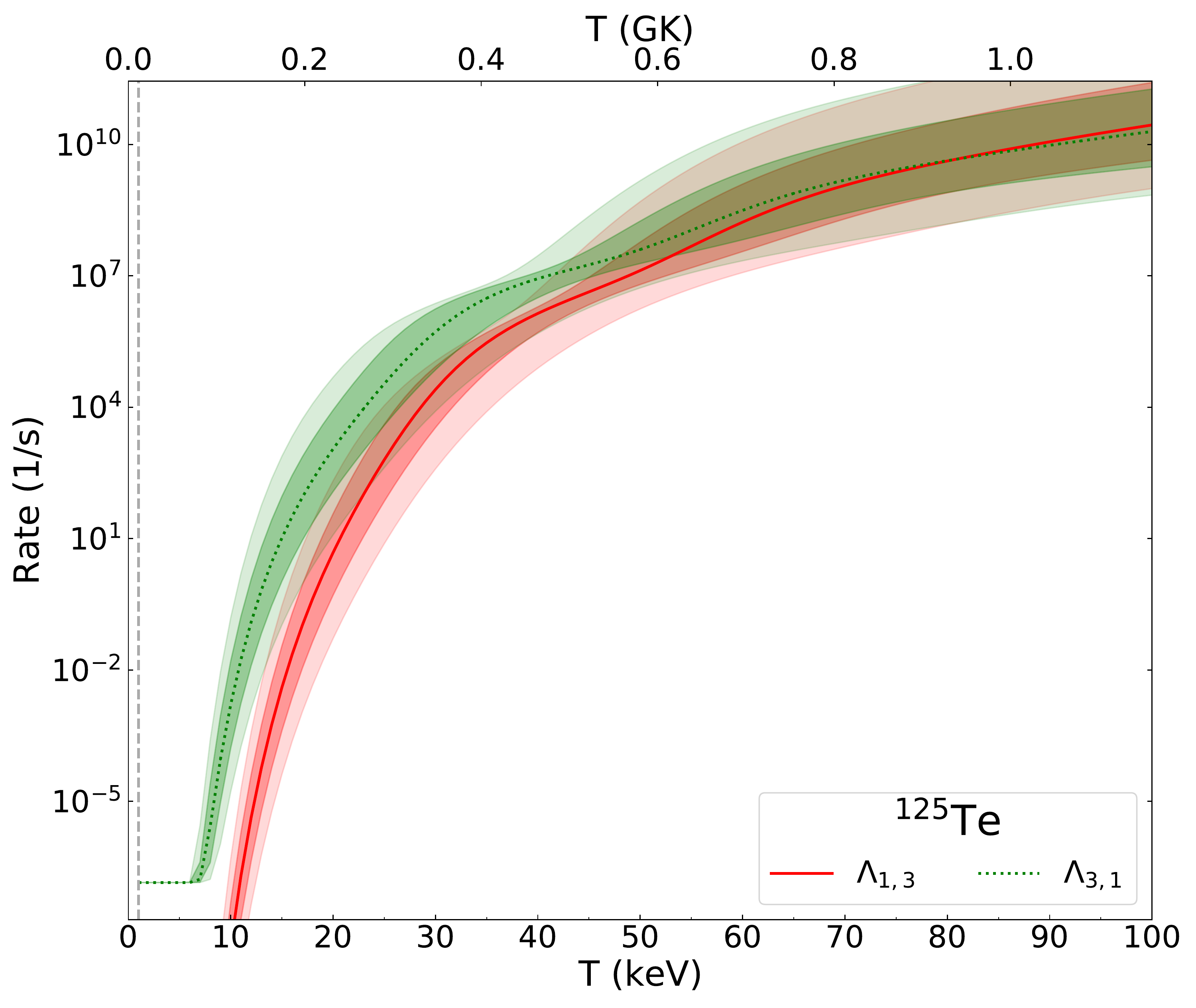}
  \includegraphics[width=\linewidth]{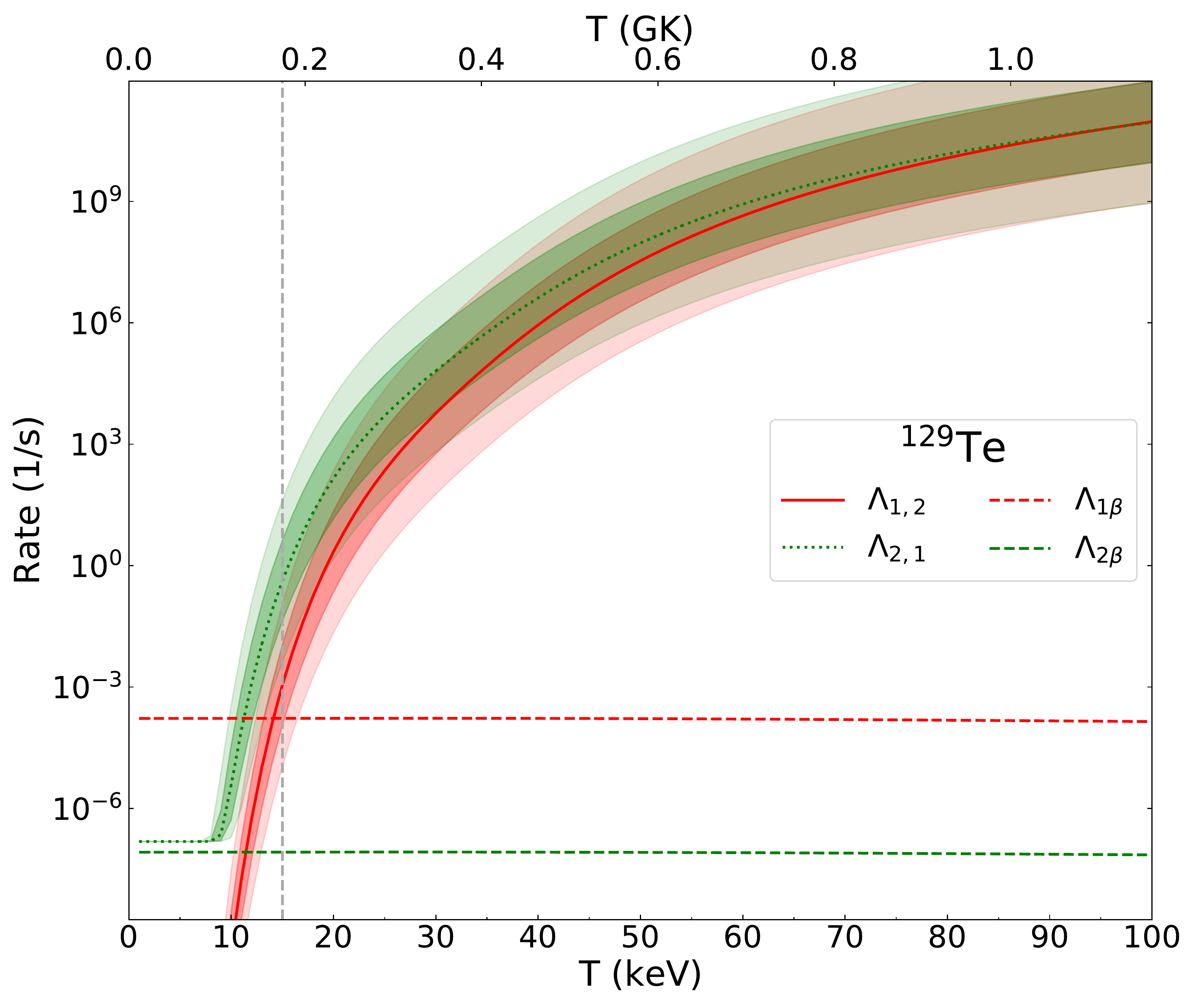}
\end{subfigure}%
\begin{subfigure}{.5\textwidth}
  \includegraphics[width=\linewidth]{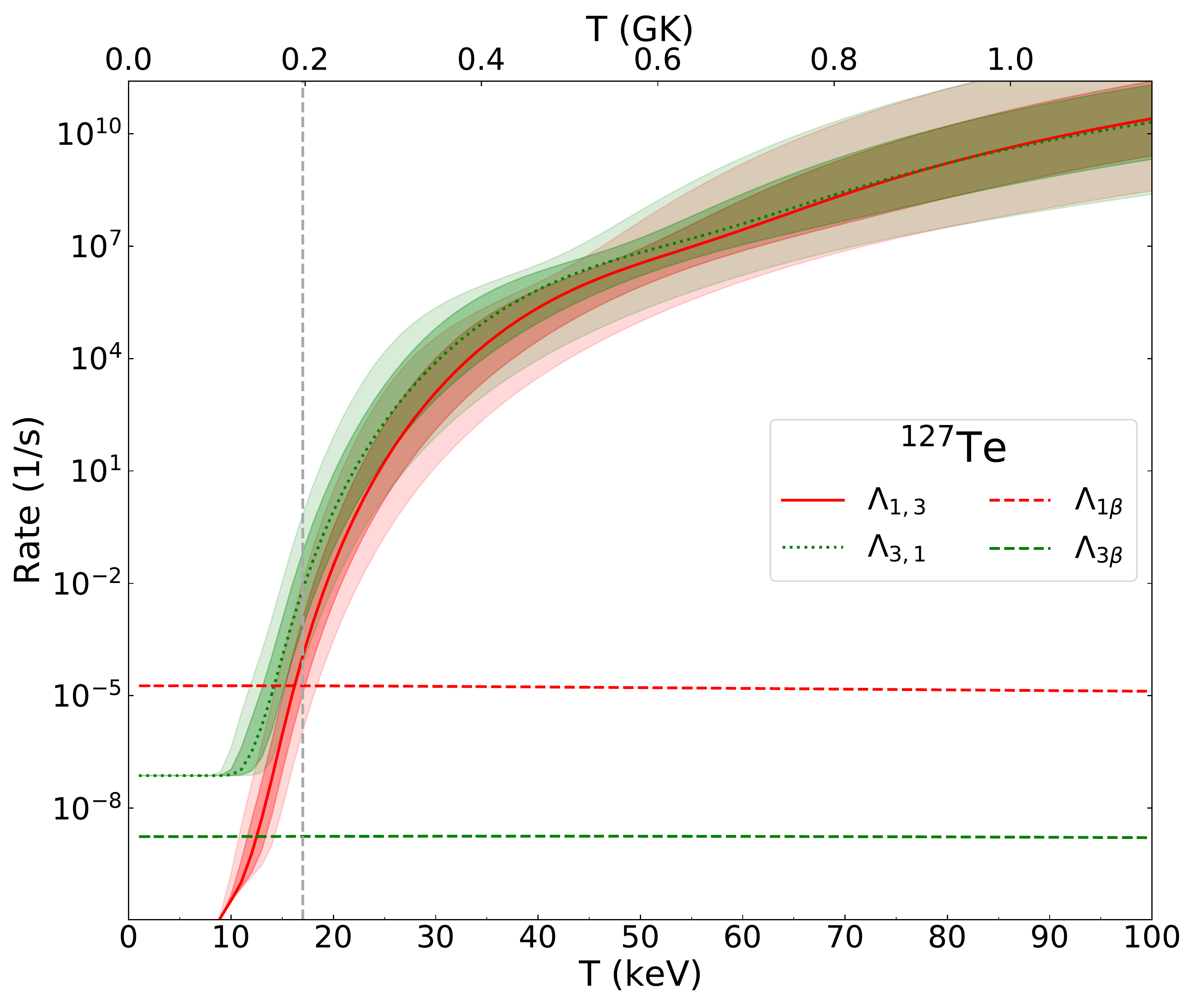}
  \includegraphics[width=\linewidth]{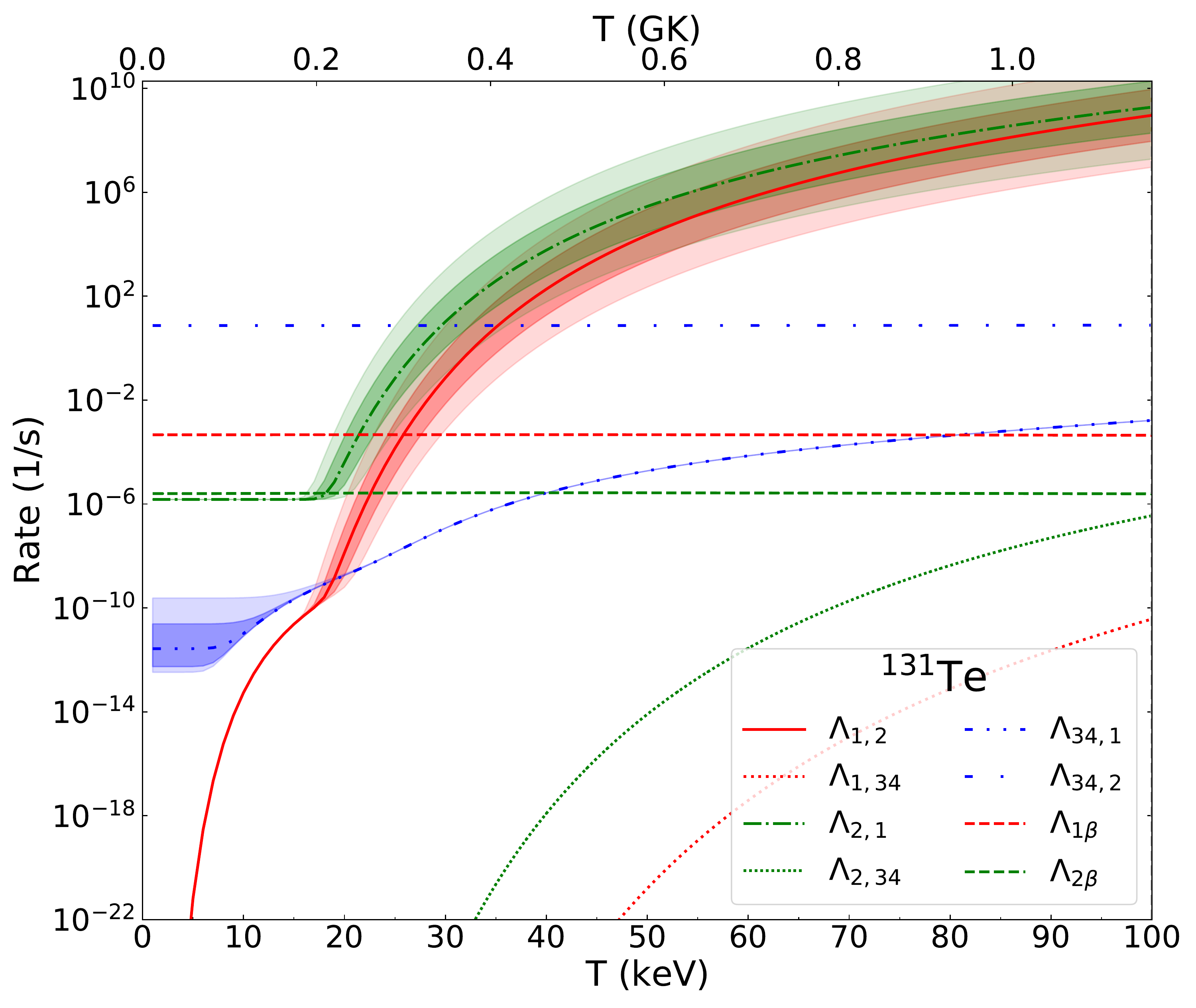}
\end{subfigure}
\begin{subfigure}{.5\textwidth}
  \includegraphics[width=\linewidth]{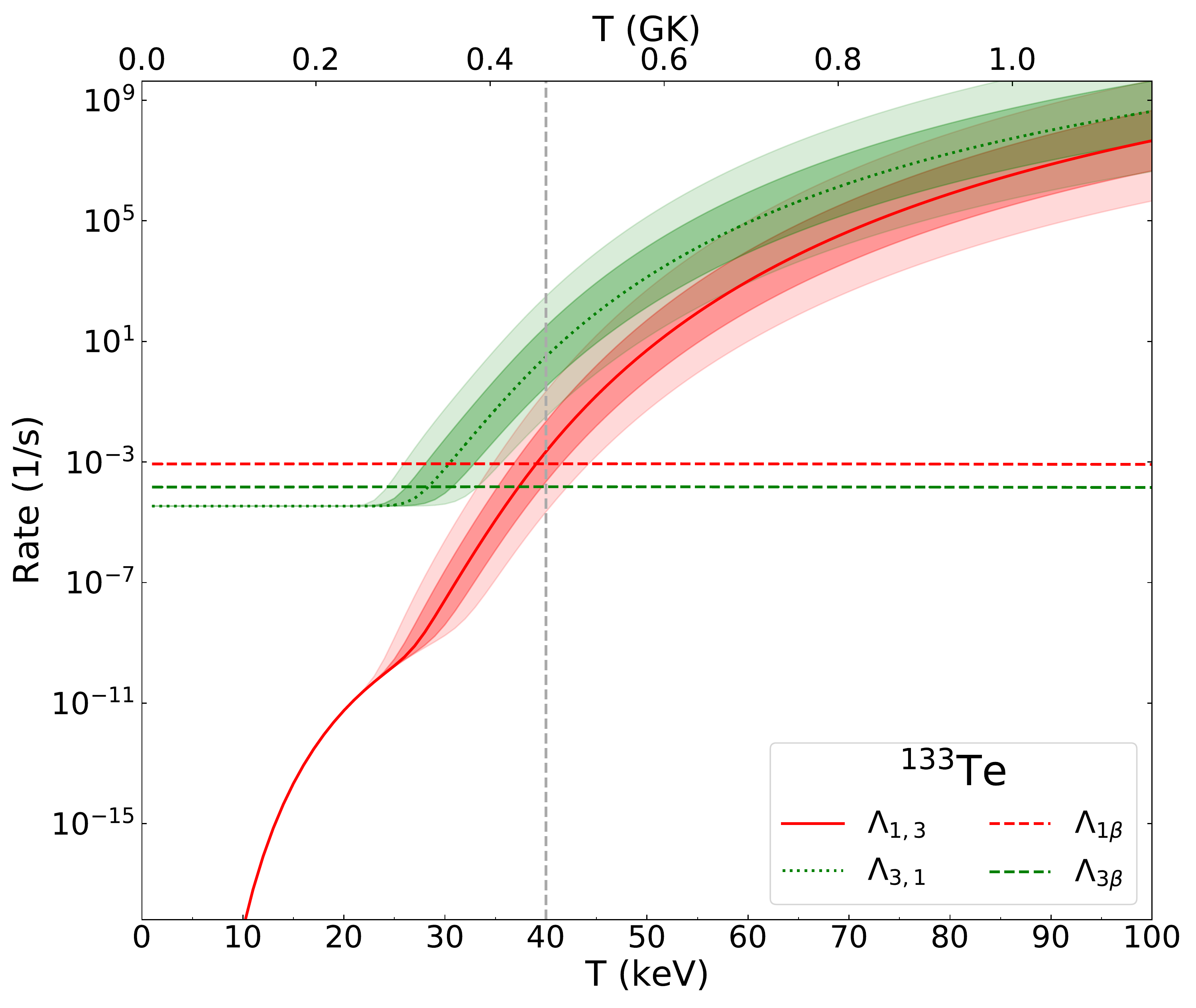}
\end{subfigure}
\caption{Effective transition rates for Te ($Z=51$) isotopes. Darkest shaded band shows unmeasured rates increased/decreased by one order of magnitude; light by two orders of magnitude. Thermalization temperature, \ttherm{}, estimated by dashed vertical grey line.}
\label{fig:it_Te}
\end{figure}

\nuc{125}{Te}: Second $r$-process peak ($A\sim130$) nuclide. 309 measured levels, 30 in this calculation.  Isomer at 144.775 keV (type N).  Known uncertainties dominated by unmeasured ($321.09 \rightarrow 35.4925$), ($402.09 \rightarrow 321.09$), ($402.09 \rightarrow 35.4925$), ($463.3668 \rightarrow 402.09$), ($525.228 \rightarrow 402.09$), ($652.9 \rightarrow 35.4925$), ($652.9 \rightarrow 402.09$), ($671.4448 \rightarrow 525.228$) transition rates.  No expected new effect on the $r$ process due to its half-life being much shorter than that of its $\beta$-decay parent.

\nuc{127}{Te}: Second $r$-process peak ($A\sim130$) nuclide. 283 measured levels, 30 in this calculation.  Isomer at 88.23 keV (type B, 13 keV).  Known uncertainties dominated by unmeasured ($340.87 \rightarrow 0.0$), ($473.26 \rightarrow 0.0$), ($631.4 \rightarrow 340.87$), ($631.4 \rightarrow 473.26$), ($685.09 \rightarrow 340.87$), ($685.09 \rightarrow 473.26$), ($685.09 \rightarrow 631.4$) transition rates.  The isomer dramatically slows $\beta$ decay, which could influence the heating curve and possibly produce an x-ray signal weeks or months after an $r$-process event.

\nuc{129}{Te}: Second $r$-process peak ($A\sim130$) nuclide. 407 measured levels, 30 in this calculation.  Isomer at 105.51 keV (type B, 10 keV).  Known uncertainties dominated by unmeasured ($180.356 \rightarrow 0.0$), ($360.0 \rightarrow 0.0$), ($360.0 \rightarrow 180.356$), ($455.0 \rightarrow 0.0$), ($455.0 \rightarrow 105.51$), ($455.0 \rightarrow 360.0$) transition rates.  The isomer dramatically slows $\beta$ decay, which could influence the heating curve and possibly produce an x- or $\gamma$-ray signal weeks or months after an $r$-process event.

\nuc{131}{Te}: Second $r$-process peak ($A\sim130$) nuclide. 319 measured levels, 44 in this calculation.  Isomers at 182.258 keV (type B, 21 keV) and 1940.0 keV (type N).  Known uncertainties dominated by unmeasured ($1267.5 \rightarrow 0.0$), ($1267.5 \rightarrow 802.214$), ($1267.5 \rightarrow 880.315$), ($642.331 \rightarrow 0.0$), ($802.214 \rightarrow 182.258$), ($880.315 \rightarrow 182.258$), ($880.315 \rightarrow 642.331$), ($880.315 \rightarrow 802.214$), ($943.43 \rightarrow 0.0$), ($943.43 \rightarrow 642.331$), ($943.43 \rightarrow 802.214$), ($943.43 \rightarrow 880.315$) transition rates.  The 1940 keV isomer decays to lower energies very quickly and likely has no effect, but the 182.258 keV isomer substantially slows $\beta$ decay, which could influence the heating curve and possibly produce an x- or $\gamma$-ray signal days after an $r$-process event.

\nuc{133}{Te}: Second $r$-process peak ($A\sim130$) nuclide. 37 measured levels, 30 in this calculation.  Isomer at 334.26 keV (type B, 29 keV).  Known uncertainties dominated by unmeasured ($1096.22 \rightarrow 0.0$), ($1096.22 \rightarrow 334.26$), ($1265.326 \rightarrow 0.0$), ($1500.56 \rightarrow 0.0$), ($1500.56 \rightarrow 1096.22$), ($1500.56 \rightarrow 334.26$), ($1639.5 \rightarrow 1096.22$), ($1639.5 \rightarrow 1265.326$), ($1639.5 \rightarrow 334.26$) transition rates.  The isomer somewhat slows $\beta$ decay in the first hours after an $r$-process event, possibly affecting the heating curve.

\subsection{Xe ($Z=54$) Isotopes}

\begin{figure}[H]
\centering
\begin{subfigure}{.5\textwidth}
  \includegraphics[width=\linewidth]{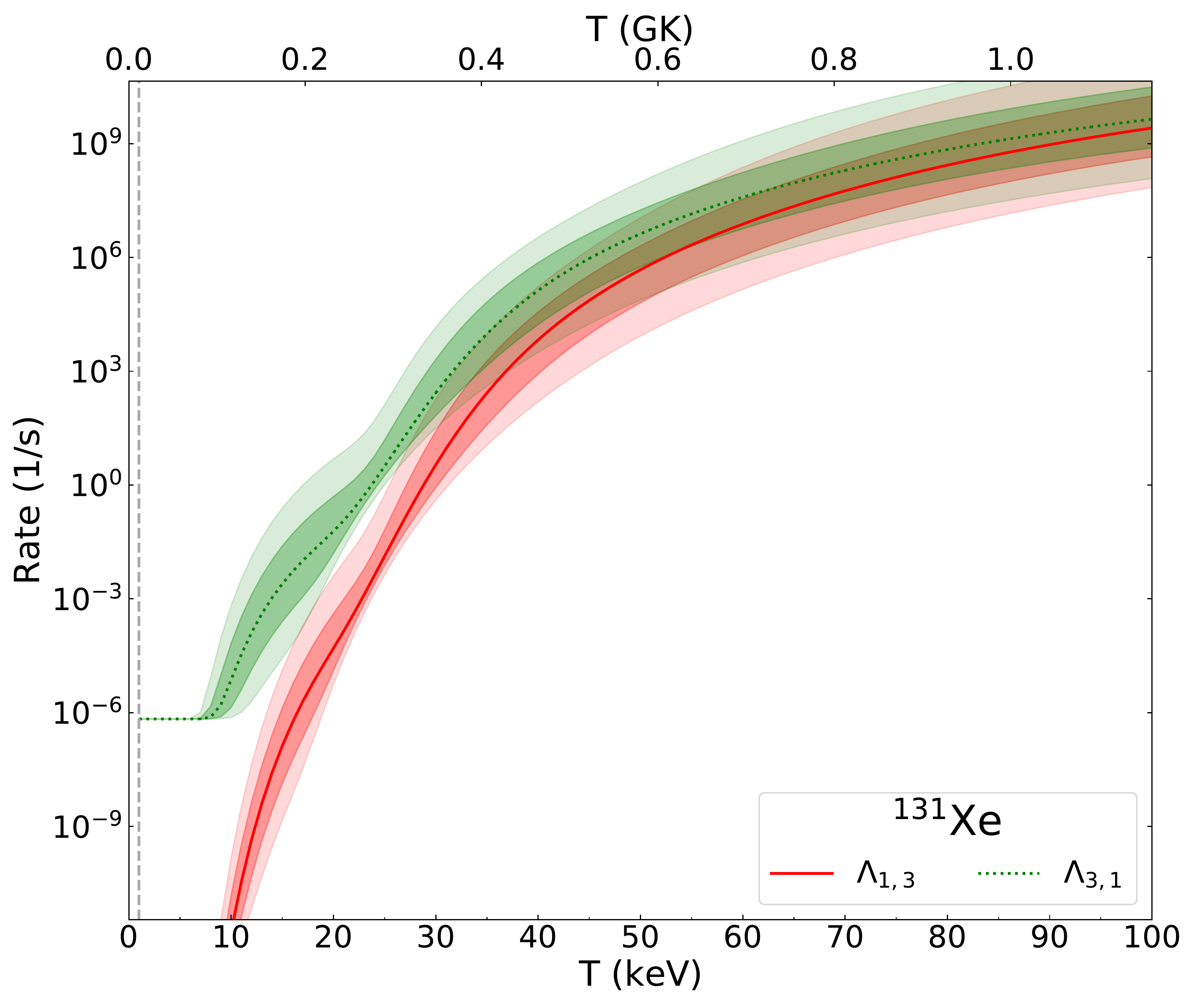}
\end{subfigure}%
\begin{subfigure}{.5\textwidth}
  \includegraphics[width=\linewidth]{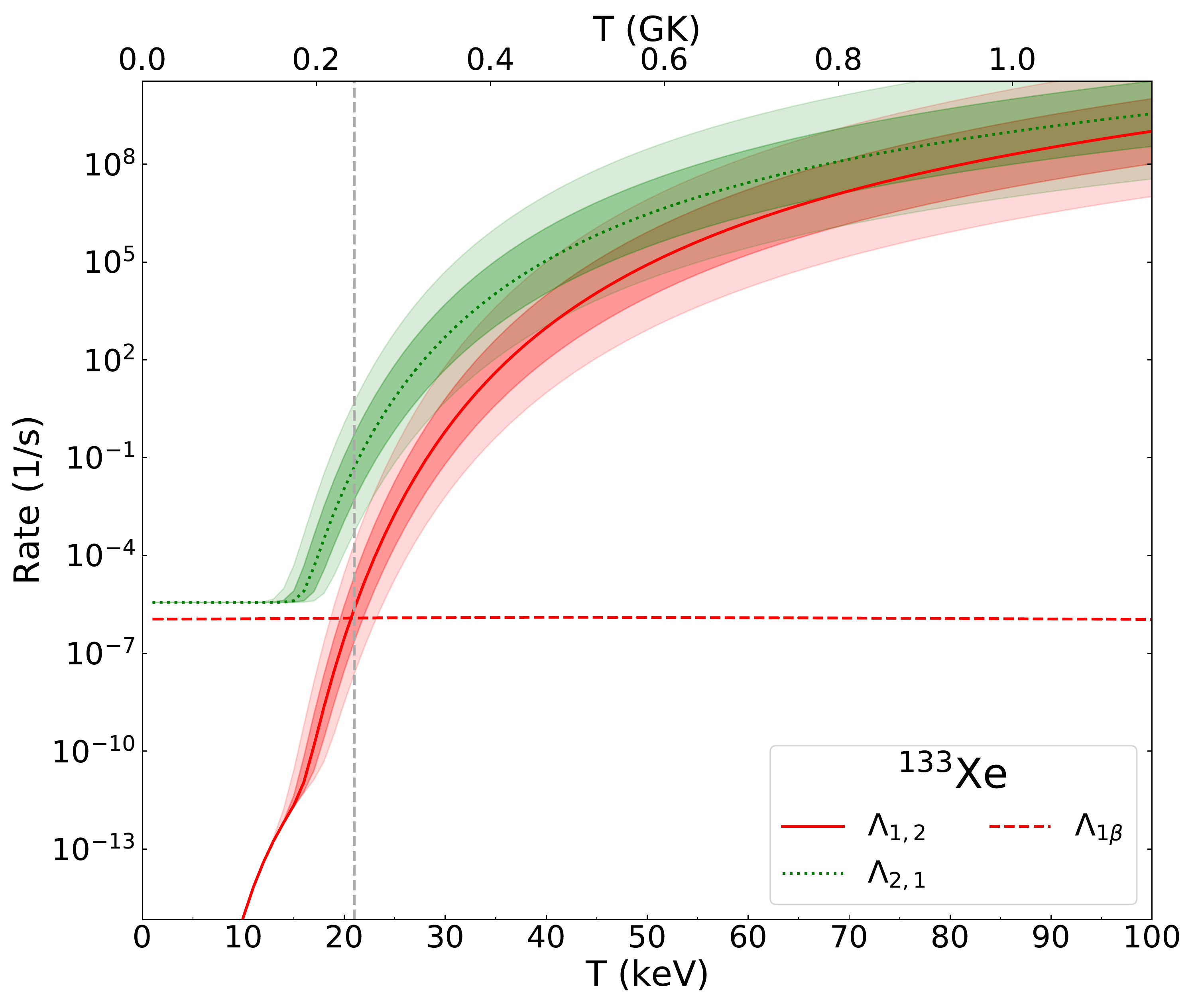}
\end{subfigure}%
\caption{Effective transition rates for Xe ($Z=54$) isotopes. Darkest shaded band shows unmeasured rates increased/decreased by one order of magnitude; light by two orders of magnitude. Thermalization temperature, \ttherm{}, estimated by dashed vertical grey line.}
\label{fig:it_Xe}
\end{figure}

\nuc{131}{Xe}: Second $r$-process peak ($A\sim130$) nuclide. 58 measured levels, 30 in this calculation.  Isomer at 163.93 keV (type N).  Known uncertainties dominated by unmeasured ($341.144 \rightarrow 0.0$), ($636.99 \rightarrow 163.93$), ($666.934 \rightarrow 636.99$), ($722.909 \rightarrow 666.934$), ($971.22 \rightarrow 163.93$), ($971.22 \rightarrow 341.144$), ($971.22 \rightarrow 666.934$), ($973.11 \rightarrow 341.144$), ($973.11 \rightarrow 666.934$) transition rates.  The isomer half-life of $\sim$12 days is a bit longer than that of its $\beta$-decay parent, and its de-excitation may produce an x-ray signal.

\nuc{133}{Xe}: Second $r$-process peak ($A\sim130$) nuclide. 29 measured levels, 29 in this calculation.  Isomer at 233.221 keV (type N).  Known uncertainties dominated by unmeasured ($529.872 \rightarrow 0.0$), ($529.872 \rightarrow 233.221$), ($607.87 \rightarrow 0.0$), ($607.87 \rightarrow 233.221$), ($743.752 \rightarrow 233.221$), ($875.328 \rightarrow 0.0$), ($875.328 \rightarrow 529.872$), ($875.328 \rightarrow 607.87$), ($875.328 \rightarrow 743.752$) transition rates.  The isomer de-excitation may produce an x-ray signal a few days after an $r$-process event.

\subsection{Ba ($Z=56$), Pr ($Z=59$), and Ho ($Z=67$) Isotopes}

\begin{figure}[H]
\centering
\begin{subfigure}{.5\textwidth}
  \includegraphics[width=\linewidth]{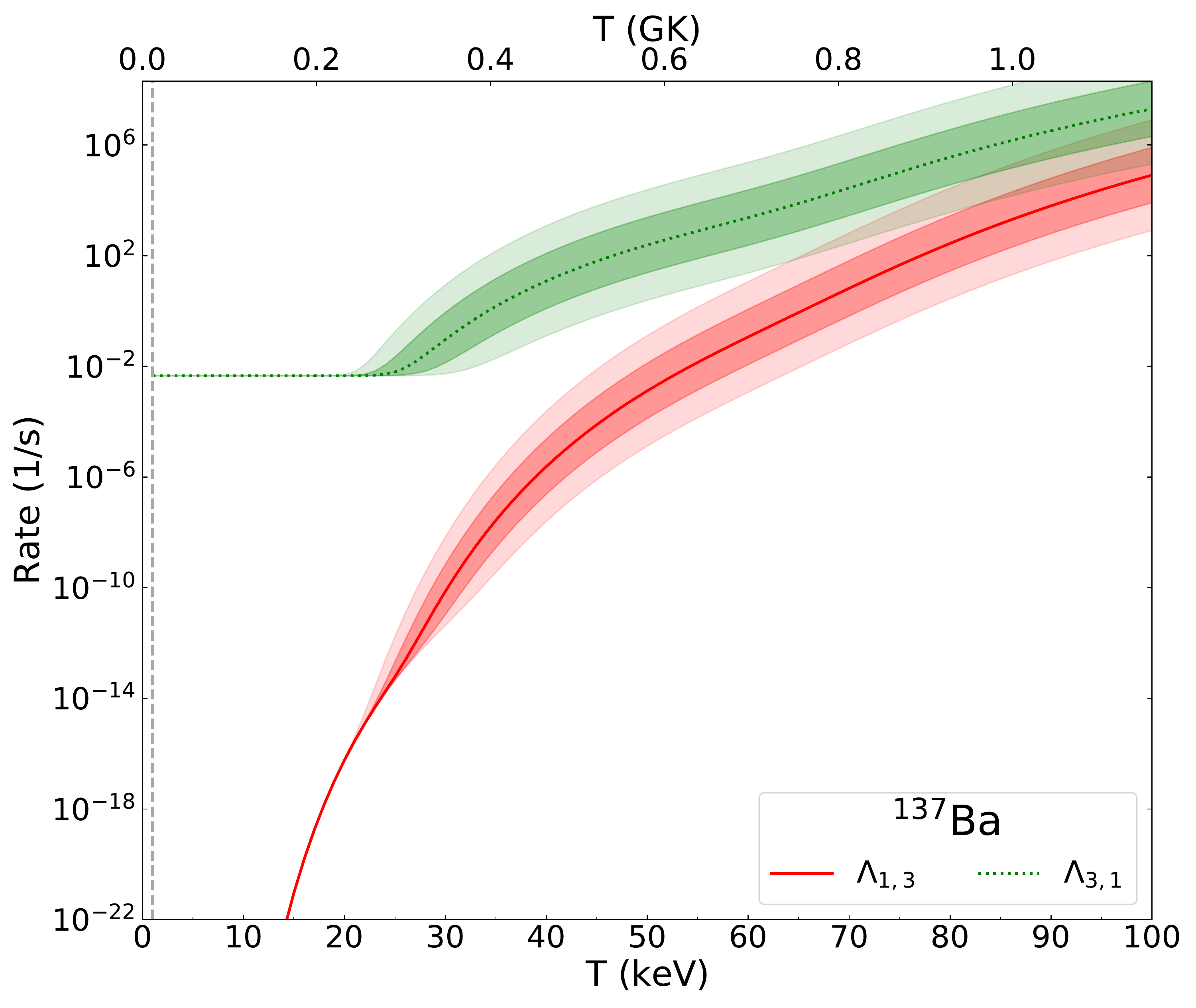}
  \includegraphics[width=\linewidth]{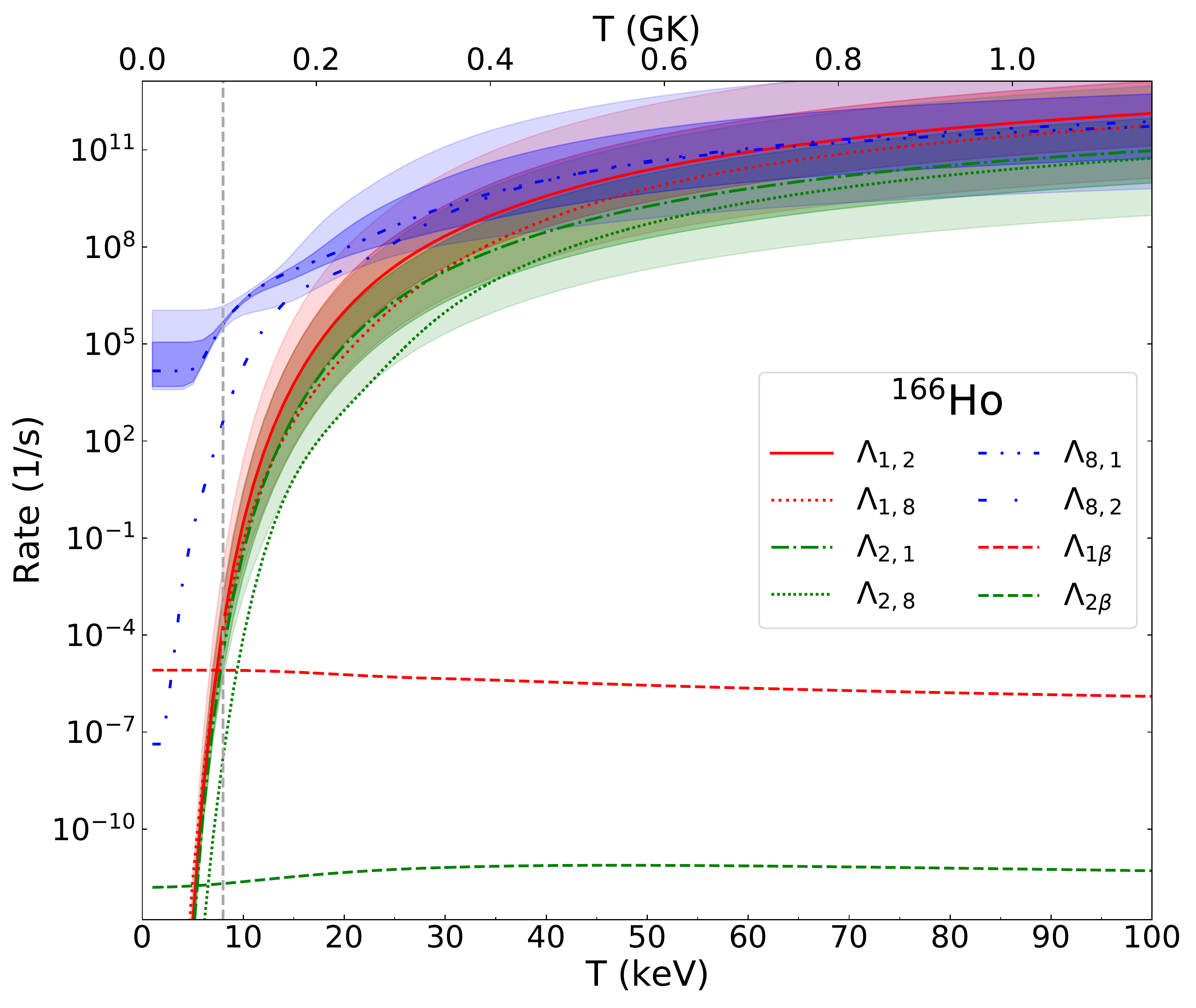}
\end{subfigure}%
\begin{subfigure}{.5\textwidth}
  \includegraphics[width=\linewidth]{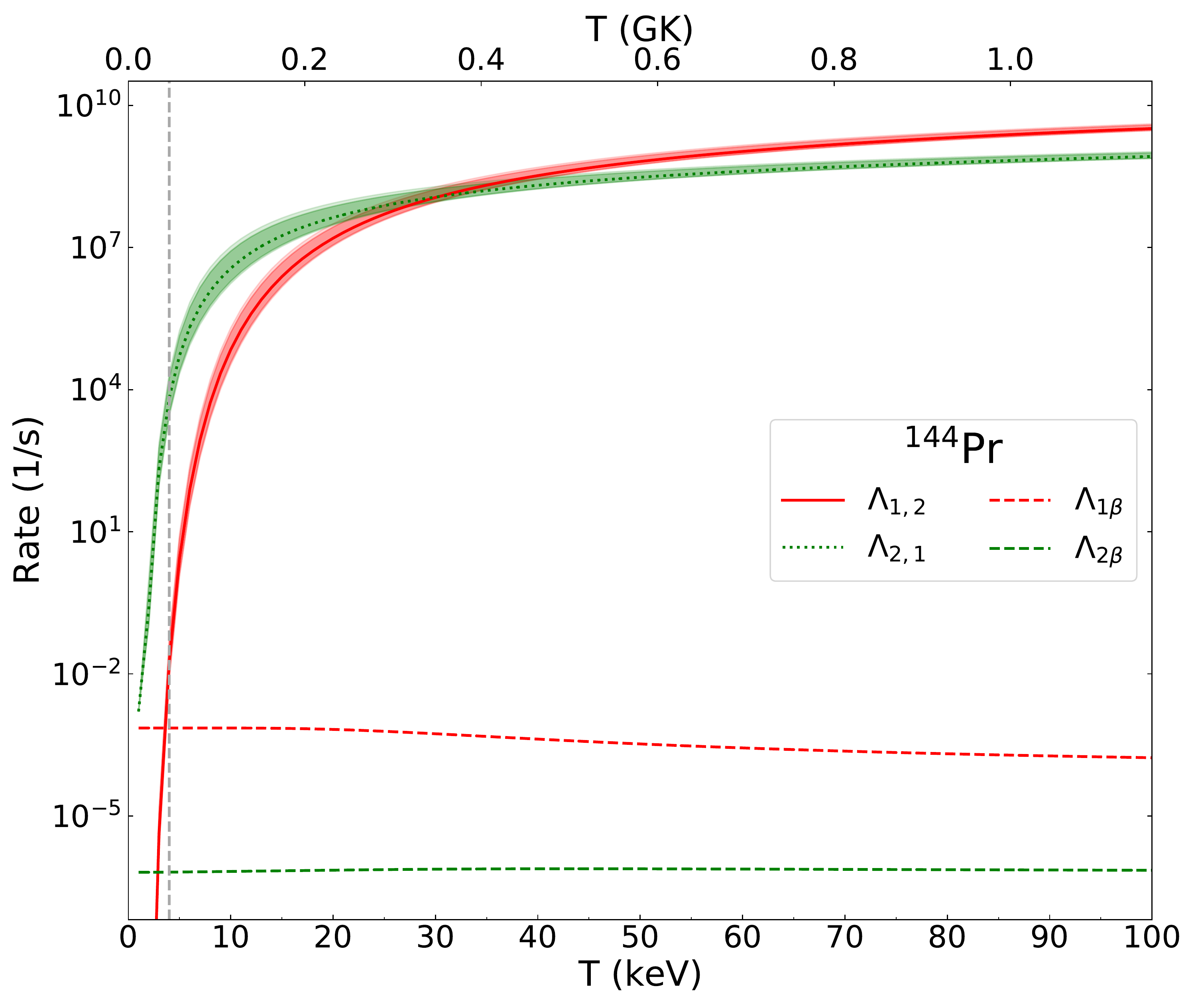}
\end{subfigure}%
\caption{Effective transition rates for Ba ($Z=56$), Pr ($Z=59$) and Ho ($Z=67$) isotopes. Darkest shaded band shows unmeasured rates increased/decreased by one order of magnitude; light by two orders of magnitude. Thermalization temperature, \ttherm{}, estimated by dashed vertical grey line.}
\label{fig:it_Ba}
\end{figure}

\nuc{137}{Ba}: Second $r$-process peak ($A\sim130$) nuclide. 96 measured levels, 30 in this calculation.  Isomer at 661.659 keV (type N).  Known uncertainties dominated by unmeasured ($1251.82 \rightarrow 661.659$), ($1293.9 \rightarrow 0.0$), ($1293.9 \rightarrow 1251.82$) transition rates.  No expected new effect on the $r$ process due to its half-life being much shorter than that of its $\beta$-decay parent.

\nuc{144}{Pr}: Rare earth nuclide. 5 measured levels, 5 in this calculation.  Isomer at 59.03 keV (type N).  Known uncertainties dominated by unmeasured ($80.12 \rightarrow 59.03$), ($99.952 \rightarrow 80.12$) transition rates.  However, the small number of measured states suggests that there are almost certainly many missing intermediate states that could dramatically change the effective rates.  Nevertheless, we do not expect new effects on the $r$ process due to the half-lives being much shorter than that of the $\beta$-decay parent.

\nuc{166}{Ho}: Rare earth nuclide. 356 measured levels, 30 in this calculation.  Isomers at 5.969 keV (type B, 6 keV) and 190.9021 keV (type N).  Known uncertainties dominated by unmeasured ($171.0738 \rightarrow 54.2391$), ($180.467 \rightarrow 171.0738$), ($180.467 \rightarrow 5.969$), ($180.467 \rightarrow 54.2391$), ($260.6625 \rightarrow 180.467$), ($263.7876 \rightarrow 180.467$), ($296.8 \rightarrow 0.0$), ($296.8 \rightarrow 263.7876$), ($296.8 \rightarrow 5.969$), ($296.8 \rightarrow 54.2391$), ($296.8 \rightarrow 82.4707$), ($54.2391 \rightarrow 5.969$) transition rates.  The 191 keV isomer decays to lower energies very quickly and likely has no effect, but the 6 keV isomer dramatically slows $\beta$ decay, which could influence the heating curve.  It also has strong $\gamma$ and x-ray lines, and the 1200 yr half-life make it a compelling candidate for observing old $r$-process remnants.

\subsection{Os ($Z=76$), Ir ($Z=77$), and Pt ($Z=78$) Isotopes}

\begin{figure}[H]
\centering
\begin{subfigure}{.5\textwidth}
  \includegraphics[width=\linewidth]{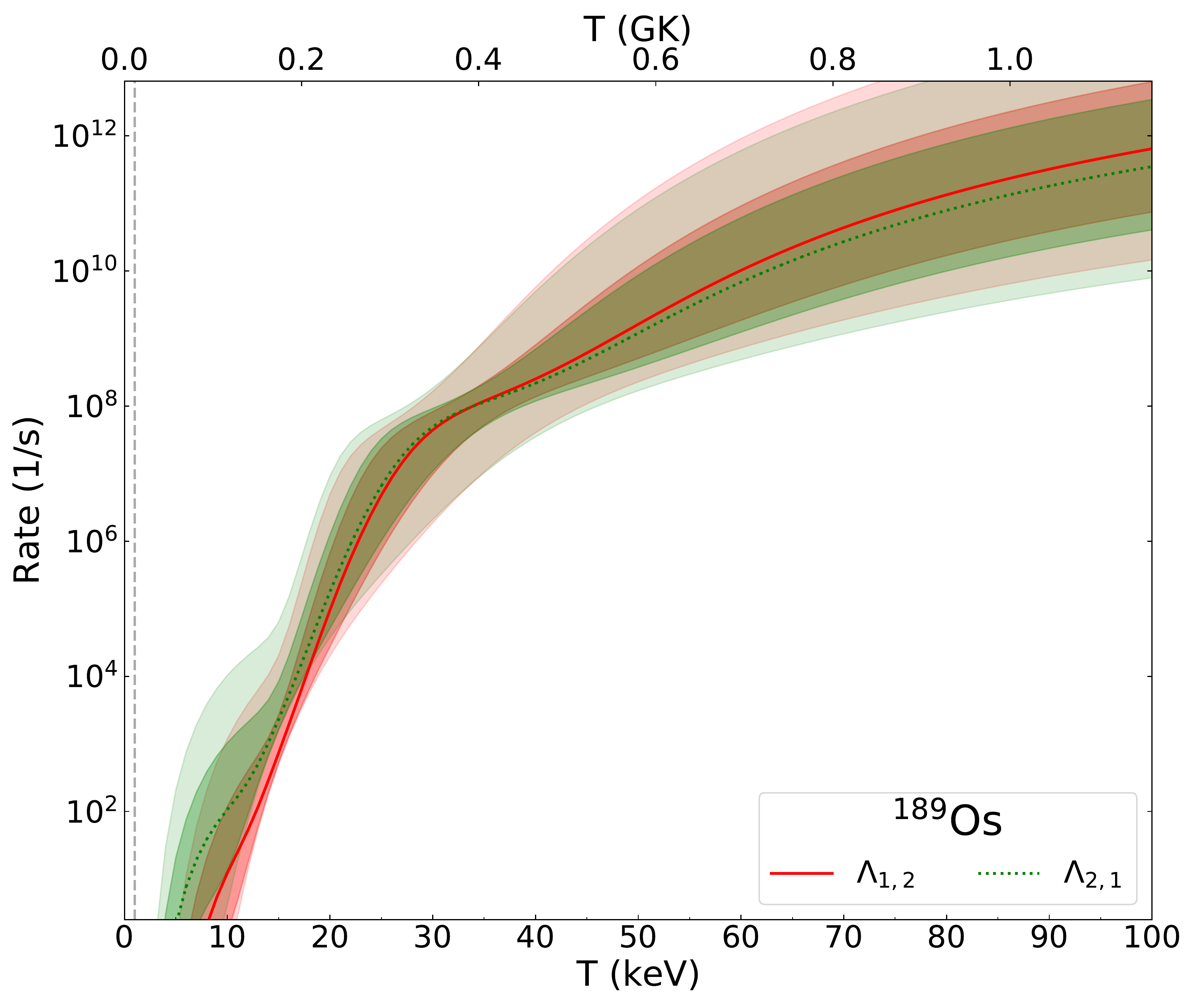}
  \includegraphics[width=\linewidth]{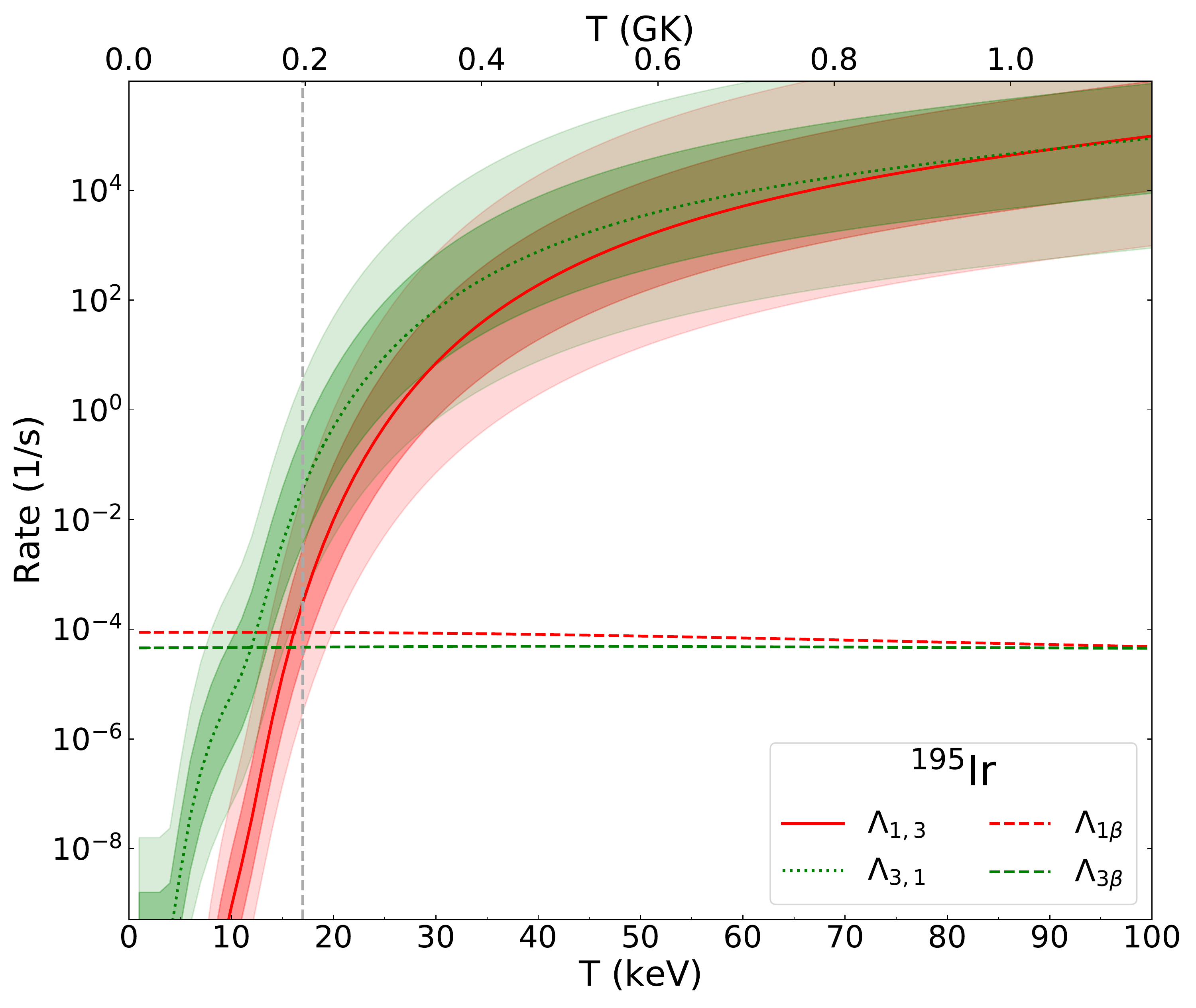}
\end{subfigure}%
\begin{subfigure}{.5\textwidth}
  \includegraphics[width=\linewidth]{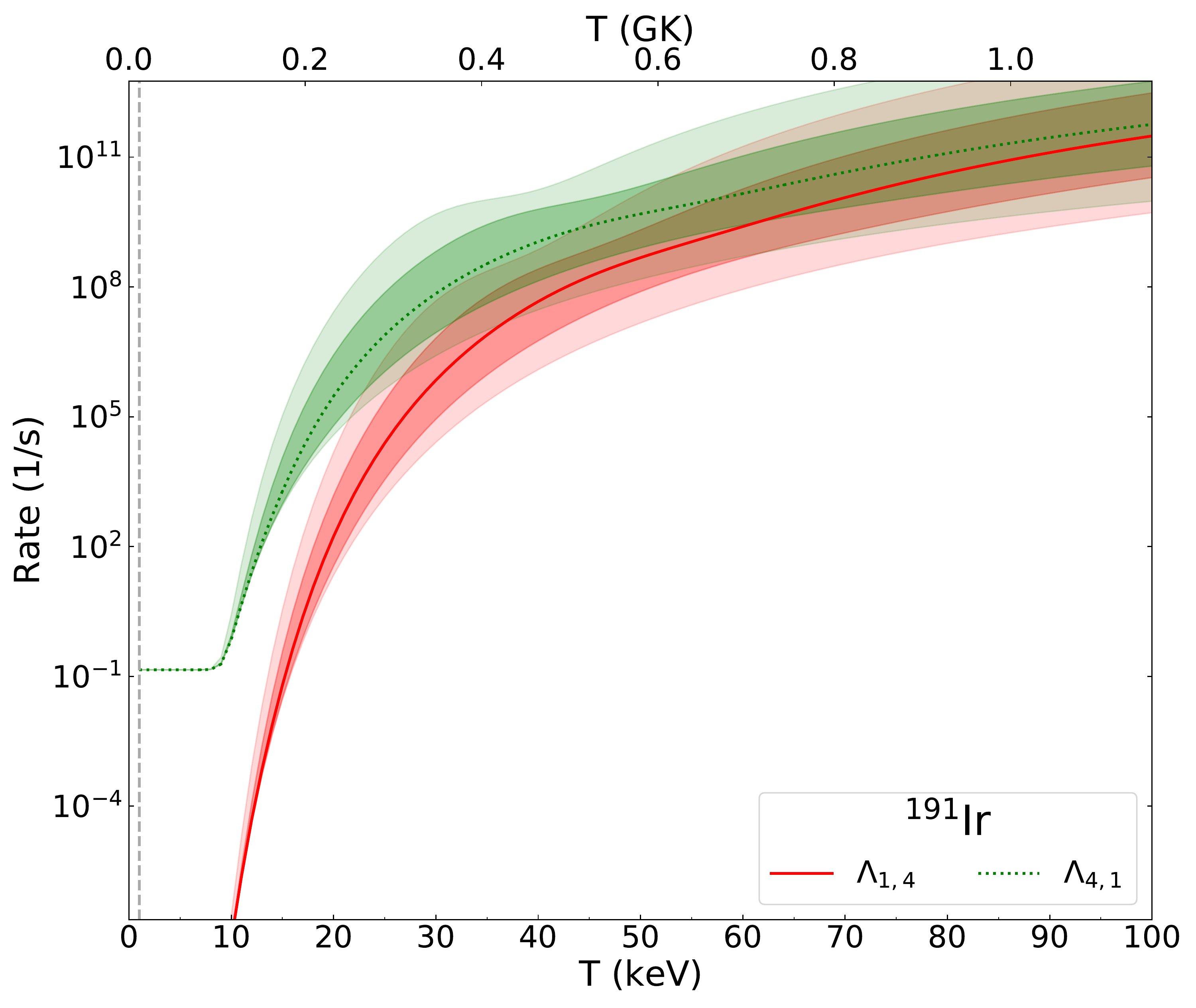}
  \includegraphics[width=\linewidth]{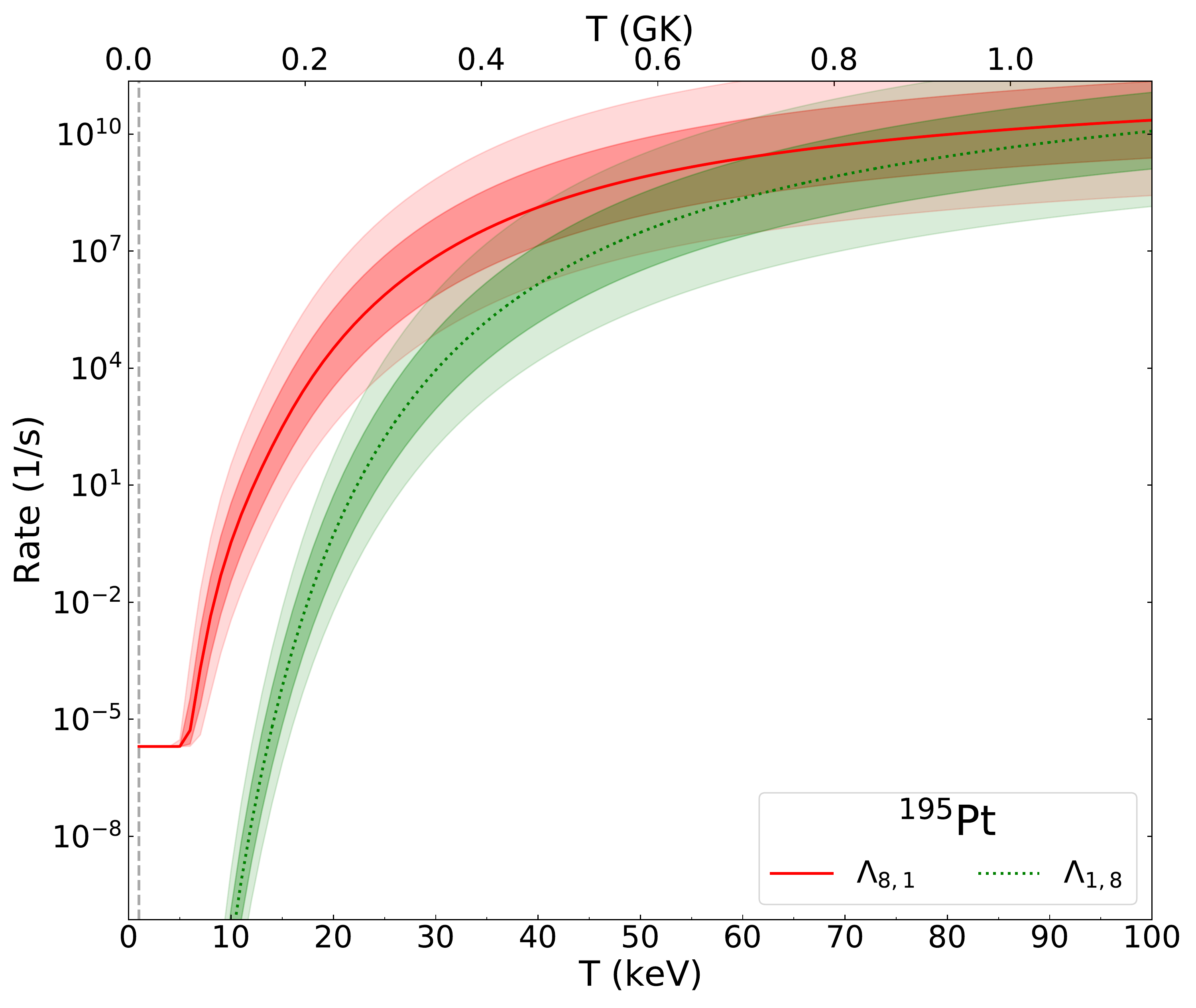}
\end{subfigure}%
\caption{Effective transition rates for Os ($Z=76$), Ir ($Z=77$) and Pt ($Z=78$) isotopes. Darkest shaded band shows unmeasured rates increased/decreased by one order of magnitude; light by two orders of magnitude. Thermalization temperature, \ttherm{}, estimated by dashed vertical grey line.}
\label{fig:it_Os}
\end{figure}

\nuc{189}{Os}: Third $r$-process peak ($A\sim195$) nuclide. 96 measured levels, 30 in this calculation.  Isomer at 30.82 keV (type N).  Known uncertainties dominated by unmeasured ($350.0 \rightarrow 216.67$), ($350.0 \rightarrow 219.39$), ($350.0 \rightarrow 97.35$), ($365.78 \rightarrow 30.82$), ($365.78 \rightarrow 69.54$), ($427.93 \rightarrow 69.54$), ($427.93 \rightarrow 95.27$), ($438.73 \rightarrow 0.0$), ($438.73 \rightarrow 69.54$), ($444.23 \rightarrow 216.67$), ($444.23 \rightarrow 219.39$), ($444.23 \rightarrow 233.58$), ($444.23 \rightarrow 30.82$), ($444.23 \rightarrow 69.54$), ($69.54 \rightarrow 30.82$), ($97.35 \rightarrow 30.82$) transition rates.  No expected new effect on the $r$ process due to its half-life being much shorter than that of its $\beta$-decay parent.

\nuc{191}{Ir}: Third $r$-process peak ($A\sim195$) nuclide. 74 measured levels, 30 in this calculation.  Isomer at 171.29 keV (type N).  Known uncertainties dominated by unmeasured ($343.23 \rightarrow 171.29$), ($390.94 \rightarrow 129.413$), ($390.94 \rightarrow 343.23$), ($502.61 \rightarrow 171.29$), ($502.61 \rightarrow 390.94$) transition rates.  No expected new effect on the $r$ process due to its half-life being much shorter than that of its $\beta$-decay parent.

\nuc{195}{Ir}: Third $r$-process peak ($A\sim195$) nuclide. 44 measured levels, 30 in this calculation.  Isomer at 100.0 keV (type N).  Known uncertainties dominated by unmeasured ($100.0 \rightarrow 0.0$), ($175.221 \rightarrow 0.0$), ($175.221 \rightarrow 100.0$), ($394.0 \rightarrow 100.0$), ($394.0 \rightarrow 175.221$) transition rates.  The isomer somewhat slows $\beta$ decay and could affect the heating curve, but it is primarily interesting because it feeds the \nuc{195}{Pt} isomer.

\nuc{195}{Pt}: Third $r$-process peak ($A\sim195$) nuclide. 92 measured levels, 30 in this calculation.  Isomer at 259.077 keV (type N).  Known uncertainties dominated by unmeasured ($199.532 \rightarrow 129.772$), ($211.406 \rightarrow 129.772$), ($211.406 \rightarrow 98.88$), ($222.23 \rightarrow 0.0$), ($222.23 \rightarrow 98.88$), ($239.264 \rightarrow 129.772$), ($431.98 \rightarrow 259.077$), ($449.65 \rightarrow 129.772$), ($449.65 \rightarrow 239.264$), ($449.65 \rightarrow 431.98$), ($507.917 \rightarrow 199.532$), ($507.917 \rightarrow 239.264$), ($507.917 \rightarrow 431.98$), ($547.16 \rightarrow 259.077$), ($547.16 \rightarrow 431.98$), ($562.8 \rightarrow 431.98$), ($667.0 \rightarrow 431.98$), ($667.0 \rightarrow 547.16$) transition rates.  Due to the high abundance of this isotope, its isomer could a strong x-ray source in the early days after an $r$-process event.

%%%%%%%%%%%%%%%%%%%%%%%%%%%%%%%%%%%%%%%%%%
\section{Discussion \& Conclusions}

\subsection{Comments on Selected Isomers and Mass Regions}

Many isomers are interesting not only for their qualities as astromers, but also for terrestrial applications and the insights they can give into nuclear structure.  We comment here on some selected isotopes covered in section \ref{sec:results}; we also briefly discuss the rare-earth region, which currently suffers from a dearth of experimental data.

\nuc{93}{Nb}: The spin-parity of the 1/2$^-$ isomer at 30.77 keV differs very much from the 9/2$^+$ ground state, and therefore this should be a good isomer terrestrially. A recent experiment at the National Superconducting Cyclotron Laboratory (NSCL) using a \nuc{93}{Nb}(t,~\nuc{3}{He}) charge-exchange reaction has successfully established electron-capture rates of the \nuc{93}{Nb} ground state \cite{gao2020gamow}. However, capture rates of the 1/2$^-$ isomer are completely unknown and must be calculated from theory \cite{tan2020novel}.

\nuc{99}{Tc}: The excited state \nuc{99m}{Tc}, produced from \nuc{99}{Mo}, is an isomer used in nuclear medicine \cite{richards1982technetium}.  It decays with a half-life of about 6 hours by emitting a 142 keV $\gamma$ ray, which is close to the energy of medical diagnostic x-rays.

Near \nuc{132}{Sn}: Around the double-magic nucleus \nuc{132}{Sn} (Z=50, N=82), isotopes discussed in the present paper with Z<50 (In, Cd, Ag, Pd) and Z>50 (Sb, Te, I, Cs)\footnote{Elements mentioned here which do not appear in section \ref{sec:results} have isomers listed in the appendix.} have been a focus for exploration in nuclear structure physics.  Generally, the structure of near-double-magic nuclei is recognized in their level spectra, which consist of two types of excitations: excitations of valence single particles and excited states formed by couplings of the valence nucleons to core excitations. Isomeric states can be found in both types of states.  However, lack of some basic experimental information has hindered understanding of both types of excitation.  The experimental single particle states are at present incomplete for this neutron-rich mass region (in particular, there is no information at all to the south-east of \nuc{132}{Sn} on the chart of nuclides); these are required inputs for reliable shell model calculations \cite{gorska2009evolution}.  Moreover, there is only limited experimental information on the sizes of the shell gaps, which are also important ingredients for understanding $r$-process nucleosynthesis \cite{jones2010magic}; the lack of knowledge on shell gaps has hindered shell-model studies on core-excited states.  Recently, progress has been made on large-scale shell model calculations, and a new type of shell model has appeared that unifies the discussion of the two aspects in a shell-model diagonalization calculation \cite{jin2011large, wang2013structure}; these theoretical advances make experimental results all the more needed.

Rare-earth nuclei: The structure of many neutron-rich rare-earth nuclei is relevant to the formation of the A$\sim$160 abundance peak in $r$-process nucleosynthesis \cite{Surman1997, mumpower2012influence, Vilen2018, Orford2018, Vilen2020, Vassh2021}.  For most of the involved nuclei, experimental information is currently very limited, and understanding of the structure must rely on theories \citep{Mumpower2016, Mumpower2017}.  However, theoretical calculations for these exotic nuclei extrapolate existing models that have only been demonstrated to work well for near-stable regions.  A recent work \cite{hartley2018masses} found a serious problem in that for the rare-earth neutron-rich nuclei (Nd, Pm, Sm, Eu, and Gd), the well-established Woods-Saxon potential, the Nilsson modified oscillator potential with ``universal'' parameters, and the folded Yukawa potential all failed to describe the neutron single-particle states.  A good understanding of the single-particle states is essential for the theoretical study of isomers, and these new results demand a careful reconsideration of mean-field models.  In a first attempt \cite{liu2020changes}, the traditional Nilsson model was extended in order to describe the deformed rare-earth nuclei in the very neutron-rich region.

\subsection{Further Refinements}

\begin{figure}[h]
    \centering
    \includegraphics[width=\textwidth]{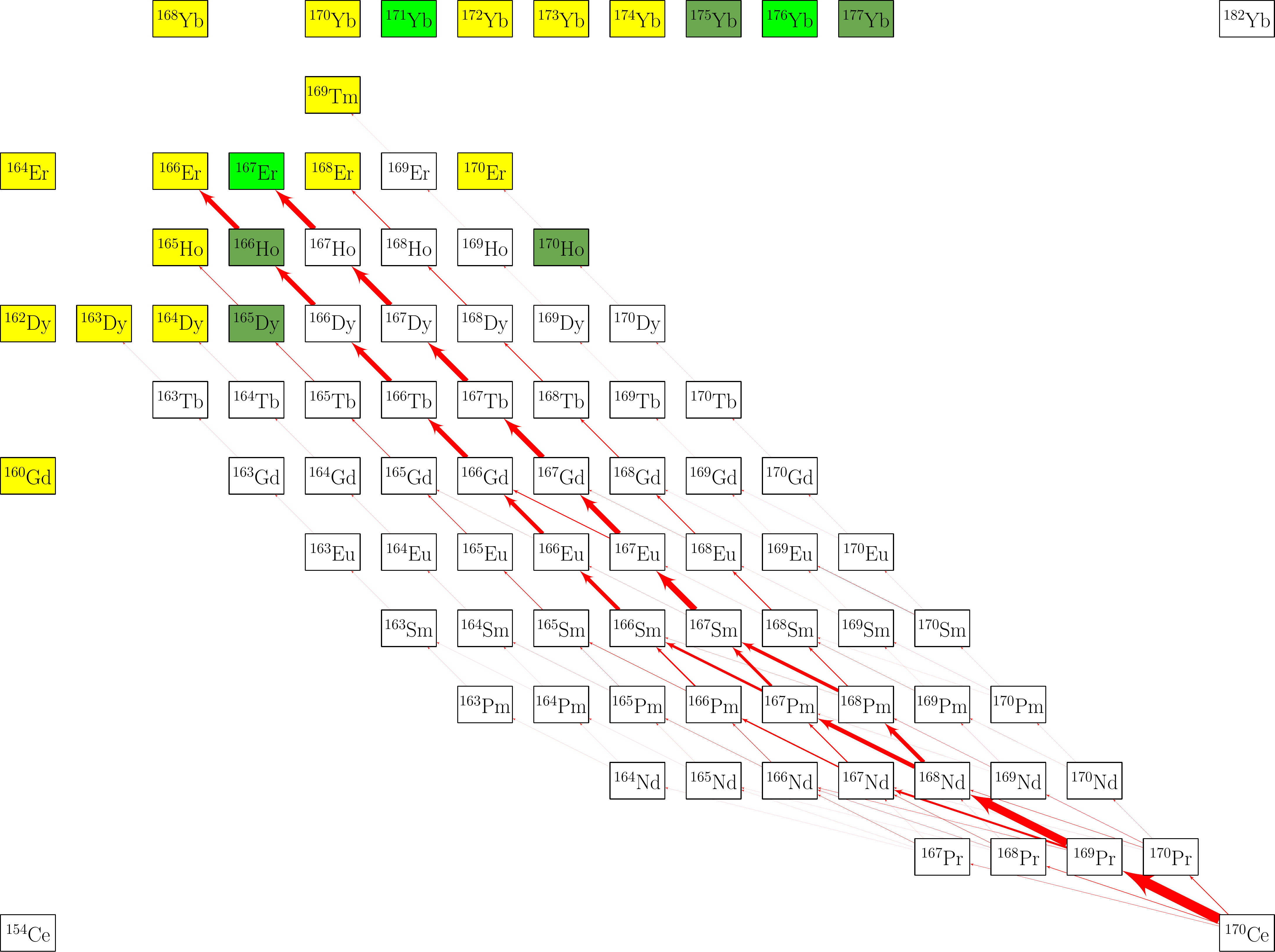}
    \caption{Section of the nuclide chart showing reaction flows integrated over time for the decay of $^{170}$Ce.  The strength of the integrated flow scales linearly with the width of the corresponding arrow.  Stable isotopes are shown in yellow or in lime green if they have known isomers.  Unstable isotopes with known isomers are in darker green.  The decay data for the calculation are from the Reaclib V2.2 database.}
    \label{fig:currents}
\end{figure}

To reiterate, our study includes only $\beta$ decays and thermally mediated internal transitions, but other reaction and decay channels can populate and destroy isomers.  We have also stressed the need for more nuclear data.  Figure \ref{fig:currents} emphasizes both of these points: further work is required to incorporate more channels, and (especially in the rare-earth region) we require many more isomer measurements.

The figure shows the reaction flow (integrated over time) of an initial population of \nuc{170}{Ce} that is allowed to undergo both $\beta$ decay and $\beta$-delayed neutron emission; species colored green have known isomers that we included in our network.  As the decay chains progress, the abundances fan out like a river delta, feeding many isotopes from a single ancestor.  The sparsity of known isomers in this delta is almost certainly due to insufficient data rather than an actual lack of isomers.  Consequently, many of the isotopes fed in this decay chain likely have isomers that contribute to the chain's dispersion, and we simply cannot adequately account for them until new measurements reveal their presence.

\subsection{Experimental Prospects}

Some experimental facilities for direct measurement of the nuclear properties needed to refine astromer predictions already exist, and the next decade will see a great expansion coming online.  The basic properties needed are 1) level energies (below Q$_{\beta}$ for most environments), 2) level spins and parities, 3) $\gamma$-ray transition strengths, \emph{particularly between ground and isomer bands}, and 4) total $\beta$-decay branching ratios to the ground state and isomeric state.  

Many of these isotopes are near stability and are already within reach of facilities such as ISAC at TRIUMF, ISOLDE at CERN, and ATLAS at Argonne.  With FRIB coming online, even more isotopes can be studied.  In addition to beam availability, spectroscopic detector systems are also needed and are available or under construction.  GAMMASPHERE has performed these types of studies for decades.  AGATA in Europe and GRETA/GRETINA are designed to perform the necessary spectroscopy while taking full advantage of next-generation facilities.  The N=126 Factory under construction at Argonne will make available beams of isotopes in the Ir/Pt region that have long been difficult to obtain.

Finally, it is worth noting that some of the necessary data will likely be taken ``for free'' as part of a larger campaign that might be focused on another isotope; $\beta$-decay branching ratios are a particularly prominent example.  As campaigns are developed for the FRIB Decay Station (FDS), for example, decays far off-stability could well decay back to stability through isotopes needed for isomer studies; simply by recognizing this additional interest, the planned experiment can have greater impact \emph{with no additional beamtime needed}.  This alone can improve $\beta$-decay branching ratio data for both Type A and Type B astromers.

\subsection{Observational Prospects}

\begin{table}
    \caption{Astromers most likely to have observable effects.  We use $g$ and $m$ to indicate the ground state and isomer, respectively.  The ``$E_m$'' column lists the isomer energy (keV), and the half-lives (``$T_{1/2}$'') of the ground state and isomer are given in their respective columns (s).  The low-temperature $\beta$-decay branching percentage for the isomer is in the ``$B_{\beta}^m$'' column.  The half-lives and branching percentages are as measured terrestrially.  ``$T_{pop}$'' is the approximate timescale on which parent nuclei decay to populate the isotope.  Finally, ``Dominant Astromer Effects'' details our interpretation of the most-likely observable consequences of the isomer as an astromer.  \textbf{T$_{pop}$} and all timescales in the last column are estimated from half-lives and therefore should be interpreted as lower bounds on the relevant timescale. \label{tab:observables}}
    
    \renewcommand\cellalign{tl}
    \begin{tabular}{ccccccl}
    \textbf{Isotope} & \textbf{E$_{m}$} & \textbf{T$_{1/2}^{g}$} & \textbf{T$_{1/2}^{m}$} & \textbf{B$_{\beta}^{m}$}   & \textbf{T$_{pop}$} & \textbf{Dominant Astromer Effects} \\ 
     & \textbf{(keV)}            & \textbf{(s)}         & \textbf{(s)}         & \textbf{(\%)} &   &  \\
    \midrule[1.2pt]
    \nuc{69}{Zn}  & 438.636 & $3.38\times 10^{3}$  & $4.95\times 10^{4}$  & 0.033  & 3 min    & \makecell{Heating slowed from 1 h to 14 h \\ 439 keV $\gamma$ ray at $\sim$14 h} \\
    \midrule
    \nuc{71}{Zn}  & 157.7   & $1.47\times 10^{2}$  & $1.43\times 10^{4}$  & 100    & 20 s     & \makecell{Heating slowed from 2 min to 4 h \\ 386 keV $\gamma$ ray at $\sim$4 h} \\
    \midrule[1.2pt]
    % \nuc{79}{Se}  & 95.77   & $1.03\times 10^{13}$ & $2.35\times 10^{2}$  & 0.056  & 9 min    & $T_{1/2~m} < T_{pop}$, no new effect \\
    % \midrule
    \nuc{81}{Se}  & 103.00  & $1.11\times 10^{3}$  & $3.44\times 10^{3}$  & 0.051  & 30 s     & Heating slowed from 20 min to 1 h \\
    \midrule[1.2pt]
    \nuc{83}{Kr}  & 41.5575 & stable               & $6.59\times 10^{3}$  & 0      & 2.5 h    & \makecell{9.4 keV $\gamma$ ray, 13 keV x ray at $\sim$2 h \\ Faint, but may be visible early} \\
    \midrule
    \nuc{85}{Kr}  & 304.871 & $3.39\times 10^{8}$  & $1.61\times 10^{4}$  & 78.8   & 3 min    & \makecell{Heating accelerated from 11 yr to 5 h \\ 151 keV $\gamma$ ray at 5 h} \\
    \midrule[1.2pt]
    % \nuc{93}{Nb}  & 30.77   & stable               & $5.09\times 10^{8}$  & 0      & 1.6 Myr  & $T_{1/2} < T_{pop}$, no new effect \\
    % \midrule
    % \nuc{95}{Nb}  & 235.69  & $3.02\times 10^{6}$  & $3.12\times 10^{5}$  & 5.6    & 64 d     & $T_{1/2} < T_{pop}$, no new effect \\
    % \midrule
    % \nuc{97}{Nb}  & 743.35  & $4.33\times 10^{3}$  & $5.87\times 10^{1}$  & 0      & 17 h     & $T_{1/2} < T_{pop}$, no new effect \\
    % \midrule[1.2pt]
    % \nuc{99}{Tc}  & 142.684 & $6.66\times 10^{12}$ & $2.16\times 10^{4}$  & 0.0037 & 66 h     & $T_{1/2~m} < T_{pop}$, no new effect \\
    % \midrule[1.2pt]
    \nuc{113}{Cd} & 263.54  & $2.54\times 10^{23}$ & $4.45\times 10^{8}$  & 99.86  & 6 h      & \makecell{Weak 264 keV $\gamma$ ray at 14 yr \\ Accelerates production of \nuc{113}{In}} \\
    \midrule
    \nuc{115}{Cd} & 181.0   & $1.92\times 10^{5}$  & $3.85\times 10^{6}$  & 100    & 20 min   & \makecell{Heating slowed from 54 h to 45 d \\ Weak 934 $\gamma$ ray at 45 d} \\
    \midrule
    \nuc{117}{Cd} & 136.4   & $8.96\times 10^{3}$  & $1.21\times 10^{4}$  & 100    & 70 s   & \makecell{Heating slowed from 2.5 h to 3.4 h \\ 1997 keV $\gamma$ ray possibly observable early} \\
    \midrule[1.2pt]
    \nuc{115}{In} & 336.244 & $1.39\times 10^{22}$ & $1.61\times 10^{4}$  & 5.0    & 54 h     & \makecell{336 keV $\gamma$ ray at 4.5 h \\ Accelerates production of \nuc{115}{Sn}} \\
    \midrule
    % \nuc{117}{In} & 315.303 & $2.59\times 10^{3}$  & $6.97\times 10^{3}$  & 52.9   & 3 h      & $T_{1/2} < T_{pop}$, no new effect \\
    % \midrule
    \nuc{119}{In} & 311.37  & $1.44\times 10^{2}$  & $1.08\times 10^{3}$  & 95.6   & 3 min    & Heating slowed from 2.5 min to 18 min \\
    % \midrule
    % \nuc{121}{In} & 313.68  & $2.31\times 10^{1}$  & $2.33\times 10^{2}$  & 98.8   & 15 s     & feeds \nuc{121}{Sn} isomer \\
    \midrule[1.2pt]
    \nuc{119}{Sn} & 89.531  & stable               & $2.53\times 10^{7}$  & 0      & 3-18 min & 24 keV $\gamma$ ray, 25 keV x ray at 293 d \\
    \midrule
    \nuc{121}{Sn} & 6.31    & $9.73\times 10^{4}$  & $1.39\times 10^{9}$  & 22.4   & 4 min    & \makecell{Decay/heating slowed from 27 h to 44 yr \\ 26 keV x ray, faint 3-4.5 keV x ray } \\
    \midrule
    \nuc{129}{Sn} & 35.15   & $1.34\times 10^{2}$  & $4.14\times 10^{2}$  & 100    & 1 s      & Heating slowed from 2.2 to 7 min \\
    \midrule[1.2pt]
    % \nuc{126}{Sb} & 17.7    & $1.07\times 10^{6}$  & $1.15\times 10^{3}$  & 86     & 230 kyr  & $T_{1/2} < T_{pop}$, no new effect \\
    % \midrule
    \nuc{128}{Sb} & 0.0+X   & $3.26\times 10^{4}$  & $6.25\times 10^{2}$  & 96.4   & 1 h      & Heating accelerated from 9 h to 10 min \\
    \midrule
    \nuc{130}{Sb} & 4.8     & $2.37\times 10^{3}$  & $3.78\times 10^{2}$  & 100    & 4 min    & Heating accelerated from 40 min to 6.3 min \\
    \midrule[1.2pt]
    % \nuc{125}{Te} & 144.775 & stable               & $4.96\times 10^{6}$  & 0      & 3 yr     & $T_{1/2~m} < T_{pop}$, no new effect \\
    % \midrule
    \nuc{127}{Te} & 88.23   & $3.37\times 10^{4}$  & $9.17\times 10^{6}$  & 2.4    & 4 d      & \makecell{Heating/decay slowed from 9.4 h to 106 d \\ 27 keV x ray at 106 d} \\
    \midrule
    \nuc{129}{Te} & 105.51  & $4.18\times 10^{3}$  & $2.9\times 10^{6}$   & 36     & 4.5 h    & \makecell{Heating/decay slowed from 70 min to 34 d \\ 27 keV x ray, faint 696 keV $\gamma$ ray at 34 d} \\
    \midrule
    \nuc{131}{Te} & 182.258 & $1.5\times 10^{3}$   & $1.2\times 10^{5}$   & 74.1   & 23 min   & \makecell{Heating slowed from 25 min to 33 h \\ 774 keV $\gamma$ ray, 29 keV x ray at 33 h} \\
    \midrule
    \nuc{133}{Te} & 334.26  & $7.5\times 10^{2}$   & $3.32\times 10^{3}$  & 83.5   & 2.5 min  & Heating slowed from 12.5 min to 1 h \\
    \midrule[1.2pt]
    \nuc{131}{Xe} & 163.930 & stable               & $1.02\times 10^{6}$  & 0      & 8 d      & 30 keV x ray at 12 d \\
    \midrule
    \nuc{133}{Xe} & 233.221 & $4.53\times 10^{5}$  & $1.9\times 10^{5}$   & 0      & 21 h     & 30 keV x ray at 2 d \\
    \midrule[1.2pt]
    % \nuc{137}{Ba} & 661.659 & stable               & $1.53\times 10^{2}$  & 0      & 30 yr    & $T_{1/2} < T_{pop}$, no new effect \\
    % \midrule[1.2pt]
    % \nuc{144}{Pr} & 59.03   & $1.04\times 10^{3}$  & $4.32\times 10^{2}$  & 0.07   & 285 d    & $T_{1/2} < T_{pop}$, no new effect \\
    % \midrule[1.2pt]
    \nuc{166}{Ho} & 5.969   & $9.66\times 10^{4}$  & $3.79\times 10^{10}$ & 100    & 82 h     & \makecell{Decay/heating slowed from 27 h to 1200 y \\ 184, 280, 712, 810 keV $\gamma$ rays, 49 keV x ray} \\
    \midrule[1.2pt]
    % \nuc{189}{Os} & 30.82   & stable               & $2.09\times 10^{4}$ & 0       & 1 d      & $T_{1/2} < T_{pop}$, no new effect \\
    % \midrule[1.2pt]
    % \nuc{191}{Ir} & 171.29  & stable               & $4.90\times 10^{0}$  & 0      & 16 d     & $T_{1/2~m} < T_{pop}$, no new effect \\
    % \midrule
    \nuc{195}{Ir} & 100     & $8.24\times 10^{3}$  & $1.32\times 10^{4}$  & 95     & 7 min    & Feeds \nuc{195}{Pt} isomer \\
    \midrule[1.2pt]
    \nuc{195}{Pt} & 259.077 & stable               & $3.46\times 10^{5}$  & 0      & 4 h      & 99 keV $\gamma$ ray, 8-14, $\sim$65, $\sim$75 keV x rays at 4d \\
    \end{tabular}

\end{table}

The isomer properties can alter the expected emission from astrophysical sources for a broad set of observations.  These properties may contribute to an explanation of some of the inconsistencies in our understanding of the emission from the neutron star merger event GW170817~\cite{2017ApJ...848L..12A}.  For example, optical emission a few days after the kilonova outburst is brighter than expected from a standard grid of neutron star merger calculations (Ristic et al., in preparation).  From table~\ref{tab:observables}, we note there are a few isomers whose half-life is either increased or decreased from the ground-state decay to a few to tens of hours: isomers of \nuc{69}{Zn}, \nuc{71}{Zn}, \nuc{85}{Kr}, \nuc{115}{In}, \nuc{131}{Te}, \nuc{133}{Xe}, and \nuc{195}{Pt}.  If these isomers are produced in sufficient abundance, they could increase the heating at these later times and, in the slow-moving ejecta, contribute enough heating to increase the optical emission at the few day timescale.  Another key timescale in kilonova observations lies in understanding the observational features of kilonova remnants.  Astromers with long-lived timescales (e.g. the 14\,y half-life of the \nuc{113}{Cd} isomer) could be observed in kilonova remnants in the Milky Way.  For such observations, the most promising are isomers whose half-lives allow contributions to old remnants (ages in the 1,000-100,000y range) and with decay lines appropriate for X-ray missions (which are more sensitive than gamma-ray missions).  \nuc{166}{Ho} fits both these criteria with a decay half-life of 1200\,y and decay lines in the gamma-rays and a 49\,keV x-ray (Table~\ref{tab:observables}).  A much more thorough study of these isomers is essential to determine the role of isomers in neutron star merger observations.

\subsection{Summary}

We studied the sensitivity of effective thermal transition rates between nuclear isomers and ground states to unmeasured internal transitions.  The most straightforward estimate of these unmeasured transitions is the Weisskopf approximation, which tends to be accurate within a factor of 100 relative to measured values.  Therefore, we varied all Weisskopf rates in our calculations up and down by factors of 10 and 100 to assess the likely range of effective transition rates.  We estimated the effect on thermalization temperatures and identified the likely most influential individual transitions.  We categorized isomers as accelerants, batteries, or neutral according to the effects they could be expected to have in the $r$-process decay back to stability.

We found that unmeasured transitions can have a very large effect on the effective rates and thermalization temperatures.  In isotopes where the effective rates are bottle-necked by the unmeasured rates, we see a strong sensitivity to our variations.  Conversely, when effective transitions flow primarily through paths with measured individual transitions, we see much less sensitivity.  Which type of transition (measured or unmeasured) limits the effective rates can vary sensitively with temperature, manifested by the variation with temperature in the widths of our uncertainty bands.  Many nuclei in section \ref{sec:results} exhibit this behavior, with measured rates dominating at low temperatures and giving way to unmeasured rates as temperature increases.  The second isomer in \nuc{131}{Te} exhibits the reverse effect: its de-excitation is controlled by unmeasured rates at low temperatures, but paths through measured transitions become accessible at higher temperatures.

Whether the measured or unmeasured individual rates throttle the effective rates can depend on whether the unmeasured rates are increased or decreased.  At the beginning of the upturn of $\Lambda_{3,1}$ in \nuc{137}{Ba}, the width of the uncertainty band arises almost entirely from turning the Weisskopf rates up.  \nuc{119}{In} shows a reverse effect, where the width at $T=45$ keV is due almost entirely to reductions in the Weisskopf rates.  \nuc{189}{Os} exhibits both behaviors: at low temperatures the width comes from turning the Weisskopf rates up, but at $T=30$ keV, the reverse is true.

The widths of the uncertainty bands can imply substantial variation in thermalization temperatures.  In \nuc{121}{In}, for example, the range of likely thermalization temperatures is about 12 keV wide.  A few of the nuclei we studied are so inadequately measured that we cannot even calculate a thermalization temperature, including the likely influential \nuc{128}{Sb} and \nuc{131}{Te} (although in the latter case, we can compute a \ttherm{} between just the GS and the first isomer).

The results of this work illuminate the limitations of existing nuclear structure and transition data data with respect to precisely understanding isomeric transitions in the $r$ process, but these are not the only nuclear uncertainties in play.  Another important quantity affected by these uncertainties is astromer $\beta$ feeding, that is, how much of a $\beta$-decay parent decays to the astromer versus the ground state.  In addition to missing direct feeding data (information about which daughter states are directly populated in decay), the unmeasured transitions can affect the subsequent (thermally mediated) $\gamma$~cascades toward long-lived states.  This, along with other reactions and decays, will be the focus of a future study.  We also plan to combine the nuclear physics unknowns with astrophysical $r$-process uncertainties.  These efforts will continue to shed light on the role of nuclear isomers in astrophysical environments and will motivate experiments to determine the key missing nuclear data quantities.

%%%%%%%%%%%%%%%%%%%%%%%%%%%%%%%%%%%%%%%%%%
\vspace{6pt}

%%%%%%%%%%%%%%%%%%%%%%%%%%%%%%%%%%%%%%%%%%
\funding{GWM, TMS and MRM were supported by the US Department of Energy through the Los Alamos National Laboratory (LANL).
LANL is operated by Triad National Security, LLC, for the National Nuclear Security Administration of U.S.\ Department of Energy (Contract No.\ 89233218CNA000001). 
GWM, MRM, and CLF were partly supported by the Laboratory Directed Research and Development program of LANL under project number 20190021DR. 
TMS was partly supported by the Fission In R-process Elements (FIRE) Topical Collaboration in Nuclear Theory, funded by the U.S. Department of Energy.
BSM was partly supported by NASA grant 80NSSC20K0338.
YS was supported by the National Natural Science Foundation of China (No. U1932206) and the National Key Program for S\&T Research and Development (No. 2016YFA0400501).
}

%%%%%%%%%%%%%%%%%%%%%%%%%%%%%%%%%%%%%%%%%%
\acknowledgments{We thank Frank Timmes and Kelly Chipps for valuable discussions.}

%%%%%%%%%%%%%%%%%%%%%%%%%%%%%%%%%%%%%%%%%%
\conflictsofinterest{The authors declare no conflict of interest. The funders had no role in the design of the study; in the collection, analyses, or interpretation of data; in the writing of the manuscript, or in the decision to publish the results.} 

%%%%%%%%%%%%%%%%%%%%%%%%%%%%%%%%%%%%%%%%%%
\abbreviations{The following abbreviations are used in this manuscript:\\

\noindent 
\begin{tabular}{@{}ll}
Astromer & Astrophysical isomer; isomer which retains its metastable characteristics in a hot environment \\
GS & Ground state \\
$r$ process & Rapid neutron capture process \\
$s$ process & Slow neutron capture process \\
$rp$ process & Rapid proton capture process \\
Type A & Astromer which accelerates decay and energy release (``accelerant'') \\
Type B & Astromer which slows decay and stores energy (``battery'') \\
Type N & Astromer which has a negligible effect on energy release (``neutral'')
\end{tabular}}

%%%%%%%%%%%%%%%%%%%%%%%%%%%%%%%%%%%%%%%%%%
\appendixtitles{no} 
\appendix

\section{}

In our study, we computed the transition rates, thermalization temperatures, and key unmeasured direct transition rates for all isomers with $T_{1/2}$>100 $\mu$s in the $r$-process region between $A = 69$ and $A = 209$.  Section \ref{sec:results} highlighted several, and here we present results for the full set of isomers.

Our results summarized in Table \ref{tab:all} include for each isomer the range of thermalization temperatures \ttherm{}, the astromer type (A, B, or N), the temperature above which type B astromers no longer function as batteries, and all unmeasured direct transitions through which at least 1\% of effective transitions flow.  This table will be an effective guide for astrophysical nucleosysthesis modelers (at what temperatures is special care needed for a particular isotope?) as well as experimenters (what would be some impactful measurements?).

\begin{landscape}
\renewcommand\theadalign{tc}
\renewcommand\theadfont{\bfseries}
\renewcommand\cellalign{tc}

\begin{longtable}[l]{>{\centering}p{0.05\linewidth} >{\centering}p{0.05\linewidth} >{\centering}p{0.05\linewidth} >{\centering}p{0.08\linewidth} >{\centering}p{0.04\linewidth} >{\raggedright\arraybackslash}p{0.05\linewidth} >{\raggedright\arraybackslash}p{0.55\linewidth}}

\caption{Results for each isomer studied in this work.  We provide the range of thermalization temperatures in the ``T$_{\rm therm}$'' column.  We give the energy of each long-lived state (``E'', in keV) along with its half-life (``T$_{1/2}$'', in seconds).  The ``B$_\beta$'' column gives the isomer low-temperature $\beta$-decay branching percentage (approximately the terrestrial branching ratio).  The ``Type'' column indicates whether each isomer is a decay accelerant (A), a decay battery (B), or neutral with respect to decay rate (N); type B isomers show in parentheses the maximum temperature (in keV) at which they behave as batteries.  We list in the ``Unmeasured Transitions'' column which unmeasured individual transitions appear along pathways that contribute at least 1\% to the effective transition rate.  We only consider paths at temperatures below the greatest value in the $T_{\rm therm}$ column, and stable isotopes use an artificial $T_{\rm therm}$ of 30 keV. \label{tab:all}} \\

\toprule
\thead{Nucleus}  & \thead{T$_{\rm therm}^\beta$ \\ keV} & \thead{E \\ keV} & \thead{T$_{1/2}$ \\ s} & \thead{B$_\beta$ \\ \%} & \thead{Type \\ (keV)} & \thead{Unmeasured Transitions} \\
\midrule[1.5pt]
\endfirsthead

\thead{Nucleus}  & \thead{T$_{\rm therm}^\beta$ \\ keV} & \thead{E \\ keV} & \thead{T$_{1/2}$ \\ s} & \thead{B$_\beta$ \\ \%} & \thead{Type \\ (keV)} & \thead{Unmeasured Transitions} \\
\midrule[1.5pt]
\endhead

\endfoot
\endlastfoot

\nuc{69}{Zn}  & 24--36 & \makecell{ 0.0 \\ 438.6 \\ } & \makecell{$3.38 \times 10^{ 3 }$ \\ $4.95 \times 10^{ 4 }$ \\ } & \makecell{ \\0.0 \\ } & \makecell{ \\B (5.0) \\ } & ($531.3 \rightarrow 0.0$), ($531.3 \rightarrow 438.636$), ($872.0 \rightarrow 438.636$), ($872.0 \rightarrow 531.3$) \\
\midrule
\nuc{71}{Zn}  & 12--17 & \makecell{ 0.0 \\ 157.7 \\ } & \makecell{$1.47 \times 10^{ 2 }$ \\ $1.43 \times 10^{ 4 }$ \\ } & \makecell{ \\100.0 \\ } & \makecell{ \\B (6.0) \\ } & ($157.7 \rightarrow 0.0$), ($286.3 \rightarrow 0.0$), ($286.3 \rightarrow 157.7$), ($465.0 \rightarrow 157.7$), ($465.0 \rightarrow 286.3$), ($489.8 \rightarrow 0.0$), ($489.8 \rightarrow 286.3$) \\
\midrule
\nuc{73}{Zn}  & 9--12  & \makecell{ 0.0 \\ 195.5 \\ } & \makecell{$2.45 \times 10^{ 1 }$ \\ $1.30 \times 10^{ -2 }$ \\ } & \makecell{ \\0.0 \\ } & \makecell{ \\N \\ } & ($307.2 \rightarrow 0.0$), ($307.2 \rightarrow 195.5$) \\
\midrule
\nuc{77}{Zn}  & 34--45 & \makecell{ 0.0 \\ 772.4 \\ } & \makecell{$2.08 \times 10^{ 0 }$ \\ $1.05 \times 10^{ 0 }$ \\ } & \makecell{ \\65.9 \\ } & \makecell{ \\N \\ } & ($1130.5 \rightarrow 772.44$), ($1130.5 \rightarrow 801.89$), ($114.721 \rightarrow 0.0$), ($1235.1 \rightarrow 772.44$), ($1235.1 \rightarrow 801.89$), ($1277.69 \rightarrow 0.0$), ($1277.69 \rightarrow 772.44$), ($1277.69 \rightarrow 801.89$), ($772.44 \rightarrow 114.721$), ($801.89 \rightarrow 114.721$), ($801.89 \rightarrow 772.44$) \\
\bottomrule
\end{longtable}

\end{landscape}

\externalbibliography{yes}
\bibliography{refs}

\begin{thebibliography}{-------}
\providecommand{\natexlab}[1]{#1}

\bibitem[{Soddy}(1917)]{Soddy1917}
{Soddy}, F.
\newblock {The Complexity of the Chemical Elements}.
\newblock {\em The Scientific Monthly} {\bf 1917}, {\em 5},~451--462.

\bibitem[Walker and Dracoulis(1999)]{walker1999energy}
Walker, P.; Dracoulis, G.
\newblock Energy traps in atomic nuclei.
\newblock {\em Nature} {\bf 1999}, {\em 399},~35--40.

\bibitem[Aprahamian and Sun(2005)]{aprahamian2005long}
Aprahamian, A.; Sun, Y.
\newblock Long live isomer research.
\newblock {\em Nature Physics} {\bf 2005}, {\em 1},~81--82.

\bibitem[{Dracoulis} \em{et~al.}(2016){Dracoulis}, {Walker}, and
  {Kondev}]{Dracoulis2016}
{Dracoulis}, G.D.; {Walker}, P.M.; {Kondev}, F.G.
\newblock {Review of metastable states in heavy nuclei}.
\newblock {\em Reports on Progress in Physics} {\bf 2016}, {\em 79},~076301.
\newblock
  doi:{\changeurlcolor{black}\href{https://doi.org/10.1088/0034-4885/79/7/076301}{\detokenize{10.1088/0034-4885/79/7/076301}}}.

\bibitem[{Hahn}(1921)]{Hahn1921}
{Hahn}, O.
\newblock {{\"U}ber ein neues radioaktives Zerfallsprodukt im Uran}.
\newblock {\em Naturwissenschaften} {\bf 1921}, {\em 9},~84--84.
\newblock
  doi:{\changeurlcolor{black}\href{https://doi.org/10.1007/BF01491321}{\detokenize{10.1007/BF01491321}}}.

\bibitem[Jain \em{et~al.}(2015)Jain, Maheshwari, Garg, Patial, and
  Singh]{jain2015atlas}
Jain, A.K.; Maheshwari, B.; Garg, S.; Patial, M.; Singh, B.
\newblock Atlas of nuclear isomers.
\newblock {\em Nuclear Data Sheets} {\bf 2015}, {\em 128},~1--130.

\bibitem[{Langanke} and {Mart{\'\i}nez-Pinedo}(2000)]{Langanke2000}
{Langanke}, K.; {Mart{\'\i}nez-Pinedo}, G.
\newblock {Shell-model calculations of stellar weak interaction rates: II. Weak
  rates for nuclei in the mass range /A=45-65 in supernovae environments}.
\newblock {\em \nphysa} {\bf 2000}, {\em 673},~481--508,
  \href{http://xxx.lanl.gov/abs/nucl-th/0001018}{{\normalfont
  [arXiv:nucl-th/nucl-th/0001018]}}.
\newblock
  doi:{\changeurlcolor{black}\href{https://doi.org/10.1016/S0375-9474(00)00131-7}{\detokenize{10.1016/S0375-9474(00)00131-7}}}.

\bibitem[Brown and Rae(2014)]{Brown2014}
Brown, B.; Rae, W.
\newblock The shell-model code NuShellX@ MSU.
\newblock {\em Nuclear Data Sheets} {\bf 2014}, {\em 120},~115--118.

\bibitem[Masuda \em{et~al.}(2019)Masuda, Yoshimi, Fujieda, Fujimoto, Haba,
  Hara, Hiraki, Kaino, Kasamatsu, Kitao, et~al.]{masuda2019x}
Masuda, T.; Yoshimi, A.; Fujieda, A.; Fujimoto, H.; Haba, H.; Hara, H.; Hiraki,
  T.; Kaino, H.; Kasamatsu, Y.; Kitao, S.; others.
\newblock X-ray pumping of the 229 Th nuclear clock isomer.
\newblock {\em Nature} {\bf 2019}, {\em 573},~238--242.

\bibitem[Zhang \em{et~al.}(2019)Zhang, Watanabe, Dracoulis, Kondev, Lane,
  Regan, Söderström, Walker, Yoshida, Kanaoka, Korkulu, Lee, Liu, Nishimura,
  Wu, Yagi, Ahn, Alharbi, Baba, Browne, Bruce, Carpenter, Carroll, Chae,
  Chiara, Dombradi, Doornenbal, Estrade, Fukuda, Griffin, Ideguchi, Inabe,
  Isobe, Kanaya, Kojouharov, Kubo, Kubono, Kurz, Kuti, Lalkovski, Lauritsen,
  Lee, Lee, Lister, Lorusso, Lotay, McCutchan, Moon, Nishizuka, Nita, Odahara,
  Patel, Phong, Podolyák, Roberts, Sakurai, Schaffner, Seweryniak, Shand,
  Shimizu, Sumikama, Suzuki, Takeda, Terashima, Vajta, Valiente-Dóbon, Xu, and
  Zhu]{zhang2019isomer}
Zhang, G.; Watanabe, H.; Dracoulis, G.; Kondev, F.; Lane, G.; Regan, P.;
  Söderström, P.A.; Walker, P.; Yoshida, K.; Kanaoka, H.; Korkulu, Z.; Lee,
  P.; Liu, J.; Nishimura, S.; Wu, J.; Yagi, A.; Ahn, D.; Alharbi, T.; Baba, H.;
  Browne, F.; Bruce, A.; Carpenter, M.; Carroll, R.; Chae, K.; Chiara, C.;
  Dombradi, Z.; Doornenbal, P.; Estrade, A.; Fukuda, N.; Griffin, C.; Ideguchi,
  E.; Inabe, N.; Isobe, T.; Kanaya, S.; Kojouharov, I.; Kubo, T.; Kubono, S.;
  Kurz, N.; Kuti, I.; Lalkovski, S.; Lauritsen, T.; Lee, C.; Lee, E.; Lister,
  C.; Lorusso, G.; Lotay, G.; McCutchan, E.; Moon, C.B.; Nishizuka, I.; Nita,
  C.; Odahara, A.; Patel, Z.; Phong, V.; Podolyák, Z.; Roberts, O.; Sakurai,
  H.; Schaffner, H.; Seweryniak, D.; Shand, C.; Shimizu, Y.; Sumikama, T.;
  Suzuki, H.; Takeda, H.; Terashima, S.; Vajta, Z.; Valiente-Dóbon, J.; Xu,
  Z.; Zhu, S.
\newblock Interplay of quasiparticle and vibrational excitations: First
  observation of isomeric states in 168Dy and 169Dy.
\newblock {\em Physics Letters B} {\bf 2019}, {\em 799},~135036.
\newblock
  doi:{\changeurlcolor{black}\href{https://doi.org/https://doi.org/10.1016/j.physletb.2019.135036}{\detokenize{https://doi.org/10.1016/j.physletb.2019.135036}}}.

\bibitem[Liu \em{et~al.}(2020)Liu, Lee, Watanabe, Nishimura, Zhang, Wu, Walker,
  Regan, S{\"o}derstr{\"o}m, Kanaoka, et~al.]{liu2020isomeric}
Liu, J.; Lee, J.; Watanabe, H.; Nishimura, S.; Zhang, G.; Wu, J.; Walker, P.;
  Regan, P.; S{\"o}derstr{\"o}m, P.A.; Kanaoka, H.; others.
\newblock Isomeric and $\beta$-decay spectroscopy of Ho 173, 174.
\newblock {\em Physical Review C} {\bf 2020}, {\em 102},~024301.

\bibitem[{Nesterenko} \em{et~al.}(2020){Nesterenko}, {Kankainen}, {Kostensalo},
  {Nobs}, {Bruce}, {Beliuskina}, {Canete}, {Eronen}, {Gamba}, {Geldhof}, {de
  Groote}, {Jokinen}, {Kurpeta}, {Moore}, {Morrison}, {Podoly{\'a}k},
  {Pohjalainen}, {Rinta-Antila}, {de Roubin}, {Rudigier}, {Suhonen},
  {Vil{\'e}n}, {Virtanen}, and {{\"A}yst{\"o}}]{nesterenko2020isomer}
{Nesterenko}, D.A.; {Kankainen}, A.; {Kostensalo}, J.; {Nobs}, C.R.; {Bruce},
  A.M.; {Beliuskina}, O.; {Canete}, L.; {Eronen}, T.; {Gamba}, E.R.; {Geldhof},
  S.; {de Groote}, R.; {Jokinen}, A.; {Kurpeta}, J.; {Moore}, I.D.; {Morrison},
  L.; {Podoly{\'a}k}, Z.; {Pohjalainen}, I.; {Rinta-Antila}, S.; {de Roubin},
  A.; {Rudigier}, M.; {Suhonen}, J.; {Vil{\'e}n}, M.; {Virtanen}, V.;
  {{\"A}yst{\"o}}, J.
\newblock {Novel Penning-trap techniques reveal isomeric states in $^{128}$In
  and $^{130}$In for the first time}.
\newblock {\em arXiv e-prints} {\bf 2020}, p. arXiv:2005.09398,
  \href{http://xxx.lanl.gov/abs/2005.09398}{{\normalfont
  [arXiv:nucl-ex/2005.09398]}}.

\bibitem[Orford \em{et~al.}(2020)Orford, Kondev, Savard, Clark, Porter, Ray,
  Buchinger, Burkey, Gorelov, Hartley, Klimes, Sharma, Valverde, and
  Yan]{orford2020isomer}
Orford, R.; Kondev, F.G.; Savard, G.; Clark, J.A.; Porter, W.S.; Ray, D.;
  Buchinger, F.; Burkey, M.T.; Gorelov, D.A.; Hartley, D.J.; Klimes, J.W.;
  Sharma, K.S.; Valverde, A.A.; Yan, X.L.
\newblock Spin-trap isomers in deformed, odd-odd nuclei in the light rare-earth
  region near $N=98$.
\newblock {\em Phys. Rev. C} {\bf 2020}, {\em 102},~011303.
\newblock
  doi:{\changeurlcolor{black}\href{https://doi.org/10.1103/PhysRevC.102.011303}{\detokenize{10.1103/PhysRevC.102.011303}}}.

\bibitem[Sikorsky \em{et~al.}(2020)Sikorsky, Geist, Hengstler, Kempf, Gastaldo,
  Enss, Mokry, Runke, D{\"u}llmann, Wobrauschek,
  et~al.]{sikorsky2020measurement}
Sikorsky, T.; Geist, J.; Hengstler, D.; Kempf, S.; Gastaldo, L.; Enss, C.;
  Mokry, C.; Runke, J.; D{\"u}llmann, C.E.; Wobrauschek, P.; others.
\newblock Measurement of the Th 229 Isomer Energy with a Magnetic
  Microcalorimeter.
\newblock {\em Physical Review Letters} {\bf 2020}, {\em 125},~142503.

\bibitem[Walker \em{et~al.}(2020)Walker, Hirayama, Lane, Watanabe, Dracoulis,
  Ahmed, Brunet, Hashimoto, Ishizawa, Kondev, et~al.]{walker2020properties}
Walker, P.; Hirayama, Y.; Lane, G.; Watanabe, H.; Dracoulis, G.; Ahmed, M.;
  Brunet, M.; Hashimoto, T.; Ishizawa, S.; Kondev, F.; others.
\newblock Properties of Ta 187 Revealed through Isomeric Decay.
\newblock {\em Physical Review Letters} {\bf 2020}, {\em 125},~192505.

\bibitem[Izzo \em{et~al.}(2021)Izzo, Bergmann, Dietrich, Dunling, Fusco,
  Jacobs, Kootte, Kripk{\'o}-Koncz, Lan, Leistenschneider,
  et~al.]{izzo2021mass}
Izzo, C.; Bergmann, J.; Dietrich, K.; Dunling, E.; Fusco, D.; Jacobs, A.;
  Kootte, B.; Kripk{\'o}-Koncz, G.; Lan, Y.; Leistenschneider, E.; others.
\newblock Mass measurements of neutron-rich indium isotopes for r-process
  studies.
\newblock {\em Physical Review C} {\bf 2021}, {\em 103},~025811.

\bibitem[Gombas \em{et~al.}(2021)Gombas, DeYoung, Spyrou, Dombos, Algora,
  Baumann, Crider, Engel, Ginter, Kwan, et~al.]{gombas2021beta}
Gombas, J.; DeYoung, P.; Spyrou, A.; Dombos, A.; Algora, A.; Baumann, T.;
  Crider, B.; Engel, J.; Ginter, T.; Kwan, E.; others.
\newblock $\beta$-decay feeding intensity distributions for Nb 103, 104 m.
\newblock {\em Physical Review C} {\bf 2021}, {\em 103},~035803.

\bibitem[Walker and Podoly{\'{a}}k(2020)]{Walker2020}
Walker, P.; Podoly{\'{a}}k, Z.
\newblock 100 years of nuclear isomers{\textemdash}then and now.
\newblock {\em Physica Scripta} {\bf 2020}, {\em 95},~044004.
\newblock
  doi:{\changeurlcolor{black}\href{https://doi.org/10.1088/1402-4896/ab635d}{\detokenize{10.1088/1402-4896/ab635d}}}.

\bibitem[{Misch} \em{et~al.}(2020){Misch}, {Ghorui}, {Banerjee}, {Sun}, and
  {Mumpower}]{Misch2020a}
{Misch}, G.W.; {Ghorui}, S.K.; {Banerjee}, P.; {Sun}, Y.; {Mumpower}, M.R.
\newblock {Astromers: Nuclear Isomers in Astrophysics}.
\newblock {\em arXiv e-prints} {\bf 2020}, p. arXiv:2010.15238,
  \href{http://xxx.lanl.gov/abs/2010.15238}{{\normalfont
  [arXiv:astro-ph.HE/2010.15238]}}.

\bibitem[{Coc} \em{et~al.}(1999){Coc}, {Porquet}, and {Nowacki}]{Coc1999}
{Coc}, A.; {Porquet}, M.G.; {Nowacki}, F.
\newblock {Lifetimes of $^{26}$Al and $^{34}$Cl in an astrophysical plasma}.
\newblock {\em \prc} {\bf 1999}, {\em 61},~015801,
  \href{http://xxx.lanl.gov/abs/astro-ph/9910186}{{\normalfont
  [astro-ph/9910186]}}.
\newblock
  doi:{\changeurlcolor{black}\href{https://doi.org/10.1103/PhysRevC.61.015801}{\detokenize{10.1103/PhysRevC.61.015801}}}.

\bibitem[Gupta and Meyer(2001)]{Gupta2001}
Gupta, S.S.; Meyer, B.S.
\newblock Internal equilibration of a nucleus with metastable states: 26 Al as
  an example.
\newblock {\em Physical Review C} {\bf 2001}, {\em 64},~025805.

\bibitem[Runkle \em{et~al.}(2001)Runkle, Champagne, and Engel]{Runkle2001}
Runkle, R.; Champagne, A.; Engel, J.
\newblock Thermal Equilibration of 26Al.
\newblock {\em The Astrophysical Journal} {\bf 2001}, {\em 556},~970.

\bibitem[Iliadis \em{et~al.}(2011)Iliadis, Champagne, Chieffi, and
  Limongi]{iliadis2011effects}
Iliadis, C.; Champagne, A.; Chieffi, A.; Limongi, M.
\newblock The effects of thermonuclear reaction rate variations on 26al
  production in massive stars: a sensitivity study.
\newblock {\em The Astrophysical Journal Supplement Series} {\bf 2011}, {\em
  193},~16.

\bibitem[Banerjee \em{et~al.}(2018)Banerjee, Misch, Ghorui, and
  Sun]{Banerjee2018}
Banerjee, P.; Misch, G.W.; Ghorui, S.K.; Sun, Y.
\newblock Effective stellar $\beta$-decay rates of nuclei with long-lived
  isomers: Al 26 and Cl 34.
\newblock {\em Physical Review C} {\bf 2018}, {\em 97},~065807.

\bibitem[Reifarth \em{et~al.}(2018)Reifarth, Fiebiger, G{\"o}bel, Heftrich,
  Kausch, K{\"o}ppchen, Kurtulgil, Langer, Thomas, and Weigand]{Reifarth2018}
Reifarth, R.; Fiebiger, S.; G{\"o}bel, K.; Heftrich, T.; Kausch, T.;
  K{\"o}ppchen, C.; Kurtulgil, D.; Langer, C.; Thomas, B.; Weigand, M.
\newblock Treatment of isomers in nucleosynthesis codes.
\newblock {\em International Journal of Modern Physics A} {\bf 2018}, {\em
  33},~1843011.

\bibitem[{Mahoney} \em{et~al.}(1982){Mahoney}, {Ling}, {Jacobson}, and
  {Lingenfelter}]{Mahoney1982}
{Mahoney}, W.A.; {Ling}, J.C.; {Jacobson}, A.S.; {Lingenfelter}, R.E.
\newblock {Diffuse galactic gamma-ray line emission from nucleosynthetic Fe-60,
  Al-26, and Na-22 - Preliminary limits from HEAO 3.}
\newblock {\em \apj} {\bf 1982}, {\em 262},~742--748.
\newblock
  doi:{\changeurlcolor{black}\href{https://doi.org/10.1086/160469}{\detokenize{10.1086/160469}}}.

\bibitem[{Diehl} \em{et~al.}(1995){Diehl}, {Dupraz}, {Bennett}, {Bloemen},
  {Hermsen}, {Knoedlseder}, {Lichti}, {Morris}, {Ryan}, {Schoenfelder},
  {Steinle}, {Strong}, {Swanenburg}, {Varendorff}, and {Winkler}]{Diehl1995}
{Diehl}, R.; {Dupraz}, C.; {Bennett}, K.; {Bloemen}, H.; {Hermsen}, W.;
  {Knoedlseder}, J.; {Lichti}, G.; {Morris}, D.; {Ryan}, J.; {Schoenfelder},
  V.; {Steinle}, H.; {Strong}, A.; {Swanenburg}, B.; {Varendorff}, M.;
  {Winkler}, C.
\newblock {COMPTEL observations of Galactic \^26\^Al emission.}
\newblock {\em \aap} {\bf 1995}, {\em 298},~445.

\bibitem[Lugaro and Karakas(2008)]{lugaro200826al}
Lugaro, M.; Karakas, A.I.
\newblock 26Al and 60Fe yields from AGB stars.
\newblock {\em New Astronomy Reviews} {\bf 2008}, {\em 52},~416--418.

\bibitem[{Parikh} \em{et~al.}(2009){Parikh}, {Faestermann}, {Hertenberger},
  {Kr{\"u}cken}, {Schafstadler}, {Wirth}, {Behrens}, {Bildstein}, {Bishop},
  {Eppinger}, {Herlitzius}, {Hinke}, {Schlarb}, {Seiler}, and
  {Wimmer}]{Parikh2009}
{Parikh}, A.; {Faestermann}, T.; {Hertenberger}, R.; {Kr{\"u}cken}, R.;
  {Schafstadler}, D.; {Wirth}, H.F.; {Behrens}, T.; {Bildstein}, V.; {Bishop},
  S.; {Eppinger}, K.; {Herlitzius}, C.; {Hinke}, C.; {Schlarb}, M.; {Seiler},
  D.; {Wimmer}, K.
\newblock {New Cl34 proton-threshold states and the thermonuclear
  S33(p,{\ensuremath{\gamma}})Cl34 rate in ONe novae}.
\newblock {\em \prc} {\bf 2009}, {\em 80},~015802,
  \href{http://xxx.lanl.gov/abs/0906.3610}{{\normalfont
  [arXiv:nucl-ex/0906.3610]}}.
\newblock
  doi:{\changeurlcolor{black}\href{https://doi.org/10.1103/PhysRevC.80.015802}{\detokenize{10.1103/PhysRevC.80.015802}}}.

\bibitem[Abia \em{et~al.}(2001)Abia, Busso, Gallino, Dom{\'\i}nguez, Straniero,
  and Isern]{Abia2001}
Abia, C.; Busso, M.; Gallino, R.; Dom{\'\i}nguez, I.; Straniero, O.; Isern, J.
\newblock The 85Kr s-process branching and the mass of carbon stars.
\newblock {\em The Astrophysical Journal} {\bf 2001}, {\em 559},~1117.

\bibitem[{Misch} \em{et~al.}(2020){Misch}, {Sprouse}, and
  {Mumpower}]{Misch2020b}
{Misch}, G.W.; {Sprouse}, T.M.; {Mumpower}, M.R.
\newblock {Astromers in the radioactive decay of r-process nuclei}.
\newblock {\em arXiv e-prints} {\bf 2020}, p. arXiv:2011.11889,
  \href{http://xxx.lanl.gov/abs/2011.11889}{{\normalfont
  [arXiv:astro-ph.HE/2011.11889]}}.

\bibitem[Novikov \em{et~al.}(2001)Novikov, Schatz, Dendooven, B{\'e}raud,
  Mieh{\'e}, Popov, Seliverstov, Vorobjev, Baumann, Borge,
  et~al.]{novikov2001isomeric}
Novikov, Y.N.; Schatz, H.; Dendooven, P.; B{\'e}raud, R.; Mieh{\'e}, C.; Popov,
  A.; Seliverstov, D.; Vorobjev, G.; Baumann, P.; Borge, M.; others.
\newblock Isomeric state of 80 Y and its role in the astrophysical rp-process.
\newblock {\em The European Physical Journal A-Hadrons and Nuclei} {\bf 2001},
  {\em 11},~257--261.

\bibitem[Grineviciute \em{et~al.}(2014)Grineviciute, Brown, and
  Schatz]{grineviciute2014role}
Grineviciute, J.; Brown, B.; Schatz, H.
\newblock The role of excited states in rp-process for sd shell nuclei.
\newblock {\em arXiv preprint arXiv:1404.7268} {\bf 2014}.

\bibitem[Chipps \em{et~al.}(2018)Chipps, Kozub, Sumithrarachchi, Ginter,
  Baumann, Lund, Lapierre, Villari, Montes, Jin, et~al.]{chipps2018k}
Chipps, K.; Kozub, R.; Sumithrarachchi, C.; Ginter, T.; Baumann, T.; Lund, K.;
  Lapierre, A.; Villari, A.; Montes, F.; Jin, S.; others.
\newblock K 38 isomer production via fast fragmentation.
\newblock {\em Physical Review Accelerators and Beams} {\bf 2018}, {\em
  21},~121301.

\bibitem[Pain(2020)]{pain2020exp}
Pain, S.
\newblock NSCL Experiment E18037 {\bf 2020}.

\bibitem[Doll \em{et~al.}(1999)Doll, B{\"o}rner, Jaag, K{\"a}ppeler, and
  Andrejtscheff]{doll1999lifetime}
Doll, C.; B{\"o}rner, H.; Jaag, S.; K{\"a}ppeler, F.; Andrejtscheff, W.
\newblock Lifetime measurement in 176 Lu and its astrophysical consequences.
\newblock {\em Physical Review C} {\bf 1999}, {\em 59},~492.

\bibitem[S{\"o}derlund \em{et~al.}(2004)S{\"o}derlund, Patchett, Vervoort, and
  Isachsen]{soderlund2004176lu}
S{\"o}derlund, U.; Patchett, P.J.; Vervoort, J.D.; Isachsen, C.E.
\newblock The 176Lu decay constant determined by Lu--Hf and U--Pb isotope
  systematics of Precambrian mafic intrusions.
\newblock {\em Earth and Planetary Science Letters} {\bf 2004}, {\em
  219},~311--324.

\bibitem[Albar{\`e}de \em{et~al.}(2006)Albar{\`e}de, Scherer, Blichert-Toft,
  Rosing, Simionovici, and Bizzarro]{albarede2006gamma}
Albar{\`e}de, F.; Scherer, E.E.; Blichert-Toft, J.; Rosing, M.; Simionovici,
  A.; Bizzarro, M.
\newblock $\gamma$-ray irradiation in the early Solar System and the conundrum
  of the 176Lu decay constant.
\newblock {\em Geochimica et Cosmochimica Acta} {\bf 2006}, {\em
  70},~1261--1270.

\bibitem[Shafer \em{et~al.}(2010)Shafer, Brandon, Lapen, Righter, Peslier, and
  Beard]{shafer2010trace}
Shafer, J.; Brandon, A.; Lapen, T.; Righter, M.; Peslier, A.; Beard, B.
\newblock Trace element systematics and 147Sm--143Nd and 176Lu--176Hf ages of
  Larkman Nunatak 06319: Closed-system fractional crystallization of an
  enriched shergottite magma.
\newblock {\em Geochimica et Cosmochimica Acta} {\bf 2010}, {\em
  74},~7307--7328.

\bibitem[Bloch and Ganguly(2015)]{bloch2015176}
Bloch, E.; Ganguly, J.
\newblock 176 Lu--176 Hf geochronology of garnet II: numerical simulations of
  the development of garnet--whole-rock 176 Lu--176 Hf isochrons and a new
  method for constraining the thermal history of metamorphic rocks.
\newblock {\em Contributions to Mineralogy and Petrology} {\bf 2015}, {\em
  169},~14.

\bibitem[Thielemann \em{et~al.}(2011)Thielemann, Arcones, K{\"a}ppeli,
  Liebend{\"o}rfer, Rauscher, Winteler, Fr{\"o}hlich, Dillmann, Fischer,
  Martinez-Pinedo, et~al.]{thielemann2011astrophysical}
Thielemann, F.K.; Arcones, A.; K{\"a}ppeli, R.; Liebend{\"o}rfer, M.; Rauscher,
  T.; Winteler, C.; Fr{\"o}hlich, C.; Dillmann, I.; Fischer, T.;
  Martinez-Pinedo, G.; others.
\newblock What are the astrophysical sites for the r-process and the production
  of heavy elements?
\newblock {\em Progress in Particle and Nuclear Physics} {\bf 2011}, {\em
  66},~346--353.

\bibitem[{Fujimoto} and {Hashimoto}(2020)]{Fujimoto2020}
{Fujimoto}, S.i.; {Hashimoto}, M.a.
\newblock {The impact of isomers on a kilonova associated with neutron star
  mergers}.
\newblock {\em \mnras} {\bf 2020}, {\em 493},~L103--L107,
  \href{http://xxx.lanl.gov/abs/2001.10668}{{\normalfont
  [arXiv:astro-ph.HE/2001.10668]}}.
\newblock
  doi:{\changeurlcolor{black}\href{https://doi.org/10.1093/mnrasl/slaa016}{\detokenize{10.1093/mnrasl/slaa016}}}.

\bibitem[{Sprouse} \em{et~al.}(2021){Sprouse}, {Misch}, and
  {Mumpower}]{sprouse2021jade}
{Sprouse}, T.M.; {Misch}, G.W.; {Mumpower}, M.R.
\newblock {Radioactive decay of $r$-process nuclei: isochronic evolution}.
\newblock {\em arXiv preprint arXiv:2102.03846} {\bf 2021}.

\bibitem[{Bhat}(1992)]{ENSDF}
{Bhat}, M.R., {Evaluated Nuclear Structure Data File (ENSDF)}.
\newblock In {\em Nuclear Data for Science and Technology. Series: Research
  Reports in Physics};  1992; pp. 817--821.
\newblock
  doi:{\changeurlcolor{black}\href{https://doi.org/10.1007/978-3-642-58113-7_227}{\detokenize{10.1007/978-3-642-58113-7_227}}}.

\bibitem[Weisskopf and Wigner(1930)]{Weisskopf1930}
Weisskopf, V.; Wigner, E.P.
\newblock Calculation of the natural brightness of spectral lines on the basis
  of Dirac's theory.
\newblock {\em Z. Phys.} {\bf 1930}, {\em 63},~54--73.

\bibitem[{Lippuner} and {Roberts}(2015)]{lippuner2015heating}
{Lippuner}, J.; {Roberts}, L.F.
\newblock {r-process Lanthanide Production and Heating Rates in Kilonovae}.
\newblock {\em \apj} {\bf 2015}, {\em 815},~82,
  \href{http://xxx.lanl.gov/abs/1508.03133}{{\normalfont
  [arXiv:astro-ph.HE/1508.03133]}}.
\newblock
  doi:{\changeurlcolor{black}\href{https://doi.org/10.1088/0004-637X/815/2/82}{\detokenize{10.1088/0004-637X/815/2/82}}}.

\bibitem[{Zhu} \em{et~al.}(2018){Zhu}, {Wollaeger}, {Vassh}, {Surman},
  {Sprouse}, {Mumpower}, {M{\"o}ller}, {McLaughlin}, {Korobkin}, {Kawano},
  {Jaffke}, {Holmbeck}, {Fryer}, {Even}, {Couture}, and {Barnes}]{zhu2018cf}
{Zhu}, Y.; {Wollaeger}, R.T.; {Vassh}, N.; {Surman}, R.; {Sprouse}, T.M.;
  {Mumpower}, M.R.; {M{\"o}ller}, P.; {McLaughlin}, G.C.; {Korobkin}, O.;
  {Kawano}, T.; {Jaffke}, P.J.; {Holmbeck}, E.M.; {Fryer}, C.L.; {Even}, W.P.;
  {Couture}, A.J.; {Barnes}, J.
\newblock {Californium-254 and Kilonova Light Curves}.
\newblock {\em \apjl} {\bf 2018}, {\em 863},~L23,
  \href{http://xxx.lanl.gov/abs/1806.09724}{{\normalfont
  [arXiv:astro-ph.HE/1806.09724]}}.
\newblock
  doi:{\changeurlcolor{black}\href{https://doi.org/10.3847/2041-8213/aad5de}{\detokenize{10.3847/2041-8213/aad5de}}}.

\bibitem[{Horowitz} \em{et~al.}(2019){Horowitz}, {Arcones}, {C{\^o}t{\'e}},
  {Dillmann}, {Nazarewicz}, {Roederer}, {Schatz}, {Aprahamian}, {Atanasov},
  {Bauswein}, {Beers}, {Bliss}, {Brodeur}, {Clark}, {Frebel}, {Foucart},
  {Hansen}, {Just}, {Kankainen}, {McLaughlin}, {Kelly}, {Liddick}, {Lee},
  {Lippuner}, {Martin}, {Mendoza-Temis}, {Metzger}, {Mumpower}, {Perdikakis},
  {Pereira}, {O'Shea}, {Reifarth}, {Rogers}, {Siegel}, {Spyrou}, {Surman},
  {Tang}, {Uesaka}, and {Wang}]{Horowitz2019}
{Horowitz}, C.J.; {Arcones}, A.; {C{\^o}t{\'e}}, B.; {Dillmann}, I.;
  {Nazarewicz}, W.; {Roederer}, I.U.; {Schatz}, H.; {Aprahamian}, A.;
  {Atanasov}, D.; {Bauswein}, A.; {Beers}, T.C.; {Bliss}, J.; {Brodeur}, M.;
  {Clark}, J.A.; {Frebel}, A.; {Foucart}, F.; {Hansen}, C.J.; {Just}, O.;
  {Kankainen}, A.; {McLaughlin}, G.C.; {Kelly}, J.M.; {Liddick}, S.N.; {Lee},
  D.M.; {Lippuner}, J.; {Martin}, D.; {Mendoza-Temis}, J.; {Metzger}, B.D.;
  {Mumpower}, M.R.; {Perdikakis}, G.; {Pereira}, J.; {O'Shea}, B.W.;
  {Reifarth}, R.; {Rogers}, A.M.; {Siegel}, D.M.; {Spyrou}, A.; {Surman}, R.;
  {Tang}, X.; {Uesaka}, T.; {Wang}, M.
\newblock {r-process nucleosynthesis: connecting rare-isotope beam facilities
  with the cosmos}.
\newblock {\em Journal of Physics G Nuclear Physics} {\bf 2019}, {\em
  46},~083001,  \href{http://xxx.lanl.gov/abs/1805.04637}{{\normalfont
  [arXiv:astro-ph.SR/1805.04637]}}.
\newblock
  doi:{\changeurlcolor{black}\href{https://doi.org/10.1088/1361-6471/ab0849}{\detokenize{10.1088/1361-6471/ab0849}}}.

\bibitem[{C{\^o}t{\'e}} \em{et~al.}(2021){C{\^o}t{\'e}}, {Eichler}, {Yag{\"u}e
  L{\'o}pez}, {Vassh}, {Mumpower}, {Vil{\'a}gos}, {So{\'o}s}, {Arcones},
  {Sprouse}, {Surman}, {Pignatari}, {Pet{\H{o}}}, {Wehmeyer}, {Rauscher}, and
  {Lugaro}]{Cote2021}
{C{\^o}t{\'e}}, B.; {Eichler}, M.; {Yag{\"u}e L{\'o}pez}, A.; {Vassh}, N.;
  {Mumpower}, M.R.; {Vil{\'a}gos}, B.; {So{\'o}s}, B.; {Arcones}, A.;
  {Sprouse}, T.M.; {Surman}, R.; {Pignatari}, M.; {Pet{\H{o}}}, M.K.;
  {Wehmeyer}, B.; {Rauscher}, T.; {Lugaro}, M.
\newblock {$^{129}$I and $^{247}$Cm in meteorites constrain the last
  astrophysical source of solar r-process elements}.
\newblock {\em Science} {\bf 2021}, {\em 371},~945--948,
  \href{http://xxx.lanl.gov/abs/2006.04833}{{\normalfont
  [arXiv:astro-ph.SR/2006.04833]}}.
\newblock
  doi:{\changeurlcolor{black}\href{https://doi.org/10.1126/science.aba1111}{\detokenize{10.1126/science.aba1111}}}.

\bibitem[Gao \em{et~al.}(2020)Gao, Zegers, Zamora, Bazin, Brown, Bender,
  Crawford, Engel, Falduto, Gade, et~al.]{gao2020gamow}
Gao, B.; Zegers, R.; Zamora, J.; Bazin, D.; Brown, B.; Bender, P.; Crawford,
  H.; Engel, J.; Falduto, A.; Gade, A.; others.
\newblock Gamow-Teller transitions to Zr 93 via the Nb 93 (t, He 3+ $\gamma$)
  reaction at 115 MeV/u and its application to the stellar electron-capture
  rates.
\newblock {\em Physical Review C} {\bf 2020}, {\em 101},~014308.

\bibitem[Tan \em{et~al.}(2020)Tan, Liu, Wang, Li, and Sun]{tan2020novel}
Tan, L.; Liu, Y.X.; Wang, L.J.; Li, Z.; Sun, Y.
\newblock A novel method for stellar electron-capture rates of excited nuclear
  states.
\newblock {\em Physics Letters B} {\bf 2020}, {\em 805},~135432.

\bibitem[Richards \em{et~al.}(1982)Richards, Tucker, and
  Srivastava]{richards1982technetium}
Richards, P.; Tucker, W.D.; Srivastava, S.C.
\newblock Technetium-99m: an historical perspective.
\newblock {\em International Journal of Applied Radiation and Isotopes} {\bf
  1982}, {\em 33},~793--799.

\bibitem[G{\'o}rska \em{et~al.}(2009)G{\'o}rska, C{\'a}ceres, Grawe,
  Pf{\"u}tzner, Jungclaus, Pietri, Werner-Malento, Podoly{\'a}k, Regan,
  Rudolph, et~al.]{gorska2009evolution}
G{\'o}rska, M.; C{\'a}ceres, L.; Grawe, H.; Pf{\"u}tzner, M.; Jungclaus, A.;
  Pietri, S.; Werner-Malento, E.; Podoly{\'a}k, Z.; Regan, P.; Rudolph, D.;
  others.
\newblock Evolution of the N= 82 shell gap below 132Sn inferred from core
  excited states in 131In.
\newblock {\em Physics Letters B} {\bf 2009}, {\em 672},~313--316.

\bibitem[Jones \em{et~al.}(2010)Jones, Adekola, Bardayan, Blackmon, Chae,
  Chipps, Cizewski, Erikson, Harlin, Hatarik, et~al.]{jones2010magic}
Jones, K.; Adekola, A.S.; Bardayan, D.W.; Blackmon, J.C.; Chae, K.; Chipps, K.;
  Cizewski, J.; Erikson, L.; Harlin, C.; Hatarik, R.; others.
\newblock The magic nature of 132 Sn explored through the single-particle
  states of 133 Sn.
\newblock {\em Nature} {\bf 2010}, {\em 465},~454--457.

\bibitem[Jin \em{et~al.}(2011)Jin, Hasegawa, Tazaki, Kaneko, Sun,
  et~al.]{jin2011large}
Jin, H.; Hasegawa, M.; Tazaki, S.; Kaneko, K.; Sun, Y.; others.
\newblock Large-scale shell-model calculation with core excitations for
  neutron-rich nuclei beyond 132 Sn.
\newblock {\em Physical Review C} {\bf 2011}, {\em 84},~044324.

\bibitem[Wang \em{et~al.}(2013)Wang, Sun, Jin, Kaneko, and
  Tazaki]{wang2013structure}
Wang, H.K.; Sun, Y.; Jin, H.; Kaneko, K.; Tazaki, S.
\newblock Structure analysis for hole-nuclei close to 132 Sn by a large-scale
  shell-model calculation.
\newblock {\em Physical Review C} {\bf 2013}, {\em 88},~054310.

\bibitem[Surman \em{et~al.}(1997)Surman, Engel, Bennett, and Meyer]{Surman1997}
Surman, R.; Engel, J.; Bennett, J.R.; Meyer, B.S.
\newblock Source of the Rare-Earth Element Peak in $\mathit{r}$-Process
  Nucleosynthesis.
\newblock {\em Phys. Rev. Lett.} {\bf 1997}, {\em 79},~1809--1812.
\newblock
  doi:{\changeurlcolor{black}\href{https://doi.org/10.1103/PhysRevLett.79.1809}{\detokenize{10.1103/PhysRevLett.79.1809}}}.

\bibitem[Mumpower \em{et~al.}(2012)Mumpower, McLaughlin, and
  Surman]{mumpower2012influence}
Mumpower, M.R.; McLaughlin, G.C.; Surman, R.
\newblock Influence of neutron capture rates in the rare earth region on the
  r-process abundance pattern.
\newblock {\em Physical Review C} {\bf 2012}, {\em 86},~035803.

\bibitem[{Vilen} \em{et~al.}(2018){Vilen}, {Kelly}, {Kankainen}, {Brodeur},
  {Aprahamian}, {Canete}, {Eronen}, {Jokinen}, {Kuta}, {Moore}, {Mumpower},
  {Nesterenko}, {Penttil{\"a}}, {Pohjalainen}, {Porter}, {Rinta-Antila},
  {Surman}, {Voss}, and {{\'n}yst{\"o}}]{Vilen2018}
{Vilen}, M.; {Kelly}, J.M.; {Kankainen}, A.; {Brodeur}, M.; {Aprahamian}, A.;
  {Canete}, L.; {Eronen}, T.; {Jokinen}, A.; {Kuta}, T.; {Moore}, I.D.;
  {Mumpower}, M.R.; {Nesterenko}, D.A.; {Penttil{\"a}}, H.; {Pohjalainen}, I.;
  {Porter}, W.S.; {Rinta-Antila}, S.; {Surman}, R.; {Voss}, A.;
  {{\'n}yst{\"o}}, J.
\newblock {Precision Mass Measurements on Neutron-Rich Rare-Earth Isotopes at
  JYFLTRAP: Reduced Neutron Pairing and Implications for r -Process
  Calculations}.
\newblock {\em \prl} {\bf 2018}, {\em 120},~262701,
  \href{http://xxx.lanl.gov/abs/1801.08940}{{\normalfont
  [arXiv:nucl-ex/1801.08940]}}.
\newblock
  doi:{\changeurlcolor{black}\href{https://doi.org/10.1103/PhysRevLett.120.262701}{\detokenize{10.1103/PhysRevLett.120.262701}}}.

\bibitem[{Orford} \em{et~al.}(2018){Orford}, {Vassh}, {Clark}, {McLaughlin},
  {Mumpower}, {Savard}, {Surman}, {Aprahamian}, {Buchinger}, {Burkey},
  {Gorelov}, {Hirsh}, {Klimes}, {Morgan}, {Nystrom}, and {Sharma}]{Orford2018}
{Orford}, R.; {Vassh}, N.; {Clark}, J.A.; {McLaughlin}, G.C.; {Mumpower}, M.R.;
  {Savard}, G.; {Surman}, R.; {Aprahamian}, A.; {Buchinger}, F.; {Burkey},
  M.T.; {Gorelov}, D.A.; {Hirsh}, T.Y.; {Klimes}, J.W.; {Morgan}, G.E.;
  {Nystrom}, A.; {Sharma}, K.S.
\newblock {Precision Mass Measurements of Neutron-Rich Neodymium and Samarium
  Isotopes and Their Role in Understanding Rare-Earth Peak Formation}.
\newblock {\em \prl} {\bf 2018}, {\em 120},~262702.
\newblock
  doi:{\changeurlcolor{black}\href{https://doi.org/10.1103/PhysRevLett.120.262702}{\detokenize{10.1103/PhysRevLett.120.262702}}}.

\bibitem[{Vilen} \em{et~al.}(2020){Vilen}, {Kelly}, {Kankainen}, {Brodeur},
  {Aprahamian}, {Canete}, {de Groote}, {de Roubin}, {Eronen}, {Jokinen},
  {Moore}, {Mumpower}, {Nesterenko}, {O'Brien}, {Perdomo}, {Penttil{\"a}},
  {Reponen}, {Rinta-Antila}, and {Surman}]{Vilen2020}
{Vilen}, M.; {Kelly}, J.M.; {Kankainen}, A.; {Brodeur}, M.; {Aprahamian}, A.;
  {Canete}, L.; {de Groote}, R.P.; {de Roubin}, A.; {Eronen}, T.; {Jokinen},
  A.; {Moore}, I.D.; {Mumpower}, M.R.; {Nesterenko}, D.A.; {O'Brien}, J.;
  {Perdomo}, A.P.; {Penttil{\"a}}, H.; {Reponen}, M.; {Rinta-Antila}, S.;
  {Surman}, R.
\newblock {Exploring the mass surface near the rare-earth abundance peak via
  precision mass measurements at JYFLTRAP}.
\newblock {\em \prc} {\bf 2020}, {\em 101},~034312,
  \href{http://xxx.lanl.gov/abs/1908.05043}{{\normalfont
  [arXiv:nucl-ex/1908.05043]}}.
\newblock
  doi:{\changeurlcolor{black}\href{https://doi.org/10.1103/PhysRevC.101.034312}{\detokenize{10.1103/PhysRevC.101.034312}}}.

\bibitem[{Vassh} \em{et~al.}(2021){Vassh}, {McLaughlin}, {Mumpower}, and
  {Surman}]{Vassh2021}
{Vassh}, N.; {McLaughlin}, G.C.; {Mumpower}, M.R.; {Surman}, R.
\newblock {Markov Chain Monte Carlo Predictions of Neutron-rich Lanthanide
  Properties as a Probe of r-process Dynamics}.
\newblock {\em \apj} {\bf 2021}, {\em 907},~98,
  \href{http://xxx.lanl.gov/abs/2006.04322}{{\normalfont
  [arXiv:nucl-th/2006.04322]}}.
\newblock
  doi:{\changeurlcolor{black}\href{https://doi.org/10.3847/1538-4357/abd035}{\detokenize{10.3847/1538-4357/abd035}}}.

\bibitem[{Mumpower} \em{et~al.}(2016){Mumpower}, {McLaughlin}, {Surman}, and
  {Steiner}]{Mumpower2016}
{Mumpower}, M.R.; {McLaughlin}, G.C.; {Surman}, R.; {Steiner}, A.W.
\newblock {The Link between Rare-Earth Peak Formation and the Astrophysical
  Site of the R Process}.
\newblock {\em \apj} {\bf 2016}, {\em 833},~282,
  \href{http://xxx.lanl.gov/abs/1603.02600}{{\normalfont
  [arXiv:nucl-th/1603.02600]}}.
\newblock
  doi:{\changeurlcolor{black}\href{https://doi.org/10.3847/1538-4357/833/2/282}{\detokenize{10.3847/1538-4357/833/2/282}}}.

\bibitem[{Mumpower} \em{et~al.}(2017){Mumpower}, {McLaughlin}, {Surman}, and
  {Steiner}]{Mumpower2017}
{Mumpower}, M.R.; {McLaughlin}, G.C.; {Surman}, R.; {Steiner}, A.W.
\newblock {Reverse engineering nuclear properties from rare earth abundances in
  the r process}.
\newblock {\em Journal of Physics G Nuclear Physics} {\bf 2017}, {\em
  44},~034003,  \href{http://xxx.lanl.gov/abs/1609.09858}{{\normalfont
  [arXiv:nucl-th/1609.09858]}}.
\newblock
  doi:{\changeurlcolor{black}\href{https://doi.org/10.1088/1361-6471/44/3/034003}{\detokenize{10.1088/1361-6471/44/3/034003}}}.

\bibitem[Hartley \em{et~al.}(2018)Hartley, Kondev, Orford, Clark, Savard,
  Ayangeakaa, Bottoni, Buchinger, Burkey, Carpenter, et~al.]{hartley2018masses}
Hartley, D.; Kondev, F.; Orford, R.; Clark, J.; Savard, G.; Ayangeakaa, A.;
  Bottoni, S.; Buchinger, F.; Burkey, M.; Carpenter, M.; others.
\newblock Masses and $\beta$-Decay Spectroscopy of Neutron-Rich Odd-Odd Eu 160,
  162 Nuclei: Evidence for a Subshell Gap with Large Deformation at N= 98.
\newblock {\em Physical review letters} {\bf 2018}, {\em 120},~182502.

\bibitem[Liu \em{et~al.}(2020)Liu, Lv, Sun, and Kondev]{liu2020changes}
Liu, Y.X.; Lv, C.J.; Sun, Y.; Kondev, F.G.
\newblock Changes of deformed shell gaps at N~ 100 in light rare-earth,
  neutron-rich nuclei.
\newblock {\em Journal of Physics G: Nuclear and Particle Physics} {\bf 2020},
  {\em 47},~055108.

\bibitem[Abbott \em{et~al.}(2017)Abbott, Abbott, Abbott, Acernese, Ackley,
  Adams, Adams, Addesso, Adhikari, Adya, et~al.]{2017ApJ...848L..12A}
Abbott, B.P.; Abbott, R.; Abbott, T.; Acernese, F.; Ackley, K.; Adams, C.;
  Adams, T.; Addesso, P.; Adhikari, R.; Adya, V.; others.
\newblock {Multi-messenger Observations of a Binary Neutron Star Merger}.
\newblock {\em \apjl} {\bf 2017}, {\em 848},~L12,
  \href{http://xxx.lanl.gov/abs/1710.05833}{{\normalfont
  [arXiv:astro-ph.HE/1710.05833]}}.
\newblock
  doi:{\changeurlcolor{black}\href{https://doi.org/10.3847/2041-8213/aa91c9}{\detokenize{10.3847/2041-8213/aa91c9}}}.

\end{thebibliography}

%%%%%%%%%%%%%%%%%%%%%%%%%%%%%%%%%%%%%%%%%%
\end{document}